\documentclass[11pt]{article}
\usepackage{appendix}
\usepackage{amssymb}
\usepackage{amsmath}
\usepackage{epsfig}
\usepackage{url}
\usepackage[]{algorithm2e}

\pdfoutput=1

\newcommand\blfootnote[1]{%
  \begingroup
  \renewcommand\thefootnote{}\footnote{#1}%
  \addtocounter{footnote}{-1}%
  \endgroup
}

\begin{document}
\newtheorem{theorem}{Theorem}[section]
\newtheorem{corollary}[theorem]{Corollary}
\newtheorem{lemma}[theorem]{Lemma}

\newtheorem{definition}{Definition}

\newcommand{\define}{\stackrel{\triangle}{=}}
\allowdisplaybreaks

\def\QED{\mbox{\rule[0pt]{1.5ex}{1.5ex}}}
\def\proof{\noindent\hspace{2em}{\it Proof: }}

\date{}
\title{Topological Interference Management\\ through Index Coding} 
\author{Syed A. Jafar\vspace{0.1cm}\\
{\small Center for Pervasive Communications and Computing}\\
{\small Department of Electrical Engineering and Computer Science}\\
{\small University of California Irvine, Irvine, California, 92697}\\
       }
\maketitle
         \blfootnote{Presented in part at IEEE GLOBECOM 2012. A preliminary version of this work appears as ArXiv:1203.2384.}    

\vspace{-1cm}
\begin{abstract}
While much recent progress on interference networks has come about under the assumption of abundant channel state information at the transmitters (CSIT), a complementary perspective is sought in this work through the study of interference networks with no CSIT except a coarse knowledge of the topology of the network that only allows a distinction between weak and significant channels and no further knowledge  of the channel coefficients' realizations. Modeled as a degrees-of-freedom (DoF) study of a partially connected interference network with no CSIT, the problem is found to have a counterpart in the capacity analysis of wired networks with arbitrary linear network coding  at intermediate nodes, under the assumption that the sources are aware only of the end to end topology of the network. The wireless (wired) network DoF (capacity)  region, expressed in dimensionless units as a multiple of the DoF (capacity) of a single point to point channel (link), is found to be bounded above by the capacity of an index coding problem where the antidotes graph is the complement of the interference graph of the original network and the bottleneck link capacity is normalized to unity. The problems are shown to be equivalent under linear solutions over the same field. An interference alignment perspective is then used to translate the existing index coding solutions into   wireless network DoF (wired network capacity) solutions, as well as to find new  and unified solutions to different classes of all three problems. For networks with $K$ messages, a study of the extremes -- when each message achieves half the cake, and when each message can achieve no more than $1/K$ of the cake, reveals the necessary and sufficient conditions for each, in terms of alignment graphs and demand graphs, respectively. Half the cake per message is achievable if and only if the alignment graph has no internal conflicts. No more than $1/K$ of the cake is achievable if and only if the network can be relaxed into a $K$-unicast setting with an acyclic demand graph, possibly by eliminating some demands. For half-rate-feasible networks, best case capacity (DoF) improvements over the best fractional orthogonal scheduling (TDMA) and fractional partition multicast (CDMA) solutions are explored for multiple groupcast and multiple unicast settings. For intermediate cases where neither half the cake, nor $1/K$ of the cake per message is capacity (DoF) optimal, the interference alignment perspective is used to characterize the symmetric capacity (DoF) of all cases where each  alignment set either does not contain a cycle or does not contain a fork. A study of linear feasible rates shows duality properties that are used to extend the scope of previous results. For wireless networks, extensions to multiple antenna networks are made in symmetric settings where all nodes are equipped with the same number of antennas. The study of certain topologies of interest, motivated by cellular networks reveals interesting aligned frequency reuse patterns.
\end{abstract}
\newpage
\section{Introduction}
Recent years have seen remarkable advances in our understanding of the information theoretic capacity limits of  interference networks, albeit primarily under the assumption  of abundant channel state information at the transmitters (CSIT). While this has revealed ingenious ways to exploit the finer aspects of CSIT, the results have been difficult to translate into practice where CSIT is rarely available to the extent that is assumed. This is a problem not only for wireless interference networks but also for wired networks with inter-session linear network coding, where the abundant CSIT assumption corresponds to the knowledge of all the transfer functions, comprised of coding coefficients inside the network. The motivation for this work comes from both wireless and wired interference network perspectives.

\subsection{Wireless Interference Networks}
Recognizing the difficulty of translating theoretical insights based on abundant CSIT assumptions into practice, researchers have started exploring settings with relaxed CSIT assumptions. The state of affairs is exemplified by the evolution of the  idea of interference alignment which was initially studied under the assumption of perfect  CSIT \cite{Cadambe_Jafar_int, Nazer_Gastpar_Jafar_Vishwanath} and has since then been studied under a variety of relaxed CSIT assumptions including  compound channels \cite{Gou_Jafar_Wang}, delayed CSIT \cite{Maddah_Tse, Maleki_Jafar_Shamai}, mixed CSIT \cite{Gou_Jafar, Sheng_Kobayashi_Gesbert_Yi},  alternating CSIT \cite{Tandon_Jafar_Shamai_Poor}, and CSIT comprised of coherence patterns \cite{Jafar_corr, Wang_Gou_Jafar}. From this research have emerged clever  interference alignment schemes that take advantage of  {channel}-variations \cite{Cadambe_Jafar_int},  complementary {channel} states \cite{Nazer_Gastpar_Jafar_Vishwanath}, outdated channel states \cite{Maddah_Tse}, quaternionic structures inherent in complex {channels} \cite{Cadambe_Jafar_Wang}, naturally existing {channel} correlations and coherence patterns \cite{Jafar_corr},  desirable {channel} coherence patterns enforced through antenna switching \cite{Wang_Gou_Jafar}, the  linear \cite{Yetis_Gou_Jafar_Kayran_TSP}, algebraic \cite{Annapureddy_ElGamal_Veeravalli} and rational independence \cite{Motahari_Gharan_Khandani_real} of {channels}, and the fundamental information dimension of given {channel} realizations \cite{Wu_Shamai_Verdu}. Nevertheless, much of the theoretical insights remain too fragile so far to be translated directly into practice. 

\begin{figure}[h]
\begin{center}
\includegraphics[width=4.5in]{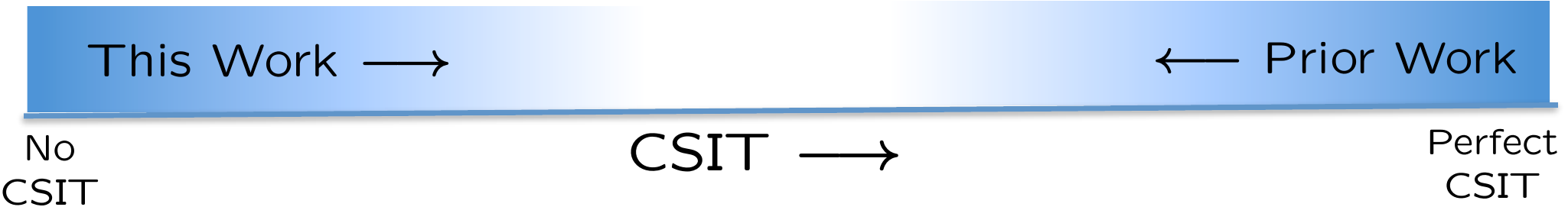}
\caption{\small \it Motivation for this work and relationship to prior work}\label{fig:direction}
\end{center}
\end{figure}

This work is motivated by a complementary perspective, illustrated in Fig. \ref{fig:direction}. Instead of starting with abundant CSIT and then incrementally relaxing the CSIT assumptions, what if one started with no CSIT and then incrementally increased the available CSIT ---  how far could one go toward the center (practical settings where CSIT is neither extremely minimal nor overly abundant) in Fig. \ref{fig:direction} before the problem becomes intractable? 

Consider a $K$ user wireless interference network, defined by input-output relationships:
\begin{eqnarray*}
y_1(n)&=&h_{11}x_1(n)+h_{12}x_2(n)+\cdots+h_{1K}x_K(n)+z_1(n)\\
y_2(n)&=&h_{21}x_1(n)+h_{22}x_2(n)+\cdots+h_{2K}x_K(n)+z_2(n)\\
\vdots&\vdots&\vdots\\
y_K(n)&=&h_{K1}x_1(n)+h_{K2}x_2(n)+\cdots+h_{KK}x_K(n)+z_K(n)
\end{eqnarray*}
where over the $n^{th}$ channel use, $x_j(n)$ is the symbol transmitted by transmitter $j$, $h_{ij}$ is the constant channel coefficient between transmitter $j$ and receiver $i$,  $z_i(n)\sim\mathcal{N}^c(0,N_o)$ is the additive white Gaussian noise (AWGN) at receiver $i$ and $y_i(n)$ is the symbol received by receiver $i$, with $i,j\in\{1,2,\cdots, K\}$.  All symbols are complex.

As a starting point, suppose absolutely no CSIT is available. It is easy to see that this is a degenerate setting, because with absolutely no CSIT, i.e., no knowledge of  even the desired channels, no guarantees of reliable communication can be made, and the capacity is zero even in the absence of interference.

Evidently, to make the problem non-degenerate, at least there must be  rate guarantees for the \emph{desired} channel in the \emph{absence of interference}. Suppose the transmit power constraints are set such that each user is able to support a certain SNR value on the desired link in the \emph{absence of interference}. That is,
\begin{eqnarray}
\frac{\left|h_{ii}\right|^2P_i}{N_o}&\geq&\mbox{SNR}, ~~~~~~\forall i\in\{1,2,\cdots, K\},\label{eq:desiredpower}
\end{eqnarray}
where $P_i$ is the average transmit power constraint for Transmitter $i$, $
\mbox{E}\left[\frac{1}{N}\sum_{n=1}^N\left|x_{i}(n)\right|^2\right]\leq P_i$. 
$N$ is the length of codewords and the expectation is over the messages. Thus, the power constraints are chosen such that, in the absence of interference, each user can achieve a rate $\log(1+\mbox{SNR})$. This is  a natural assumption if all the users desire similar rates. Suppose beyond the  interference-free SNR guarantee of (\ref{eq:desiredpower}),  no CSIT is available. The transmitters have absolutely no knowledge of the strengths of the channels to undesired receivers. 

What is the  capacity of this interference network? As shown in Section \ref{sec:1/K}, it turns out that the  capacity of this interference network can be precisely determined, and corresponds to the simple time-division-multiple-access (TDMA) scheme, where the users take turns so that each user transmits for a fraction $1/K$ of the time, and achieves a rate $\frac{1}{K}\log\left(1+K\mbox{SNR}\right)$ which is the symmetric capacity of this network. While it is remarkable that the exact capacity can be found in this case, the optimality of TDMA is perhaps not very interesting. 

Next, let us move further to the right in terms of Fig. \ref{fig:direction} by incrementally increasing CSIT. In addition to the  interference-free SNR guarantees for the desired channels, as in (\ref{eq:desiredpower}), let us allow just one bit of CSIT about the interference channel strengths. A natural choice for this one bit CSIT could be as follows. The receivers compare the nominal received power from the undesired links to a pre-chosen threshold value, which is effectively the  acceptable noise floor,  and assign a `$0$' to all those (weak) links whose collective contribution is below the  noise floor. All other (significant) links are assigned the value `1'. These assignments comprise the 1-bit CSIT  of the undesired channel strengths. Note that this is 1-bit CSIT for the entire duration of communication, and not 1-bit per channel use, i.e.,  the assignments are permanent, based on the nominal values of average received signal strengths, and are not dependent on the actual transmit powers which may vary with time, e.g., if the optimal scheme is such that the users do not transmit all the time. So, with this 1-bit CSIT of interference carrying links, what is the capacity of this network? Unlike the previous cases which turned out to be  straightforward, we will see that this question turns out to be most interesting and highly non-trivial in general. 

\subsubsection{Example}
\begin{figure}[h]
\begin{centering}
\includegraphics[width=4.5in]{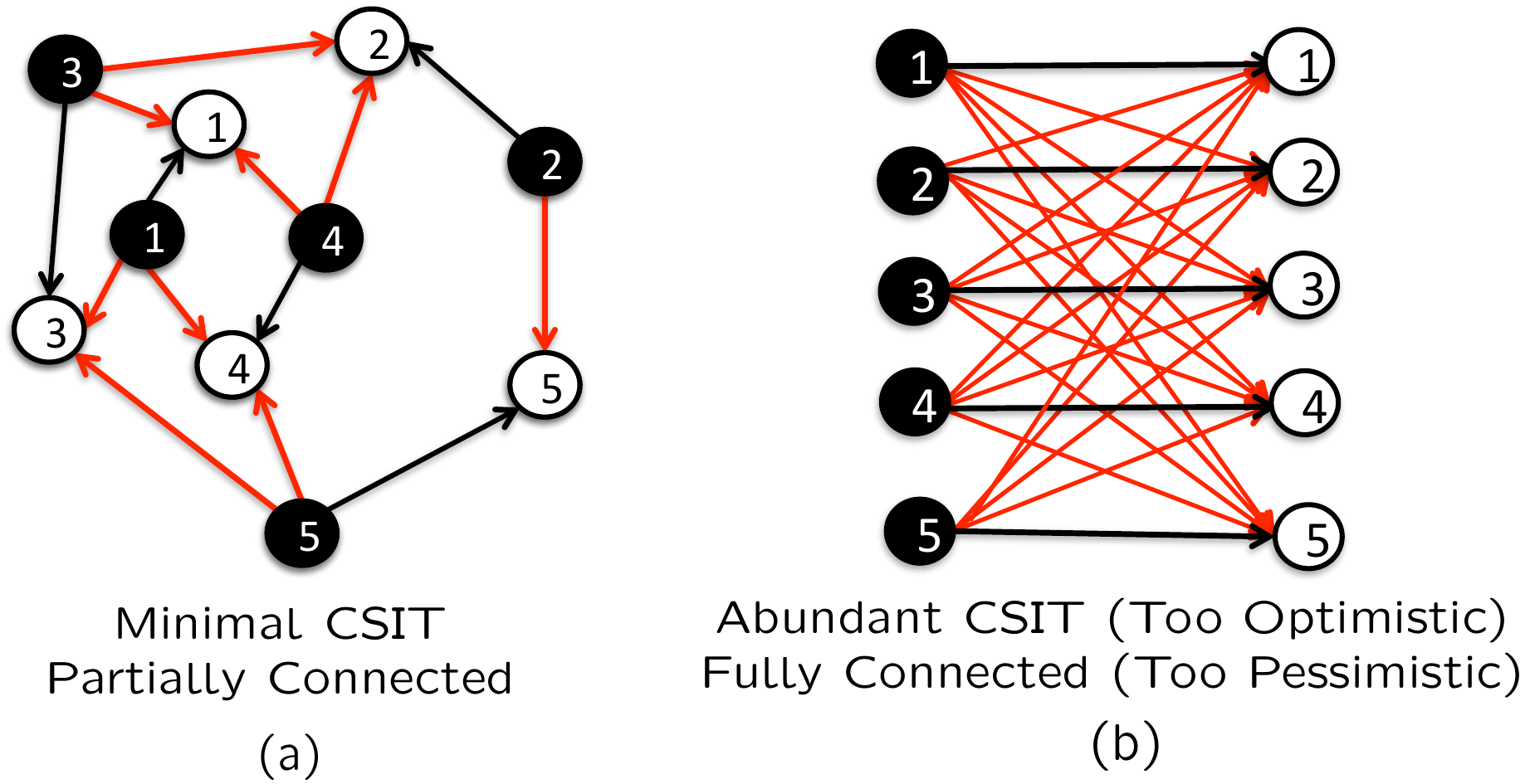}
\caption{\small\it  (a)Partially connected network with no CSIT (except topology).  (b) Fully connected network with abundant CSIT.}\label{fig:contrast}
\end{centering}
\end{figure}

Under the assumptions stated above, the interference network is comprised of three kinds of channels --- desired channels,  significant interference channels, and weak interference channels. Graphically, if we represent all sources as black nodes, all destinations as white nodes, all desired channels as black edges, all significant interference channels as red edges, and omit the weak interference channels, both to avoid cluttering the graph, and to emphasize their `insignificant' character, then the resulting network might look  like the simple example shown in Fig. \ref{fig:contrast}(a), comprised of $K=5$ users. From the transmitters' perspective, the CSIT consists of interference-free SNR guarantees of (\ref{eq:desiredpower}) for the desired channels, and the following information about undesired channels
\begin{eqnarray*}
\left|h_{12}\right|^2P_2+\left|h_{15}\right|^2P_5&\leq&N_o\\
\left|h_{21}\right|^2P_1+\left|h_{25}\right|^2P_5&\leq&N_o\\
\left|h_{32}\right|^2P_2+\left|h_{34}\right|^2P_4&\leq&N_o\\
\left|h_{42}\right|^2P_2+\left|h_{43}\right|^2P_3&\leq&N_o\\
\left|h_{51}\right|^2P_1+\left|h_{53}\right|^2P_3+\left|h_{54}\right|^2P_4&\leq&N_o
\end{eqnarray*}
which identifies the weak interference channels. All other interference channels are considered significant and are shown as red edges in Fig. \ref{fig:contrast}(a).

We will show that the capacity  for this network can be characterized to within a constant gap as $\frac{1}{2}\log(1+3\mbox{SNR}/8)\leq C \leq \frac{1}{2}\log(1+2\mbox{SNR})$ bits/channel-use per user. The gap between the two bounds is at most 1.2 bits. 
While the details of this example will be explained later, what is important at this point is to highlight the core  problem: In order to obtain the constant gap result, we  will need to first find the degrees of freedom (DoF) of the underlying \emph{partially} connected interference network (where the weak channels are set to zero) with no CSIT except the knowledge of the topology of the network. This is what we call the topological interference management problem for wireless networks.

\subsubsection{Topological Interference Management  for Wireless Networks}
The topological interference management problem, in the wireless setting, refers to the DoF analysis of an interference network where all weak interference channels are set to zero, and where the only CSIT available to the transmitters is the resulting topology of the network, i.e., which channels are zero and which ones are non-zero.   The topological interference management problem corresponding to the example network of Fig. \ref{fig:contrast}(a), is defined by the the input output equations:
\begin{eqnarray}
y_1(n)&=&h_{11}x_1(n)+h_{13}x_3(n)+h_{14}x_4(n)+z_1(n)\\
y_2(n)&=&h_{22}x_2(n)+h_{23}x_3(n)+h_{24}x_4(n)+z_2(n)\\
y_3(n)&=&h_{31}x_1(n)+h_{33}x_3(n)+h_{35}x_5(n)+z_3(n)\\
y_4(n)&=&h_{41}x_1(n)+h_{44}x_4(n)+h_{45}x_5(n)+z_4(n)\\
y_5(n)&=&h_{52}x_2(n)+h_{55}x_5(n)+z_5(n)
\end{eqnarray}
Note that all weak interference channels are eliminated (set to zero) here. For the remaining non-zero channels $h_{ij}$ their values are known to be bounded away from zero but the values themselves are not known to the transmitters. Thus, the only CSIT is comprised of the connectivity, i.e., the topology of the network.

A DoF characterization for such partially connected interference networks with no CSIT (beyond topology knowledge)   is the central question addressed in this paper. For the simple example of Fig. \ref{fig:contrast}(a), the DoF value will turn out to be 1/2 per user, which will lead us to a constant gap capacity characterization. Note that  orthogonal schemes  such as scheduling of non-interfering groups of users, cannot achieve a symmetric DoF higher than 1/3 per user. This is easy to see because User 2 interferes with User 5, User 5 interferes with User 4 and User 4 interferes with User 2. Indeed, even for this simple example, it turns out that interference alignment is needed to achieve the optimal symmetric DoF value of 1/2 per user.

\subsubsection{Practical Significance and Relationship to Prior Work}\label{sec:practical}
Since the topological interference management problem for wireless networks is a DoF problem, it is worthwhile to contrast it with previous DoF studies, and to understand the practical implications. The contrast is drawn pictorially  in Fig. \ref{fig:contrast}. Most prior work is based on 
DoF studies of canonical models like Figure \ref{fig:contrast}(b), where the network is fully connected and abundant CSIT is available. This is problematic on both counts. First, of course, the abundant CSIT assumption is too optimistic, and therefore likely to produce fragile results. Second, the fully connected assumption, which means that all channel coefficients are non-zero, is also problematic. It may seem reasonable at first because technically one could argue that the signal strength never decays to absolutely zero. However, the DoF metric, in addition to its coarse and asymptotic character which makes it useful primarily for first order analysis, implicitly \emph{treats all  non-zero channels as equally strong}, in the sense that any channel with a non-zero constant channel coefficient value can carry exactly 1 DoF, regardless of the magnitude of the constant. Thus, the fully connected model in conjunction with the DoF metric trivializes the underlying topology of the network.

 Having all interference strength  comparable to desired signals is a worst-case  scenario \cite{Jafar_ergodic} rarely encountered in practice. It is not surprising then that the DoF results for fully connected network models, under realistic channel uncertainty, tend to be overly pessimistic, often predicting a dramatic collapse of the DoF \cite{ lozano2012limitations, xu2011accuracy}, when in fact practical networks work fine with even smaller amounts of CSIT.  
Universal physical phenomena such as propagation path loss, shadowing, and fading  render  transmitters and receivers essentially \emph{disconnected} beyond a point. The localized nature of connectivity is the enabling premise for spatial frequency re-use, which is the attribute responsible more than anything else, for the success of existing cellular networks.  To the extent that the differences in  strong vs weak signal strengths have been considered through generalized degrees of freedom (GDoF) studies \cite{Jafar_Vishwanath_GDoF, Niesen_Maddah_Ali_X}, ADT style layered deterministic  models \cite{Avestimehr_Diggavi_Tse, Etkin_Tse_Wang},  partially connected Wyner type models \cite{Shamai_Wyner, Gesbert_cluster, Lapidoth_Levy_Shamai_Wigger}, and low-complexity achievable schemes \cite{Charafeddine_Sezgin_Paulraj,Zhao_Tan_Avestimehr_Diggavi_Pottie}, they are mostly limited to  small networks, multicast settings,  assume abundant CSIT, or make no claims of information theoretic optimality.

The topological aspects are expected to become even more interesting in the future, as supported by the following observations. 1) Increasingly dense  and indoor environments and trends toward customer deployed networks, pico and femto cells, mesh networks, peer-to-peer networks, and the use of directional antennas, all point to increasingly complex connectivity patterns. 2) Spectrum shortage is driving the push toward higher frequency bands (e.g., 60GHz) which experience not only significantly higher path loss and shadowing effects and shorter range of communication, but also a greater variance of the signal strengths with distance, again  highlighting the increasing topological complexity of wireless networks. Evidently, there is a need to explore information theoretically optimal ways to exploit the knowledge of the topology of a wireless  network without relying on finer forms of CSIT. 

\subsection{Wired Networks with Linear Network Coding}
The preceding observations are not restricted to wireless networks. Analogous to the accuracy versus overhead tradeoff of acquiring CSIT in wireless networks, there is a similar  accuracy versus overhead tradeoff in wired network topology discovery, e.g., when the topology discovery is achieved through end-to-end probes, i.e., through network tomography.  For example, layer 3 topology discovery infers the end-to-end connectivity, which is less accurate and carries less overhead than layer 1 topology which attempts to infer the internal link-level topology of the network, which is more accurate and carries much more overhead. While such topological considerations are centerstage in wired networks, traditionally the topology is inferred in terms of orthogonal data pipes, for routing purposes. 

Recently, the advent of network coding has blurred the dichotomy of wired\footnote{By wired networks,  we mean networks of non-interfering capacitated (noise-less) links. Note that while the original wired network is comprised of non-interfering links, because of network coding operations at intermediate nodes, the resulting network does  generally experience interference between flows.} and wireless. With network coding, and linear network coding in particular, performed at the intermediate nodes, an end-to-end wired network behaves  much the same way as a wireless interference network, albeit over finite fields or packets rather than real or complex signals. Interference is introduced by inter-session coding at the relay nodes, which transmit linear combinations of  their incoming symbols on their outgoing links, thus creating an end-to-end linear interference network. It is then natural to try to transfer the emerging interference management principles from wireless to wired networks, and there is recent work aimed at doing just that \cite{Das_Vishwanath_Jafar_Markopoulou, Ramakrishnan_Das_Maleki_Markopoulou_Jafar_Vishwanath, Ganesan_Bavirisetti_Rajan, Cadambe_Jafar_Maleki_Ramchandran_Suh}. However, in transferring insights from wireless networks to wired networks, a key limitation is that the interference alignment schemes for wireless networks are developed for simple topologies, fully connected settings, focusing on channel realizations, whereas wired networks present critical topological features --- nodes are absolutely disconnected if there exists no path between them --- and are often less tied to `channel realizations', since the channel is simply a manifestation of the network coding operations. Thus, the emphasis on \emph{channel} rather than \emph{topology} is also a hurdle in extending the interference management principles from wireless to wired networks. 

\subsubsection{Topological Interference Management for Wired Networks}
The topological interference management problem for wired networks is the natural counterpart of the wireless case: it refers to the capacity analysis of wired single input single output (SISO) networks based on optimal coding/decoding operations at the original source and final destination nodes, while the intermediate network performs linear network coding operations, creating an end-to-end linear interference network. Note that a SISO classification requires that each source has only one outgoing edge and each destination has only one incoming edge. All edges have the same capacity $\log|\mathbb{GF}|$ which allows each edge to carry one symbol from a finite field, i.e., a Galois Field,  denoted as $\mathbb{GF}$, per channel use. The linear network coding operations are performed over the same finite field. Analogous to the assumption of constant channels in the wireless setting, the network coding coefficients are assumed to remain constant throughout the duration of communication. The sources and destinations are not necessarily restricted to perform linear encoding or decoding operations. Indeed, they may use whatever encoding and decoding schemes are information theoretically optimal. Most importantly, the CSIT consists of only the topology of the network, i.e., the transmitters are only aware whether the end-to-end channel coefficients are zero or non-zero values. The following example illustrates the topological interference management problem for wired networks.

\subsubsection{Example}
\begin{figure}[h]
\begin{centering}
\includegraphics[width=3in]{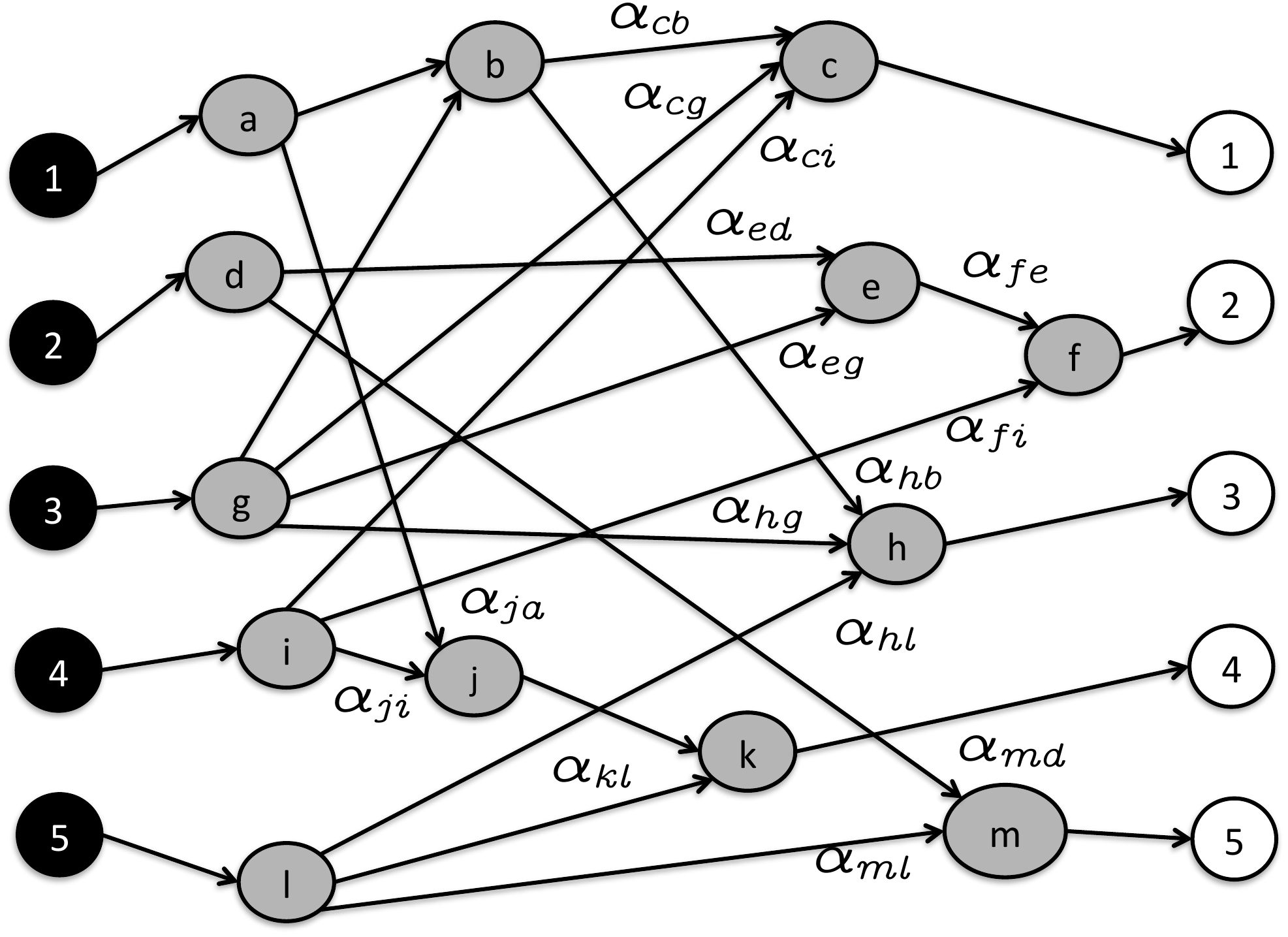}
\caption{\small\it Wired multiple unicast network with linear network coding}\label{fig:wired}
\end{centering}
\end{figure}

Consider the multiple unicast setting illustrated in Fig. \ref{fig:wired} where five sources, shown as black nodes on the left, want to communicate with their corresponding destinations on the right, through a network of intermediate nodes, connected with noise-less links capable of carrying one finite field $\mathbb{GF}$ symbol (packet) per channel use. The intermediate nodes employ linear network coding, so that the end-to-end transfer functions are represented as:
\begin{eqnarray}
y_1(n)&=&h_{11}x_1(n)+h_{13}x_3(n)+h_{14}x_4(n)\\
y_2(n)&=&h_{22}x_2(n)+h_{23}x_3(n)+h_{24}x_4(n)\\
y_3(n)&=&h_{31}x_1(n)+h_{33}x_3(n)+h_{35}x_5(n)\\
y_4(n)&=&h_{41}x_1(n)+h_{44}x_4(n)+h_{45}x_5(n)\\
y_5(n)&=&h_{52}x_2(n)+h_{55}x_5(n)
\end{eqnarray}
where over the $n^{th}$ channel use, $x_i(n)$ is the symbol transmitted by source $i$, $h_{ji}$ is the non-zero and constant channel coefficient between transmitter $i$ and receiver $j$ and $y_j(n)$ is the symbol received by receiver $j$, with $i,j\in\{1,2,\cdots, 5\}$.  The channel coefficients are comprised of the network coding coefficients, e.g., $h_{12}=\alpha_{bg}\alpha_{cb}+\alpha_{cg}$. All symbols and linear operations are over $\mathbb{GF}$. Most importantly the transmitters are only aware of which $h_{ij}$ are non-zero, i.e., the knowledge of the  end-to-end topology of the network. We are interested in the  capacity of this network.

The observant reader may have noticed that the wired network of Fig. \ref{fig:wired} and the partially connected  wireless network of Fig. \ref{fig:contrast}(a) have the same logical end-to-end topology. Indeed, these two problems are solved simultaneously and share a common solution. The DoF of the partially connected wireless network of Fig. \ref{fig:contrast}(a) and the capacity of the wired network of Fig. \ref{fig:wired} are both equal to 0.5 per user in their respective normalized units. The normalized unit here represents the DoF (capacity) of one wireless (wired) link by itself, i.e.,  $\log|\mathbb{GF}|$ corresponds to 1 DoF.

\subsection{Unified View: Capacity of Linear Networks}
It is not merely a coincidence that the partially connected wireless network of Fig. \ref{fig:contrast}(a) has the same DoF, i.e., first order capacity approximation, as the exact capacity of the wired network of Fig. \ref{fig:wired}, in their respective normalized units. There is a larger message here which is worth highlighting, and it has to do with the capacity of \emph{linear} communication networks (wired and wireless).

The capacity of communication networks is often regarded as the holy grail of network information theory. While little progress has been made in the classical sense, e.g., with the classical discrete memoryless network model, there has been remarkable and rapid progress on the smaller but widely prevalent class of (essentially) linear communication networks, where the channel outputs are linear combinations of channel inputs. This includes wired networks with linear network coding at intermediate nodes, often modeled as purely linear networks over finite fields, as in this work. It also includes wireless networks where the interaction between signals is linear, albeit corrupted by additive noise. Here the capacity of the underlying linear communication network corresponds to the degrees of freedom (DoF) metric, which naturally de-emphasizes the additive noise, returning the focus to the linear model of signal interaction. The essential property here is the linearity of the communication network, so the distinction between wired and wireless settings is often of little significance. Since DoF studies are often viewed narrowly through the lens of wireless networks, it is worth highlighting this under-appreciated aspect. It is true that DoF results provide a first order approximation to the capacity of wireless networks. However, what the DoF value really represents is the capacity of the underlying \emph{linear} communication network where the received signals are simply linear combinations of transmitted signals.  This translates into a first order capacity approximation, DoF, for wireless networks because of the presence of additive noise, a necessary evil because without it the infinite resolution of the complex alphabet would make the capacity undefined (infinite). For corresponding linear wired networks, where noise is not an issue, indeed the DoF results often (not always, because the choice of field can be significant\footnote{The exceptions have not been explored in the literature, and are also beyond the scope of this work.}) translate directly into exact capacity results. So the significance of DoF studies extends broadly to the entire class of linear communication networks. We will also highlight this theme throughout this paper: how the capacity results for wired networks directly translate into DoF results for wireless networks, and vice-versa. 

From a unified perspective, the central question in this work is to determine the capacity (DoF) of partially connected SISO linear communication networks for arbitrary topologies, with no CSIT except the topology of the network. We formally state the problem next, before proceeding to an overview of the results.

\section{Problem Statement}
The topological interference management problem for wired and wireless settings, is  specified in terms of the following parameters.
\begin{enumerate}
\item A topology matrix $\mathcal{T}=[t_{ij}]_{D\times S}\in\{0,1\}_{D\times S}$.
\item $S$ message sets $\mathcal{W}(S_j)$, $j\in\{1,2,\cdots, S\}$.
\item $D$ message sets $\mathcal{W}(D_i)$, $i\in\{1,2,\cdots, D\}$. 
\item A field $\mathbb{F}$.
\end{enumerate}

The parameters define a linear communication network with $S$ source nodes, labeled $S_1, S_2, \cdots, S_S$ and $D$ destination nodes, labeled $D_1, D_2, \cdots, D_D$. The only parameter that distinguishes the wired and wireless settings is the field $\mathbb{F}$, taken to be a Galois Field $\mathbb{GF}$ for wired settings and the field of complex numbers $\mathbb{C}$ for wireless networks. We will refer to an instance of the topological interference management problem for wireless networks as $${\bf TIM}(\mathcal{T}, \mathcal{W}(S), \mathcal{W}(D), \mathbb{C})$$ and for wired networks as $${\bf TIM}(\mathcal{T}, \mathcal{W}(S), \mathcal{W}(D), \mathbb{GF}).$$

\bigskip
\noindent{\bf Channel:} The channel input-output relationships are defined as:
\begin{eqnarray}
\left[\begin{array}{c}
y_1(n)\\y_2(n)\\ \vdots \\y_D(n)
\end{array}
\right]
&=&\left[\begin{array}{cccc}
h_{11}&h_{12}&\cdots&h_{1S}\\
h_{21}&h_{22}&\cdots&h_{2S}\\
\vdots&\vdots&\vdots&\vdots\\
h_{D1}&h_{D2}&\cdots&h_{DS}\\
\end{array}\right]\left[\begin{array}{c}
x_1(n)\\x_2(n)\\ \vdots \\x_S(n)
\end{array}
\right]
+\left[\begin{array}{c} z_1(n)\\z_2(n)\\ \vdots \\z_D(n) \end{array}\right]\label{eq:channelmodel}
\end{eqnarray}
where, over the $n^{th}$ channel use, $x_j(n)$ is the transmitted symbol from Source $S_j$, $y_i(n)$ is the received symbol at Destination $D_i$, $z_i(n)$ is the additive noise at Destination $D_i$, and $h_{ij}$ is the constant channel coefficient between Source $S_j$ and Destination $D_i$. All symbols belong to the field $\mathbb{F}$.

\bigskip
\noindent {\bf Message Sets:} Source node $S_j$ has a set of independent messages, $\mathcal{W}({S_j})$, that it wants to send to their desired destinations. Destination node $D_i$ has a set of independent messages $\mathcal{W}(D_i)$ that it desires. The set of all messages is denoted as $\mathcal{W}$.
\begin{eqnarray}
\mathcal{W}&=&\bigcup_{i=1}^D\mathcal{W}(D_i)=\bigcup_{j=1}^S\mathcal{W}(S_j)
\end{eqnarray}
and  $K=|\mathcal{W}|$ is the total number of messages. Each message has a unique source, i.e., $\mathcal{W}(S_j)\cap\mathcal{W}(S_{j'})=\phi$ if $j\neq j'$. A message must have at least one desired destination, and may have more than one desired destinations.  If every message has a unique destination, it is called a multiple unicast setting. The setting where every message is desired by all destinations is often called the multicast setting. The general case where every message is desired by at least one, possibly more than one, but not necessarily all destinations, is called the multiple groupcast setting \cite{Maleki_Cadambe_Jafar}. When it is important to highlight the number of messages, the multiple unicast or multiple groupcast settings will be referred to as $K$-unicast and $K$-groupcast, respectively.



\bigskip
\noindent{\bf Achievable Rates:} Each message $W\in\mathcal{W}$ is an independent random variable that  takes values $\{1,2,\cdots, 2^{NR(W)}\}$, each with equal probability. Source $S_j$ uses an encoder, which is a mapping from the set of messages $\mathcal{W}(S_j)$ to a sequence of transmitted symbols $x_j(1), x_j(2), \cdots, x_j(N)$ over $N$ channel uses. Destination $D_i$ uses a decoder, which is a mapping from the sequence of received symbols $y_i(1), y_i(2), \cdots, y_i(N)$ to a set of decoded values for its desired messages. An error occurs if at any destination, the decoded value of a desired message is not the same as the  value of that  transmitted message. Here, $N$ is the length of the codebook and $R(W)$ is the rate associated with message $W$. A rate allocation  $R(W)$, that assigns rates to all messages $W\in\mathcal{W}$, is said to be achievable if there exists a sequence of source encoders and destination decoders, indexed by $N$, such that the probability of error approaches zero as $N$ approaches infinity. The closure of achievable rate allocations is known as the capacity region, denoted as $\mathcal{C}$. The symmetric capacity $C_{\mbox{\small sym}}$, of the network is the highest value $R_0$, such that the rate allocation $R(W)=R_o, \forall W\in\mathcal{W}$, is inside the capacity region.

\bigskip
\noindent{\bf Channel State Information (CSI):} \noindent We will assume the following throughout this work.
\begin{enumerate}
\item The channel coefficient values are assumed to be fixed throughout the duration of communication.
\item The topology of the network, $\mathcal{T}$,  is known to all sources and destinations. 
\item Besides the topology information, there is no CSIT.
\item Besides the topology information, the CSIR only includes the knowledge of the \emph{desired} channel coefficients at each receiver.
\end{enumerate}

So far, the problem description is identical for both wired and wireless networks. The remaining assumptions, specialized to these networks, are stated next.

\bigskip
\subsection{\bf Wireless Networks}
 For wireless networks, the field $\mathbb{F}=\mathbb{C}$, the field of complex numbers, so that all symbols are complex, and the  $z_i(n)$ terms represent additive white Gaussian noise,  independent identically distributed $\sim\mathcal{N}^c(0,N_o)$. The average transmit power constraint at  source $S_j$ is  set as $P_j$, i.e.,  $\frac{1}{N}\mbox{E}\left[\sum_{n=1}^N\left|x_{j}(n)\right|^2\right]\leq P_j$, to ensure the following nominal interference-free SNR guarantees for all desired links:
\begin{eqnarray}
\frac{\left|h_{ij}\right|^2P_j}{N_o}&\geq&\mbox{SNR}, ~~~~~~\forall i\in\mathcal\{1,2,\cdots,D\}, j\in\{1,2,\cdots,S\}, \mathcal{W}(D_i)\cap\mathcal{W}(S_j)\neq\phi.\label{eq:desiredpowers}
\end{eqnarray}
Thus, the power constraints are chosen such that, in the absence of all other messages, each message by itself can achieve a rate $\log(1+\mbox{SNR})$. Since (\ref{eq:desiredpowers}) is the only information available to the transmitters about desired channels, $\log(1+\mbox{SNR})$  is also the individual capacity of each message if all other messages are eliminated, i.e., allocated zero rates.

\subsubsection{DoF}
While our ultimate goal for wireless networks is the capacity characterization within a constant gap, a useful intermediate step toward this ultimate goal will be a first order analysis, i.e., a DoF analysis, of a partially connected network where the \emph{weak} channels are set to zero. 
\begin{eqnarray}
\mbox{Partially connected model:  } \forall i\in\{1,2,\cdots,D\}, j\in\{1,2,\cdots,S\}, \mbox{ if } t_{ij}=0 \mbox{ then } h_{ij}=0.\label{eq:partial2}
\end{eqnarray}
In this partially connected model, we let SNR approach infinity (by increasing the transmit power for every source proportionately), and evaluate the achievable rates normalized by $\log(SNR)$. If there exists a sequence of achievable rate allocations $R(W)$, such that the limit  $R(W)/\log(SNR)$ exists  for all $W\in\mathcal{W}$ as SNR$\rightarrow\infty$,  then these limiting values are said to be an achievable DoF allocation.
\begin{eqnarray}
\mbox{DoF}(W)=\lim_{\mbox{\tiny SNR}\rightarrow\infty} \frac{R(W)}{\log(SNR)}, \forall W\in\mathcal{W}
\end{eqnarray}
The closure of the set of achievable DoF allocations is called the DoF region and denoted as $\mathcal{DOF}$. The symmetric DoF value, $\mbox{DoF}_{\mbox{\small sym}}$ of the network is the largest value DoF$_o$, such that the DoF allocation DoF$(W)=$ DoF$_o$, $\forall W\in\mathcal{W}$, is inside the DoF region. As a fair and compact metric, we are especially interested in the symmetric DoF. 

Note that for wireless networks, it is the DoF problem for the partially connected network that we denote as ${\bf TIM}(\mathcal{T}, \mathcal{W}(S), \mathcal{W}(D), \mathbb{C})$.

\subsubsection{Capacity within a constant gap for the original wireless network}
While the majority of this paper will focus on the DoF of  partially connected networks it is important to remember that the original wireless network is not partially connected. For the original wireless network, the topology matrix identifies the \emph{weak} channels as:
\begin{eqnarray}
\sum_{j:t_{ij}= 0}\left|h_{ij}\right|^2P_j&\leq&N_o\label{eq:partial1}
\end{eqnarray}
so the average received power contribution at destination $D_i$, from all \emph{weak} interferers $S_j$, i.e., sources for which $t_{ij}=0$, can be no more than the noise floor. The topological interference management problem  is intended as a stepping stone to obtaining a capacity approximation for the original wireless network. In particular, we will find capacity approximations accurate to within a constant gap that does not depend on SNR. 

{\it Remark:} Note that the ``topological interference management problem" refers to the partially connected network model of (\ref{eq:partial2}), and the ``original network" refers to the weakly connected model of (\ref{eq:partial1}).

One might wonder, since the conventional DoF formulation involves sending transmit powers to infinity while the noise floor is fixed, how can (\ref{eq:partial1}) continue to hold for a given channel as the transmit powers approach infinity? Indeed, this is precisely the problem with the conventional DoF formulation that we alluded to in Section \ref{sec:practical}. Our goal is not to understand the network in the limit as all transmit powers approach infinity, because such a limit  says  little  about the original finite SNR setting of interest. It is even misleading to use this limit for insights into the finite SNR behavior because the conventional DoF limit treats all non-zero channels as essentially equally strong, thereby fundamentally changing the nature of the interference management problem relative to the original finite SNR setting where weak interferers fall below the noise floor and can be ignored at little cost.  Our goal is to understand the network at the original, given, \emph{finite} SNR values. When we send the transmit powers to infinity, we do so by associating to each SNR value, a corresponding network realization that  satisfies (\ref{eq:partial1}). We study this class of networks together, because these networks have the same fundamental character, the same underlying topology, and indeed the \emph{same} (normalized by log(SNR)) capacity within a constant gap. So when we obtain a capacity characterization, e.g., for the network of Fig. \ref{fig:contrast} within 1.2 bits, regardless of SNR, we have established the capacity of each member of a class of networks, such that for any SNR value, our capacity characterization is within 1.2 bits of the capacity of the network  corresponding to that SNR. This class includes the original network when the SNR value is taken to be the original, given, SNR value.  This  distinction is extremely important and worth repeating. The conventional DoF formulation fixes the channel realizations and scales all the transmit powers proportionately and finds the limiting (normalized by log(SNR)) value of capacity at infinite SNR, which has  little to do with the capacity of the original network at the given original finite SNR. Our formulation, however, proceeds through a sequence of network realizations indexed by SNR values, such that every network in this class has the \emph{same} capacity (each normalized by its corresponding SNR) within a constant gap as the original network at its original SNR. In this formulation, the connection to the original network and the original finite SNR topology is maintained throughout as the SNRs approach infinity, so that finding the normalized capacity in the infinite SNR limit also determines the capacity within a constant gap for every network realization along the way, at its corresponding SNR value. For a deeper understanding of this aspect, we point the reader to the discussion on the Generalized Degrees of Freedom (GDoF) metric presented by Bresler and Tse in \cite{Bresler_Tse}, where also the GDoF metric is instrumental in characterizing the capacity region of a 2 user interference channel at all finite SNR values within a constant gap. Indeed, our formulation of the problem as the DoF of a partially connected network can also be equivalently seen as a GDoF formulation where the SNR exponents of the weak channels have been set to zero.

\subsection{\bf Wired Networks} For wired networks, all symbols are from a finite field $\mathbb{GF}$, and the noise terms $z_i(n)$ are all zero, i.e., there is no noise. Like the DoF in the wireless case, the topology of a wired network also identifies the channels that take zero values:
\begin{eqnarray}
\forall i\in\{1,2,\cdots,D\}, j\in\{1,2,\cdots,S\},  t_{ij}=0 \mbox{ iff } h_{ij}=0.
\end{eqnarray}
However, in the wired case, this is the actual channel model from which we expect exact capacity results, and not merely an idealization for first order analysis. We are interested in the  capacity of the network normalized by the capacity of a single link, i.e., a unit capacity represents $\log|\mathbb{GF}|$ bits/channel-use. As a fair and compact metric, we are especially interested in the symmetric capacity. Analogous to the high SNR approximations in the wireless case, we will be content with capacity characterizations over ``sufficiently large  finite fields" $\mathbb{GF}$, i.e., fields with sufficiently large characteristic. Some discussion on the significance of field size restrictions is provided in Section \ref{sec:half-rate-feasibility}.

\section{Rates and DoF achievable through Linear Schemes}\label{sec:linear}
While the channel model is linear, note that the achievable rates defined in the previous section are not restricted to linear schemes. Indeed, the information theoretic notions of capacity and DoF allow arbitrary encoding schemes, and therefore provide the strongest guarantees, not constrained by complexity. However, linear schemes (also known as  vector linear schemes, signal vector space schemes, or beamforming schemes) are often of interest, not only for their simplicity, but also because they are very often optimal for linear communication networks from a capacity (wired) or DoF (wireless) perspective. Since the notion is important to this work, we will define a linear scheme next.

\bigskip
\noindent{\bf Linear Scheme:} {\it A linear scheme over $N$ channel uses achieving, in the wired case, the rates 
\begin{eqnarray}
R(W)=\frac{L(W)}{N}, \forall W\in\mathcal{W}
\end{eqnarray}
and, in the wireless case, the DoF,
\begin{eqnarray}
\mbox{DoF}(W)=\frac{L(W)}{N}, \forall W\in\mathcal{W}
\end{eqnarray}
where $L(W)$ are non-negative integer values, consists of 
\begin{enumerate}
\item precoding matrices ${\bf V}(W)\in\mathbb{F}^{N\times L(W)}$, $\forall W\in\mathcal{W}$, and
\item receiver combining matrices ${\bf U}_i(W)\in\mathbb{F}^{L(W)\times N}$, $\forall  W\in \mathcal{W}(D_i), i \in\{1,2,\cdots, D\},$
\end{enumerate}
such that the following properties are satisfied
\begin{eqnarray}
\mbox{Property 1:} &&{\bf U}_i(W){\bf V}(\tilde W)=0,\label{eq:timproperty1}\\
&&\forall i\in\{1,2,\cdots, D\}, j\in\{1,2,\cdots, S\},W\in\mathcal{W}(D_i), \tilde W\in\mathcal{W}(S_j),\nonumber\\
&& \mbox{ such that } W\neq \tilde W \mbox{ and }  t_{ij}=1.\nonumber\\
\mbox{Property 2:}&& det({\bf U}_i(W){\bf V}(W))\neq 0, ~\forall W\in\mathcal{W}(D_i), \forall i\in\{1,2, \cdots, D\}.\label{eq:timproperty2}
\end{eqnarray}
}
Evidently, property 1 ensures that all interference is eliminated and Property 2 ensures that the desired signal is recovered. 

Thus,  each message $W$ is split into $L(W)$ independent scalar streams, collectively represented by the column vector ${\bf X}(W)=(x_1(W), x_2(W), \cdots,x_{L(W)}(W))^T\in\mathbb{F}^{L(W)\times 1}$, each of which carries one symbol from $\mathbb{F}$, and is transmitted along the corresponding column vectors (the ``beamforming" vectors) of the precoding matrix  ${\bf V}(W)$. In the wired case, the symbols $x_l(W)$ are uniformly distributed over the finite field $\mathbb{F}$, each carrying $\log|\mathbb{F}|$ bits of information. In the wireless case, the $x_l(W)$ are independent Gaussian codebooks, each with power $\frac{P_j}{|\mathcal{W}(S_j)|L(W)}$ where $W\in\mathcal{W}(S_j)$, and the columns of ${\bf V}(W)$ are scaled to have unit norm (which does not affect Property 1 or 2), so that the power constraints are satisfied. Note that properties 1 and 2 do not involve SNR or the values of the non-zero channel realizations $h_{ij}$, and therefore the existence of  ${\bf U},{\bf V}$ matrices that satisfy these properties, does not depend on SNR or the non-zero channel coefficient values. It only depends on the message sets, the topology, and possibly the field $\mathbb{F}$.

\noindent Over the $N$ channel uses, Source $S_j$ sends,
\begin{eqnarray}
{\bf X}_j&=&\sum_{W\in\mathcal{W}(S_j)}{\bf V}(W){\bf X}(W).
\end{eqnarray}
Destination $D_i$ receives the $N\times 1$ vector,
\begin{eqnarray}
{\bf Y}_i&=&\sum_{j:t_{ij}=1}\sum_{W\in\mathcal{W}(S_j)}h_{ij}{\bf V}(W){\bf X}(W)+{\bf Z}_i,
\end{eqnarray}
and for each desired message $W\in\mathcal{W}(D_i)\cap\mathcal{W}(S_j)$, projects the received signal vector ${\bf Y}_i$ into the ${\bf U}_i(W)$ space to obtain,
\begin{eqnarray}
\overline{{\bf Y}}_i(W)&=&{\bf U}_i(W){\bf Y}_i=h_{ij}{\bf U}_i(W){\bf V}(W){\bf X}(W)+{\bf U}_i(W){\bf Z}_i,
\end{eqnarray}
where the contributions from all other messages are eliminated  due to Property 1. Now, as stated previously, the channel coefficient $h_{ij}$ is assumed to be  known to  Destination $i$ and is non-zero (otherwise there would be no path for this desired message and the problem would be degenerate), and according to Property 2, ${\bf U}_i(W){\bf V}(W)$ is an invertible matrix. The following non-interfering channels are obtained for each desired symbol stream. 
\begin{eqnarray}
\overline{\overline{{\bf Y}}}_i=\frac{1}{h_{ij}}\left[{\bf U}_i(W){\bf V}(W)\right]^{-1}\overline{{\bf Y}}_i(W)&=&{\bf X}(W)+\underbrace{\frac{1}{h_{ij}}\left[{\bf U}_i(W){\bf V}(W)\right]^{-1}{\bf U}_i(W){\bf Z}_i}_{\overline{\overline{{\bf Z}}}_i}\\
\Rightarrow \overline{\overline{y}}_{i,l}(W)&=&x_l(W)+\overline{\overline{z}}_{i,l}, ~~~~l\in\{1,2,\cdots,L(W)\}.\label{eq:interferencefree}
\end{eqnarray}
Thus, in the wireless case, each non-interfering channel  contributes $1/N$ DoF (it contributes 1 DoF, but because $N$ channel uses are required by the linear coding scheme, the normalized value is $1/N$ per channel use), so that DoF of $L(W)/N$ is achieved for each message $W$. Note that the additive noise power for each non-interfering channel is bounded by a constant, away from zero and infinity, even as SNR approaches infinity, so it is inconsequential for the DoF metric. In the wired case, there is no noise, and a rate of $L(W)/N$ is achieved for each message $W$. This is, of course, subject to the existence of the precoding and receiver combining matrices ${\bf U}, {\bf V}$ that satisfy properties 1 and 2. The largest achievable rates (wired) or DoF (wireless) through linear schemes, therefore, correspond to the largest values of $L(W)/N$ for which such precoding and combining matrices exist.

\section{Results}
We are interested in the capacity and DoF of partially connected linear wired and wireless networks with arbitrary connectivity and arbitrary message sets, when there is no CSIT beyond the network topology. The first set of results reveals the essence of the problem  by translating it, somewhat surprisingly,  into a  previously studied problem, the \emph{index coding problem}. The index coding problem was introduced in 1998 by Birk and Kol \cite{Birk_Kol} and  has been studied extensively in the computer-science community. The index coding problem is described in detail in Appendix \ref{sec:indexcoding}.

Both the topological interference management problem and the index coding problem are comprised of a number of sources $S$, a number of destinations $D$, and message sets $\mathcal{W}(D_i)$, $\mathcal{W}(S_j)$ associated with each destination $D_i$ and each source $S_j$. However, while a topological interference management problem is defined by a topology matrix $\mathcal{T}\in\{0,1\}^{D\times S}$, the index coding problem is defined by an antidote matrix $\mathcal{A}\in\{0,1\}^{D\times S}$. Intuitively, the roles of the antidote matrix and the topology matrix are the opposite of each other. For a source and destination pair $S_j, D_i$ which have no desired messages between them, $\mathcal{W}(D_i)\cap\mathcal{W}(S_j)=\phi$, the presence of an interference link, $t_{ij}=1$, can only hurt by exposing the destination node $D_i$ to  interference from undesired source $S_j$, but the presence of an antidote link, $a_{ij}=1$, can only help by providing the destination $D_i$ with antidotes for $\mathcal{W}(S_j)$. As it turns out, the relationship between the two problems maps the antidote matrix $\mathcal{A}$ to the \emph{complement} of the topology matrix $\mathcal{T}$. For a given $S\times D$ topology matrix $\mathcal{T}$, define the complementary topology matrix $\overline{\mathcal{T}}$, also an $S\times D$ matrix, as
\begin{eqnarray}
\overline{t}_{ij}&=&\left\{\begin{array}{ll}
0&\mbox{if } t_{ij}=1\\
1&\mbox{if } t_{ij}=0
\end{array}\right.
\end{eqnarray}
In this section we present the statements of the results as theorems and corollaries. The proofs appear in Section \ref{sec:proofs}. 

\subsection{Topological Interference Management as an Index Coding Problem}
The first set of results is stated in the following two theorems.
\begin{theorem}\label{theorem:one}
The  capacity (DoF) region of the topological interference management problem \\{\bf TIM}$(\mathcal{T},\mathcal{W}(S),\mathcal{W}(D),\mathbb{F})$, for  wired (wireless) networks, is bounded above by the  capacity region of the corresponding index coding problem {\bf IC}$(\mathcal{A},\mathcal{W}(S),\mathcal{W}(D))$, where  $\mathcal{A}=\overline{\mathcal{T}}$. Specifically,
\begin{eqnarray}
\mbox{Wireless: }\mathcal{DOF}\left({\bf TIM}(\mathcal{T},\mathcal{W}(S),\mathcal{W}(D),\mathbb{C})\right)&\subset& \mathcal{C}\left({\bf IC}(\overline{\mathcal{T}},\mathcal{W}(S),\mathcal{W}(D))\right)\\
\mbox{Wired: }\mathcal{C}\left({\bf TIM}(\mathcal{T},\mathcal{W}(S),\mathcal{W}(D),\mathbb{GF})\right)&\subset& \mathcal{C}\left({\bf IC}(\overline{\mathcal{T}},\mathcal{W}(S),\mathcal{W}(D))\right)
\end{eqnarray}
\end{theorem}
 Note the relationship between all three problems (TIM(wired), TIM(wireless), IC). They have the same sets of sources and destinations, and the same message sets. Note that TIM problem requires a field specification ---   a wired network is associated with a finite field $\mathbb{GF}$, and a wireless network with  the field of complex numbers, $\mathbb{C}$, and the capacity may change depending on the field. However, the index coding \emph{capacity} problem does not require a field specification\footnote{Field specification is important if the class of achievable schemes is restricted, say, to linear schemes.}. Yet, the index coding problem provides an outer bound for the normalized capacity  or DoF of  wired and wireless networks. The relationship becomes even stronger, an equivalence instead of an outer bound, when restricted to linear solutions.
\begin{theorem}\label{theorem:linear}
The achievable rate (DoF) region for {\bf TIM}$(\mathcal{T},\mathcal{W}(S),\mathcal{W}(D),\mathbb{F})$ through linear schemes is the same as the achievable rate (DoF) region of  {\bf IC}$(\overline{\mathcal{T}},\mathcal{W}(S),\mathcal{W}(D))$ through linear schemes over the same field $\mathbb{F}$.
\end{theorem}
Theorem \ref{theorem:linear} is quite powerful because linear solutions are often capacity optimal for the index coding problem. In particular, linear solutions will be shown to be capacity optimal for all cases solved in this paper.  It should also be noted that linear solutions are not \emph{always} optimal for the index coding problem, as shown  by Blasiak et al. in \cite{Blasiak_Kleinberg_Lubetzky_2011} for the general index coding problem, and by Maleki et al. in \cite{Maleki_Cadambe_Jafar} for the multiple unicast index coding problem, based on counterexamples inspired by matroid theory that were  originally used to show insufficiency of linear codes for the general network coding problem by Dougherty et al. in \cite{dougherty1}. Interestingly, none of the counterexamples applies directly to the topological interference management problem, for the wired or wireless case, especially for the multiple unicast setting. As such, it remains an intriguing possibility that linear codes may yet be sufficient for the topological interference management problem. 

For wireless networks, the topological interference management problem (which models a network as partially connected by removing weak interference) was motivated as a stepping stone to obtain constant gap capacity approximations for the original network. The following theorem formalizes this relationship.
\begin{theorem}\label{theorem:gap}
Whenever a non-asymptotic linear scheme is  DoF optimal for an index coding problem over $\mathbb{F}=\mathbb{C}$, then a constant gap  capacity approximation is available for the original wireless network.
\end{theorem}
The definition of a DoF optimal linear scheme is presented in Appendix \ref{sec:indexcoding}. Thus, according to Theorem \ref{theorem:gap} solving the topological interference management problem for partially connected wireless networks guarantees a constant gap capacity approximation for the original wireless network whenever linear solutions are DoF optimal for the corresponding index problem over $\mathbb{C}$. Since linear solutions are DoF optimal for all instances considered in this work, constant gap approximations for the original wireless networks are automatically implied. For the first few examples we will study the constant gap approximations in some detail to see how the gap can be made smaller. In general, however, we will be content with a constant gap guarantee, i.e., non-asymptotic linear solution to the topological interference management problem that is DoF optimal.

The solution to the motivating examples is presented next to  illustrate the application of theorems \ref{theorem:one} and \ref{theorem:linear}. A detailed treatment of the constant gap approximation for the wireless motivating example is presented in Section \ref{sec:1.2bits}.

\subsection{Example}
\begin{figure}[h]
\begin{center}
\includegraphics[width=4.5in]{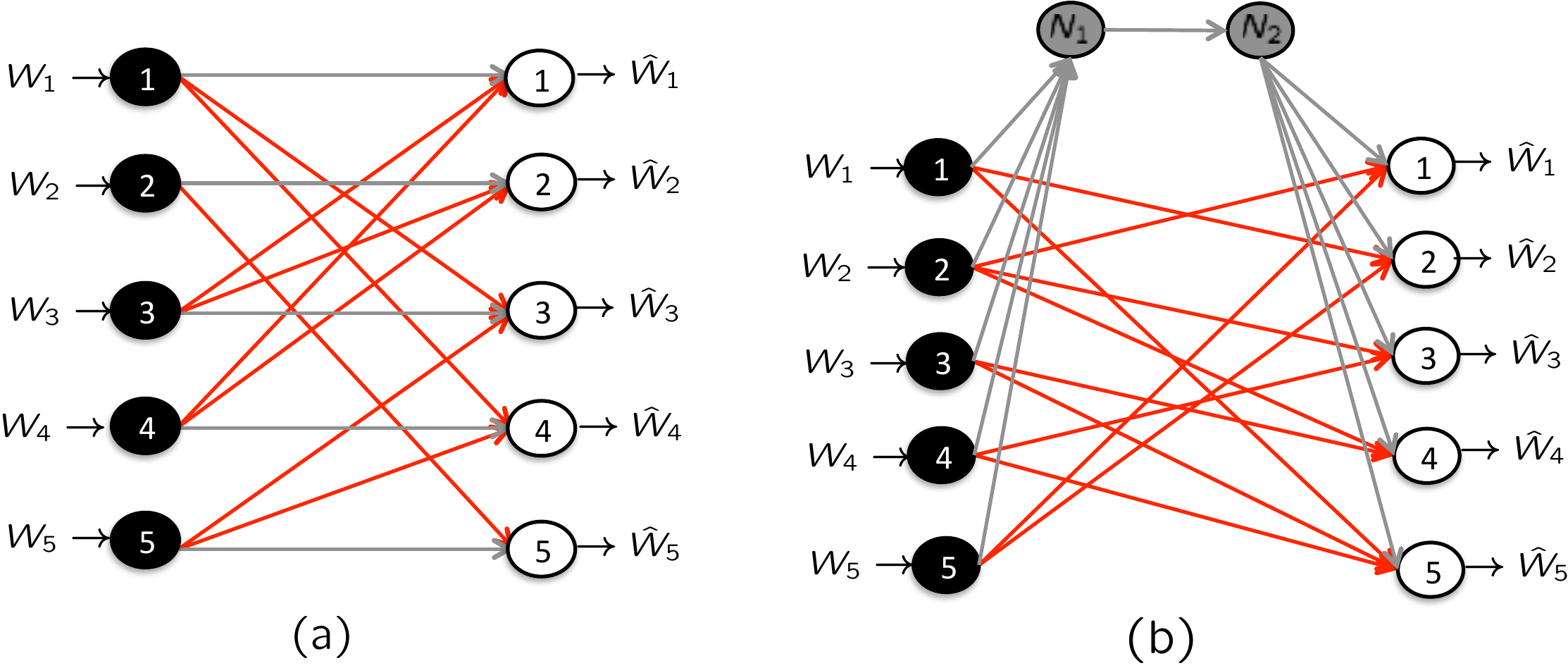}
\caption{\small \it (a) Topological Interference Management Problem, (b) Associated Index Coding Problem}\label{fig:equivalence}
\end{center}
\end{figure}

The partially connected wireless network of Fig. \ref{fig:contrast}(a) and the partially connected wired network of Fig. \ref{fig:wired} have the same end-to-end topology. The topology and the message sets are illustrated in Fig. \ref{fig:equivalence}(a), as a unified representative of both wireless and wired settings. Note that this is a multiple unicast setting with $S=D=5$ and $\mathcal{W}(S_i)=\mathcal{W}(D_i)=W_i$. 

According to Theorem \ref{theorem:one}, the capacity region of the wired network and the DoF region of the wireless network represented by Fig. \ref{fig:equivalence}(a), is bounded above by the capacity region of the index coding problem shown in Fig. \ref{fig:equivalence}(b), which has the same source and destination nodes, the same message sets, and antidote links that complement the topology of Fig. \ref{fig:equivalence}(a) -- whenever there is (is not) an edge between $S_j$ and $D_i$ in Fig. \ref{fig:equivalence}(a), then there is not (is) an antidote edge between $S_j$ and $D_i$ in Fig. \ref{fig:equivalence}(b). Recall, that all edges in the index coding problem have infinite capacity, with the exception of the bottleneck edge between nodes $N_1$ and $N_2$, which is normalized to have unit capacity.

The examples of Fig. \ref{fig:contrast}(a) and Fig. \ref{fig:wired} are chosen not because they are particularly challenging (indeed we will deal with more challenging instances of the index coding problem later in this paper) but rather for their historical significance. As it turns out, the specific index coding problem of Fig. \ref{fig:equivalence}(b) is  considered originally by Birk and Kol in \cite{Birk_Kol}, and is the first known example of interference alignment. Limiting our discussion to symmetric rates, Birk and Kol have shown that the symmetric capacity of the index coding problem is 0.5 per message, where the unit is the capacity of the bottleneck link, and furthermore linear schemes suffice to achieve the symmetric capacity  over any finite field. It is also easy to see that linear schemes achieve the symmetric DoF of the index coding problem over  the field of complex numbers. Therefore Theorems 1 and 2 imply that the symmetric capacity of the wired network of Fig. \ref{fig:wired}, and the symmetric DoF of the wireless network of Fig. \ref{fig:contrast}(a), are both equal to 0.5 per message, where the unit is the individual capacity or DoF of a non-zero link, respectively. 

For the sake of completeness let us summarize the proof of Birk and Kol's result that the symmetric capacity of the index coding problem of Fig. \ref{fig:equivalence}(b) is 0.5 per user. Let us start with the achievable scheme. Suppose the bottleneck link carries a  finite field $\mathbb{GF}$ symbol per channel use, i.e., it has capacity $\log|\mathbb{GF}|$ per channel use. In order to achieve a symmetric capacity of 0.5 per message, we will successfully transmit one $\mathbb{GF}$ symbol, $x_i$, per message $W_i$, such that each symbol can be recovered at its desired receiver, by using the bottleneck link twice.  Birk and Kol's solution is to send the following two transmissions on the bottleneck link:
\begin{eqnarray}
\mbox{First transmission: } && x_1+x_2+x_5\\
\mbox{Second transmission: } && x_2+x_3+x_4
\end{eqnarray}
Each destination node sees these two transmissions in addition to its own antidotes, from which it must decode its desired message. To see how this works, as an example consider destination $D_5$, which already has antidotes for messages $W_1, W_3, W_4$, so it can remove $x_1, x_3, x_4$ from its received signals. This leaves $D_5$ with $x_2+x_5$ from the first transmission and $x_2$ from the second transmission. Clearly, subtracting the latter from the former, it can recover its desired symbol $x_5$. Since one $\mathbb{GF}$ symbol is communicated per two channel uses, the symmetric rate achieved is $0.5\log|\mathbb{GF}|$, i.e., a normalized rate of 0.5 per message.

For the outer bound on the symmetric capacity, eliminate all messages except $W_4, W_5$. In the remaining index coding problem, destination $D_4$ sees only the output of the bottleneck link and has no antidotes, while destination $D_5$ also sees the output of the bottleneck link and in addition has $W_4$ as antidote. Any reliable coding scheme must allow $D_4$ to decode $W_4$ from just the output of the bottleneck link. However, after decoding $W_4$,  destination node $D_4$ has all the information available to destination $D_5$, so it must also be able to decode $W_5$. Since $D_4$ is able to decode both $W_4, W_5$ from just the output of the bottleneck link, the sum-rate of $W_4$ and $W_5$ cannot be more than the capacity of the bottleneck link. Therefore, the symmetric rate cannot be more than half of the bottleneck link capacity. Including other messages cannot improve the symmetric rate. Thus, we have the information theoretic outer bound that the symmetric capacity of the index coding problem of Fig. \ref{fig:equivalence}(b) cannot be more than 0.5. Since this is achievable as described above, the symmetric capacity is 0.5 per message.

\subsection{Challenges and opportunities}
The index coding problem is an intriguing problem in its own right, as evident from the following three observations that are almost paradoxical in nature. 
\begin{enumerate}
\item It is arguably the simplest problem to describe, because it has only one link of finite capacity, yet it remains an open problem in general.
\item It is obviously a special case of the general network coding problem, yet for every instance of the network coding problem, there is an equivalent index coding problem \cite{Rouayheb_Sprintson_Georghiades, Effros_Rouayheb_Langberg}.
\item It is the earliest known setting for interference alignment, yet it remains almost unexplored from an interference alignment perspective.
\end{enumerate}
These observations highlight both the challenges and opportunities associated with the index coding problem, and by virtue of our results, associated with the topological interference management problem. Observation 2 shows that through its association with the index coding problem, the topological interference management problem is associated with every network coding problem, and the association is the strongest (an equivalence) from the perspective of linear solutions. Solving the topological interference management problem in its entirety is therefore as hard as solving network coding problems in their entirety. Conversely, network coding applications such as distributed storage repair \cite{Dimakis_survey}, caching \cite{Maddah_Ali_Niesen} etc., can be seen as instances of the topological interference management problem. The challenging nature of the topological interference management problem is therefore quite evident. At the same time, Observation 3, points out an opportunity to make progress on this challenging problem --- by using the interference alignment perspective. The interference alignment perspective and the relationship to topological interference management will be used in this work in the following ways.
\begin{enumerate}
\item To simplify and interpret existing solutions of various instances of index coding problems from the interference alignment perspective.
\item To use solved instances of index coding problems to solve the corresponding instances of topological interference management problems.
\item To obtain systematic solutions to previously unsolved classes of index coding problems.
\item To identify and solve new classes of index coding problems corresponding to physically motivated topologies.
\end{enumerate}
We begin by explaining the solution to the motivating example from an interference alignment perspective.
\subsection{Interference Alignment Perspective}
Let us present the solution to our running example, from an interference alignment perspective. Recall that  interference alignment involves two objectives.
\begin{enumerate}
\item Align undesired signals (interference) as much as possible.
\item Keep desired signals separate from interference.
\end{enumerate}
\begin{figure}[!h]
\begin{center}
\includegraphics[width=1.5in]{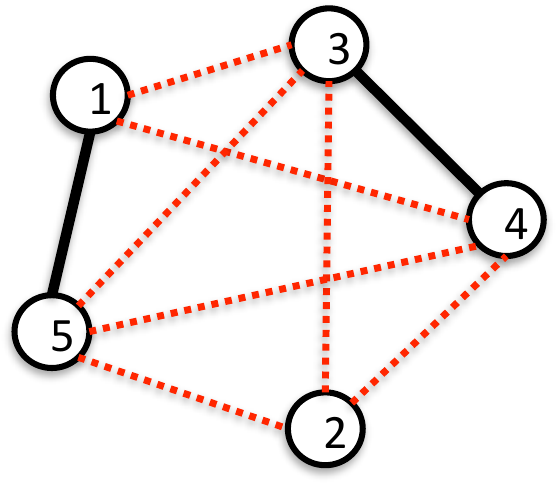}
\caption{\small \it Alignment Graph and Conflict Graph for the Network of Fig. \ref{fig:equivalence}(a)}\label{fig:alignedconflict}
\end{center}
\end{figure}
To represent these two objectives, let us introduce the notion of an alignment graph and a conflict graph, respectively. Starting with a node for each message, we construct the alignment graph and conflict graphs as follows.

\begin{enumerate}
\item{\bf Topological Interference Management}: The following definitions are for the topological interference management problem.
\begin{enumerate}
\item {\bf Alignment Graph:} Messages $W_i$ and $W_j$ are connected with a solid black edge if the source(s) of both these messages are heard by a destination that desires message $W_k\notin\{W_i, W_j\}$. 
\item{\bf Conflict Graph:} Each message $W_i$ is connected by a dashed red edge to  all other messages whose sources are heard by a destination that desires message $W_i$.  
\end{enumerate}
\item {\bf Index Coding}: The following definitions are for the index coding problem.
\begin{enumerate}
\item {\bf Alignment Graph:} Messages $W_i$ and $W_j$ are connected with a solid black edge if the source(s) of both these messages are not available as antidotes to a destination that desires message $W_k\notin\{W_i, W_j\}$. 
\item{\bf Conflict Graph:} Each message $W_i$ is connected by a dashed red edge to  all other messages whose sources are not available as antidotes to a destination that desires message $W_i$.  
\end{enumerate}
\end{enumerate}
The solid black edges comprising the alignment graph identify the messages that should be aligned as much as possible, and the dashed red edges comprising the conflict graph identify messages that need to be kept separate. The alignment graph and conflict graph for the network of Fig. \ref{fig:equivalence}(a) are shown in Fig. \ref{fig:alignedconflict}. For example, there is a solid black edge between message nodes $1$ and $5$ because sources $S_1$ and $S_5$, where these two messages originate, are heard by destination $D_3$ (also  $D_4$) where neither of these two messages is desired. Message node $1$ is connected by  dashed red edges to message nodes $3$ and $4$, because the sources $S_3$ and $S_4$ are heard by destination $D_1$ that desires $W_1$.

Let us also define the notions of alignment sets and internal conflicts which will be useful later in this work.
\begin{enumerate}
\item{\bf Alignment Set:} Each connected component (through solid black edges) of an alignment graph is called an alignment set.
\item{\bf Internal Conflict:} If two messages that belong to the same alignment set have a conflict (dashed red) edge between them, it is called an internal conflict.
\end{enumerate}
 Using this terminology, in Fig. \ref{fig:alignedconflict} we have three alignment sets: $\{W_1,W_5\}, \{W_3,W_4\}, \{W_2\}$ and there are no internal conflicts. 

\subsubsection{Symmetric Capacity within 1.2 bits}\label{sec:1.2bits}
While the symmetric DoF value of the partially connected wireless setting is already settled by Theorem \ref{theorem:one} and the index coding result of Birk and Kol, as 1/2 per message, here we will go further to obtain a constant gap approximation to the symmetric capacity of the original wireless network. 

Starting with the partially connected network, first, let us consider the outer bound. Note that the proof of Theorem \ref{theorem:one} already provides a symmetric capacity outer bound value of $1/2\log(1+|S|^2\mbox{SNR})=1/2\log(1+25\mbox{SNR})$ per message. This is good enough for a constant gap approximation, but we can obtain a smaller gap by improving the outer bound. This outer bound can be improved by considering only two messages at a time, say $W_4, W_5$, in the proof of Theorem \ref{theorem:one} which gives a symmetric capacity outer bound $1/2\log(1+4\mbox{SNR})$ per message. This can be further improved by considering the $Z$ interference channel between Users $4$ and $5$, where all non-zero links have strength $\mbox{SNR}$, which gives us a symmetric capacity bound $1/2\log(1+2\mbox{SNR})$ per message. Next, we will obtain an  inner bound on the partially connected network, using the DoF and the interference alignment perspective.

Based on the alignment and conflict graphs, we will align $W_1, W_5$, along a vector, say ${\bf V}_{1}$, we will align $W_3, W_4$ along a vector, say ${\bf V}_2$, and message $W_2$ will be sent along a different vector, say ${\bf V}_3$. All these vectors are $2\times 1$ vectors, so that they carry one symbol over two channel uses, to achieve 0.5 DoF. To avoid conflicts, i.e., to keep desired signals separate from interference, ${\bf V}_1, {\bf V}_2, {\bf V}_3$ must be pairwise linearly independent. With this choice, the received signals at each receiver over two time slots are represented as the following $2\times 1$ vectors:
\begin{eqnarray}
{\bf y}_1&=& h_{11}{\bf V}_1x_1+h_{13}{\bf V}_2x_3+h_{14}{\bf V}_2x_4+{\bf z}_1\\
{\bf y}_2&=&h_{22}{\bf V}_3x_2+h_{23}{\bf V}_2x_3+h_{24}{\bf V}_2x_4+{\bf z}_2\\
{\bf y}_3&=&h_{31}{\bf V}_1x_1+h_{33}{\bf V}_2x_3+h_{35}{\bf V}_1x_5+{\bf z}_3\\
{\bf y}_4&=&h_{41}{\bf V}_1x_1+h_{44}{\bf V}_2x_4+h_{45}{\bf V}_1x_5+{\bf z}_4\\
{\bf y}_5&=&h_{52}{\bf V}_3x_2+h_{55}{\bf V}_1x_5+{\bf z}_5
\end{eqnarray}
Note that the interference at each receiver is aligned into one dimension that is linearly independent of the desired signal dimension, and the two can be separated in the two dimensional space available to each destination over two channel uses. For example, consider destination $D_1$, which sees the desired symbol $x_1$ along ${\bf V}_1$ and both undesired symbols $x_3, x_4$ along the same vector ${\bf V}_2$. To recover its desired signal, it simply projects the received signal vector ${\bf y}_1$ along a $2\times 1$ vector ${\bf V}^{\perp}_2$, i.e., a vector orthogonal to ${\bf V}_2$. 
\begin{eqnarray}
\left({\bf V}_2^\perp\right)^T{\bf y}_1&=&h_{11}\left({\bf V}_2^\perp\right)^T{\bf V}_1x_1+\left({\bf V}_2^\perp\right)^T{\bf z}_1
\end{eqnarray}
which is an interference channel on which a rate $R(W_1)=\log(1+|\left({\bf V}_2^\perp\right)^T{\bf V}_1|^2\mbox{SNR})$ is achievable. Proceeding similarly, the achievable rate for each message is expressed as follows.
\begin{eqnarray}
R(W_1)&=&\frac{1}{2}\log\left(1+\left|\left({\bf V}_2^\perp\right)^T{\bf V}_1\right|^2\mbox{SNR}\right)\\
R(W_2)&=&\frac{1}{2}\log\left(1+\left|\left({\bf V}_2^\perp\right)^T{\bf V}_3\right|^2\mbox{SNR}\right)\\
R(W_3)&=&\frac{1}{2}\log\left(1+\left|\left({\bf V}_1^\perp\right)^T{\bf V}_2\right|^2\mbox{SNR}\right)\\
R(W_4)&=&\frac{1}{2}\log\left(1+\left|\left({\bf V}_1^\perp\right)^T{\bf V}_2\right|^2\mbox{SNR}\right)\\
R(W_5)&=&\frac{1}{2}\log\left(1+\left|\left({\bf V}_3^\perp\right)^T{\bf V}_1\right|^2\mbox{SNR}\right)
\end{eqnarray}
The factor of $1/2$ outside the log is because two overall channel uses are required to create one interference free channel use for every message. Since we are interested in symmetric rates, we would like to maximize the minimum of these rates. This corresponds to a Grassmannian packing \cite{Grassmannian} of ${\bf V}_1, {\bf V}_2, {\bf V}_3$ in a two dimensional space, leading to the choice ${\bf V}_1=[1,0]^T, {\bf V}_2=[1/2, \sqrt{3}/2]^T, {\bf V}_3=[-1/2,\sqrt{3}/{2}]$. With this choice of precoding vectors, the rate achievable by each message is the same: 
\begin{eqnarray}
R(W_1)=R(W_2)=R(W_3)=R(W_4)=R(W_5)&=&\frac{1}{2}\log(1+3\mbox{SNR}/4)
\end{eqnarray}
However, this rate is achieved in the partially connected network where weak channels are set to zero. Now let us go back to the original network and include the non-zero weak channels. Since, the weak channels collectively contribute no more than the noise floor $N_o$ at each receiver, their worst case impact is to double the noise floor to $2N_o$, i.e., reduce the SNR by half. Thus, in the original network, the following symmetric rate is achievable.
\begin{eqnarray}
R(W_1)=R(W_2)=R(W_3)=R(W_4)=R(W_5)&=&\frac{1}{2}\log(1+3\mbox{SNR}/8)
\end{eqnarray}
We now have the symmetric capacity bounded above and below as $\frac{1}{2}\log(1+3\mbox{SNR}/8)\leq C_{\mbox{\small sym}}\leq \frac{1}{2}\log(1+2\mbox{SNR})$. These bounds are within 1.2 bits of each other, regardless of the value of SNR. Thus, we have obtained a symmetric capacity approximation that is accurate to within 1.2 bits.

Note that any pairwise linearly independent choices of ${\bf V}_1, {\bf V}_2, {\bf V}_3$ yield a constant gap approximation in the wireless case and the exact capacity in the wired case. In particular, choosing ${\bf V}_1=[1,0]^T$, ${\bf V}_2=[0,1]^T$, ${\bf V}_3=[1,1]^T$ gives us Birk and Kol's solution described earlier, which is capacity optimal for the finite field case, and within a constant gap for the wireless case (using normalized ${\bf V}_3 = [1/\sqrt{2}, 1/\sqrt{2}]^T) $ although the gap would be  larger. 

\subsection{Half-rate-feasibility: When can everyone get half-the-cake?}\label{sec:half-rate-feasibility}
A remarkable result from the DoF study of fully connected wireless interference networks with perfect CSIT, is that in a $K$ user interference channel every user can  access 0.5 DoF, regardless of the number of users, $K$ \cite{Cadambe_Jafar_int}. In the absence of interference, i.e., if a user has the channel entirely to himself, the DoF value is 1. Therefore, the main result of \cite{Cadambe_Jafar_int} is often paraphrased as ``everyone gets half the cake". As positive as this result may be, it relies critically on perfect CSIT, and is therefore  difficult to translate into practice. As we explore the opposite extreme of no CSIT (except topology), a natural question is to identify in this setting the conditions that allow everyone to achieve half the cake, i.e., a symmetric capacity of 0.5 per user in the wired setting, and a symmetric DoF value of 0.5 in the wireless setting. 
Fortunately, the relationship to the index coding problem established in theorems \ref{theorem:one} and \ref{theorem:linear} and the half-rate feasibility solution for the index coding problem in \cite{Blasiak_Kleinberg_Lubetzky_2010, Maleki_Cadambe_Jafar}, allow us to directly settle this question. We state the result in the form of the following theorem.

\begin{theorem}\label{theorem:feasible}
A symmetric rate of 0.5 per user in the wired case and a symmetric DoF of 0.5 per user in the wireless case is achievable in the topological interference management problem if and only if there are no internal conflicts. 
\end{theorem}
Recall that an internal conflict exists between two message nodes if they belong to the same alignment set and one of them causes interference to one of the intended destinations of the other. Based on Theorem \ref{theorem:feasible} we will call an instance of the topological interference management problem `\emph{half-rate-feasible}' if its alignment graph has no internal conflicts, and  `\emph{half-rate-infeasible}' otherwise.

\begin{figure}[!h]
\begin{center}
\includegraphics[width=4.5in]{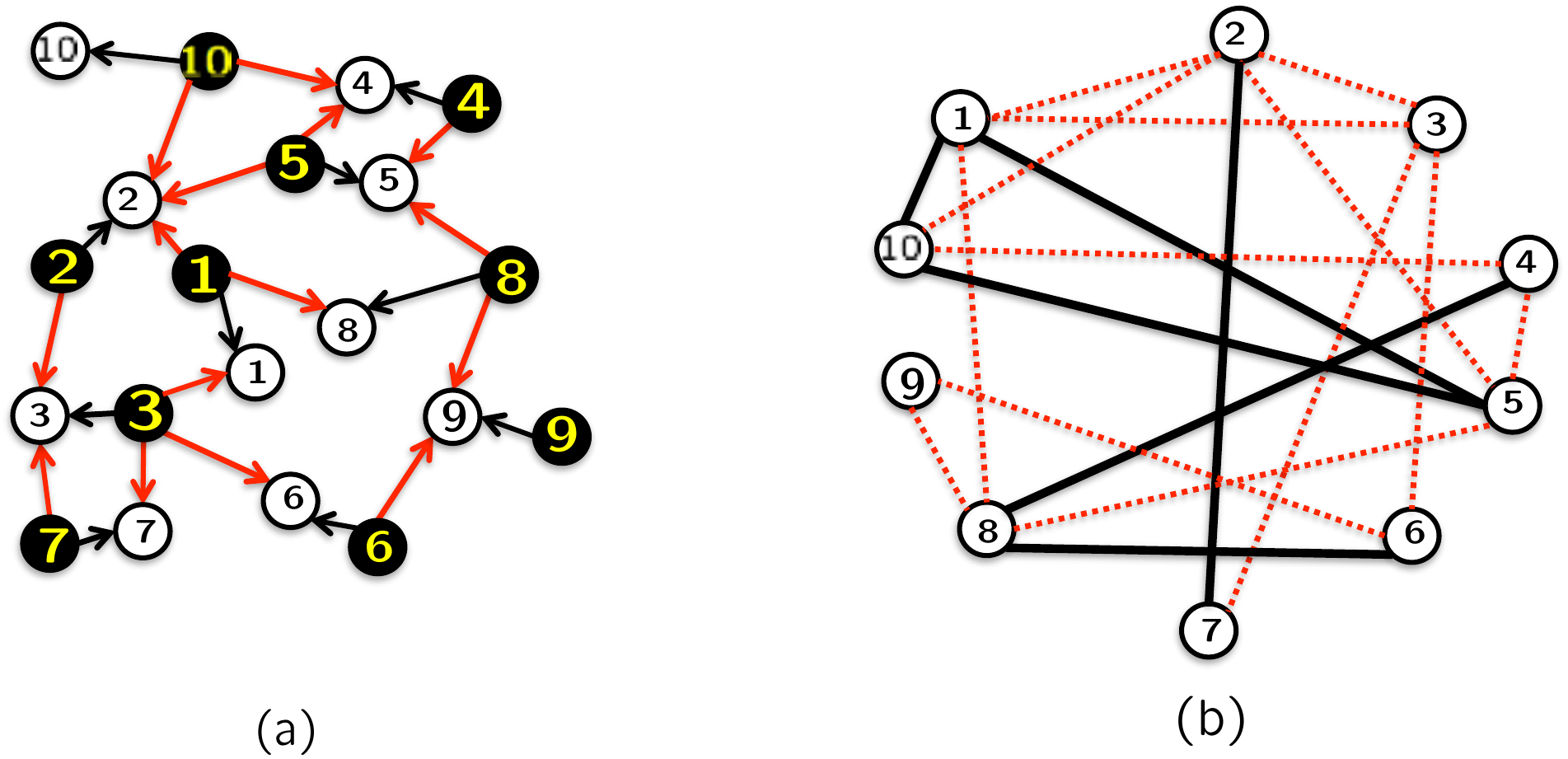}
\caption{\small \it (a) A topological interference management problem, (b) Alignment and Conflict graphs }\label{fig:feasible}
\end{center}
\vspace{-0.5cm}
\end{figure}

The motivating example studied so far is a network that contains no internal conflict, as evident from Fig. \ref{fig:alignedconflict}, which is why rate (DoF) 0.5 is achievable in that network. Another such example is shown in Fig. \ref{fig:feasible}.
 This is a 9 user interference channel, where source $S_i$ (shown in black) has a message $W_i$ for destination $D_i$, the network is partially connected according to the topology shown in Fig. \ref{fig:feasible}(a), and can represent a wireless or wired network as described previously in the unified problem statement. The alignment graph and the conflict graph are shown in Fig. \ref{fig:feasible}(b) with solid black edges and dashed red edges, respectively. Evidently, there are 5 alignment sets: $\{W_1, W_5, W_{10}\}, \{W_2, W_7\}, \{W_3\}, \{W_4, W_6, W_8\}, \{W_9\}$. Since there are no internal conflicts, according to Theorem \ref{theorem:feasible}  the symmetric rate (DoF) value of 0.5 per user is feasible for this partially connected wired (wireless) network with no CSIT beyond the network topology. To achieve this 0.5 DoF per user, it suffices to assign a $2\times 1$ vector to each of the alignment sets, such that any two vectors are pairwise linearly independent. 
 
Going beyond DoF, to find a good constant gap capacity approximation, the vectors should be made as close to orthogonal as possible. For the 5 alignment sets one could choose vectors that are $180^o/5=36^o$ apart. However, the gap can be further improved by noticing that the vectors assigned to the alignment sets need to be linearly independent only for conflicting sets. In particular, note that the alignment set $\{W_9\}$ has no conflict with the alignment set $\{W_1, W_5, W_{10}\}$, so these two alignment sets could be assigned the same vector. Equivalently, alignment sets that have no conflicts can be merged into one alignment set.  Doing so leaves us with $4$ alignment sets, and by assigning vectors that are $180^o/4=45^o$ apart, we obtain the symmetric rate inner bound $R(W_i)=\frac{1}{2}\log(1+\mbox{SNR}/2)$, $i\in\{1,2,\cdots,9\}$, for the partially connected channel model. Now we include weak interference links, bringing us back to the original wireless network. Since including weak interference links can at worst reduce SNR by a factor of 1/2, the symmetric rate achieved is $\frac{1}{2}\log(1+\mbox{SNR}/4)$. The symmetric capacity outer bound, obtained by considering any two messages that cause interference to each other (e.g., $W_1, W_2$) as in the previous example, is again obtained as $\frac{1}{2}\log(1+2\mbox{SNR})$. Thus, the symmetric capacity is bounded as $\frac{1}{2}\log(1+\mbox{SNR}/4)\leq C_{\mbox{\small sym}}\leq\frac{1}{2}\log(1+2\mbox{SNR})$ giving us an approximation that is accurate within 1.5 bits regardless of the value of SNR. On the other hand, if we see the topological interference management problem of Fig. \ref{fig:feasible}(a) as a wired network, then the symmetric capacity is 0.5. 
 
{\bf Field size considerations:} In general, the number of alignment sets can be arbitrarily large, even when rate 0.5  per user is feasible. For wireless networks that operate over complex numbers, one can generate any number of $2\times 1$ vectors that are pairwise linearly independent, so this is not an issue from a DoF perspective. However, for  wired networks, if the operational field size is too small, there may not exist sufficiently many pairwise linearly independent vectors in a 2 dimensional space. For example, over GF(2), there are only 3  vectors of size $2\times 1$ that are pairwise linearly independent. However, this does not affect the capacity result. One can use symbol extensions, e.g., use the channel 4 times, i.e., operate over a 4 dimensional space instead of a 2 dimensional space and assign  2 dimensional pairwise non-intersecting vector subspaces to each of the alignment subgraphs, along which 2 symbols are sent by each transmitter in that subgraph, to equivalently achieve rate 2/4 = 0.5 per user. For example, it is known from \cite{Oggier_Sloane_Diggavi_Calderbank} that the number of pairwise non-intersecting $2$-dimensional subspaces of a $4$ dimensional vector space over GF(2) is $\frac{2^4-1}{2^2-1}=5$, which suffices for our current example. Instead of symbol extensions, one can also use a larger field. Over a larger field, 2 channel uses would still suffice. For example in $\mathbb{GF}(3)$, we can choose $5$ pairwise linearly independent vectors $[1 ~~0]^T, [0~~1]^T, [1 ~~1]^T, [2 ~~1]^T, [1 ~~2]^T$, and assign one to each alignment set of Fig. \ref{fig:feasible}(b).  Finally, one can also try to merge the alignment subgraphs as long as it does not introduce internal conflicts. For example, as mentioned earlier, it is possible to merge $\{W_9\}$ with $\{W_1,W_5\}$ without introducing internal conflicts, which would reduce the number of required pairwise linearly independent subspaces, and is useful if smaller field sizes or less symbol extensions are preferred. 

It is worthwhile to mention that while the half rate (DoF)   feasibility question can be resolved within polynomial complexity, finding the minimum number of alignment sets that can be obtained by merging non-conflicting alignment sets, is NP hard \cite{Dau_Skachek_Chee}. In the  wired setting, this is relevant to the problem of finding if  a solution exists over a given finite field \emph{and} within a given number of channel extensions. In the wireless setting, this is relevant to the problem of finding the best inner bound for wireless networks, because the number of alignment sets determines the spacing between the vectors assigned to the alignment sets, which determines the SNR offset term.

We conclude this section with a simple corollary of Theorem \ref{theorem:feasible}.
\addtocounter{theorem}{-1}
\begin{corollary}
For the topological interference management problem, except the trivial case where there is no interference at all, whenever half rate (DoF) is feasible, it is also the symmetric  capacity (DoF) of the network.
\end{corollary}
In other words, if the symmetric capacity (DoF) of a network is greater than 0.5, then it must be 1, i.e., there can be no interference and every message will have the entire channel to itself. The proof is trivial because whenever two messages interfere, we have either a $Z$ interference channel, an interference channel, a multiple access channel,  a broadcast channel, or a point to point channel, none of which can achieve a symmetric rate (DoF) greater than 0.5 per message. 

\subsection{Optimal versus Conventional Access: Best Case Improvement}
The conventional approach to dealing with interferers that are too strong to be treated as noise, is to avoid them by assigning them orthogonal time/frequency slots (TDMA/FDMA) or precoding sequences (CDMA). In a fully connected $K$ user interference channel, conventional schemes such as TDMA/FDMA/CDMA can only achieve a symmetric DoF of $1/K$ per user. With perfect CSIT, as shown in \cite{Cadambe_Jafar_int}, interference alignment  allows (for almost all channel realizations) $1/2$ DoF per user, regardless of the number of users $K$. Thus, the improvement of the optimal scheme over orthogonal schemes is a factor of $K/2$, which grows linearly with the number of users. In a fully connected $K$-user interference network, this is also the \emph{best case} improvement factor, i.e., a greater improvement cannot be obtained for any realization of the channel coefficients.

Our present setting allows no CSIT beyond the knowledge of the topology of a partially connected network. Because the network is partially connected, it is possible in certain instances to still achieve $1/2$ DoF per user even without CSIT, as seen in the examples of Fig. \ref{fig:equivalence}(a) and Fig. \ref{fig:feasible}(a), and such instances are completely characterized in Theorem \ref{theorem:feasible}. However,  even orthogonal access schemes may be capable of achieving more than $1/K$ DoF per user in a partially connected network. Therefore, a natural question is to determine how much improvement, if any, can be achieved by optimal topological interference management solutions as compared to more conventional alternatives. In particular, in this section we will try to find the best case improvement over conventional alternatives. Since the CSIT (and therefore the DoF) are only  functions of the network topology, by best case is meant the  topology that leads to the highest improvements. 

In terms of network models, in this section we will study the multiple unicast setting as well as the general multiple groupcast setting. Within each setting we will especially consider half-rate-feasible networks since such networks are completely characterized in Theorem \ref{theorem:feasible}. For conventional alternatives, we will consider orthogonal  and multicast schemes which correspond to TDMA and CDMA, respectively, in their various generalized forms such as (fractional) orthogonal scheduling and (fractional)  partition multicast. The definitions of these terms are summarized in Appendix \ref{sec:conventionalaccess}. 

Note that while our discussion in this section will often be framed in the context of the DoF of wireless networks, as described earlier the results can equivalently be interpreted in terms of the capacity of wired networks, and in terms of the corresponding instances of the index coding problem.

\subsubsection{Multiple Unicasts: Half-rate-feasible networks}
Since we have a complete characterization of half-rate-feasible networks for the topological interference management problem, let us first compare the DoF optimal solution versus the best orthogonal (TDMA) or multicast (CDMA) solution for such networks. The result is stated in the following theorem.

\begin{theorem}\label{theorem:hrfcompare}
For every multiple unicast topological interference management problem where symmetric  rate (DoF)   of 0.5 per message is achievable, a fractional orthogonal scheduling scheme (TDMA) can achieve at least the symmetric rate (DoF)  of 0.25 per message, and so can fractional partition multicast (CDMA). Therefore, for this class of networks, the best case improvement of optimal versus fractional orthogonal scheduling or fractional partition multicast schemes is no more than a factor of 2.
\end{theorem}
\addtocounter{theorem}{-1}
\begin{corollary}\label{corollary:hrfindex}
For every multiple unicast index coding problem where symmetric  rate of 0.5 per message is achievable, a fractional clique cover scheme can achieve at least the symmetric rate of 0.25 per message, and so can fractional partition multicast.
\end{corollary}

The proof of Theorem \ref{theorem:hrfcompare} is presented in Appendix \ref{proof:hrfcompare}.  Corollary \ref{corollary:hrfindex} is a consequence of the equivalence between topological interference alignment and index coding problems established in theorems \ref{theorem:one} and \ref{theorem:linear}.

In light of this result, a question of immediate interest is whether this result can be tightened further. For example, is it possible that a fractional orthogonal scheduling (or fractional partition multicast) scheme can always achieve a symmetric DoF value of 0.3 per message whenever a symmetric DoF value of 0.5 per message is information theoretically feasible?   The following theorem shows that indeed 0.25 symmetric DoF is the best universal guarantee that can be made.

\begin{theorem}\label{theorem:hrfoptimal}
For every $\epsilon>0$, there exists a multiple unicast topological interference management problem where the symmetric rate (DoF) of 0.5 per message is achievable, and the best fractional orthogonal scheduling and the best fractional partition multicast scheme cannot achieve a symmetric rate (DoF) higher than $0.25+\epsilon$ per message.
\end{theorem}
\addtocounter{theorem}{-1}
\begin{corollary}\label{corollary:hrfoptimalindex}
For every $\epsilon>0$, there exists a multiple unicast index coding problem where the symmetric rate of 0.5 per message is achievable, and the best fractional clique cover and the best fractional  partition multicast scheme cannot achieve a symmetric rate higher than $0.25+\epsilon$ per message.
\end{corollary}
Theorem \ref{theorem:hrfoptimal} is proved in Appendix \ref{proof:hrfoptimal} and the corollary follows from the equivalence between topological interference management and index coding problems established in theorems \ref{theorem:one} and \ref{theorem:linear}.

To paraphrase theorems \ref{theorem:hrfcompare} and \ref{theorem:hrfoptimal}, in a multiple unicast setting with no CSIT except knowledge of the partially connected network topology, whenever it is possible for everyone to get half the cake, it is  possible for everyone to get a quarter of the cake through orthogonal access, i.e., TDMA, as well as through partition multicast, i.e., CDMA. Further, for this class of networks, the factor of 2 improvement is the best  guarantee that can be made over orthogonal and partition multicast schemes.

To conclude this section, let us consider the non-fractional versions of orthogonal scheduling and partition multicast schemes. An interesting observation from Theorem \ref{theorem:hrfcompare} (Corollary \ref{corollary:hrfindex}) is that a half-rate-feasible $K$-unicast topological interference management (index coding) problem always contains a set of non-interfering messages, i.e., an independent set (clique) of size at least $\lceil K/4\rceil$. This follows because if a fractional orthogonal  scheme achieves a symmetric rate (DoF) of $0.25$ then a sum-rate (DoF) of $K/4$ must be achievable by an orthogonal scheme, but the sum-rate (DoF) of an orthogonal scheme is simply the size of the largest independent set (clique number).
 
 Based on this observation, we  obtain an interesting achievability result for  orthogonal scheduling and partition multicast schemes \emph{without fractionalization}. In a half-rate-feasible $K$-unicast setting, an orthogonal scheduling scheme can serve  $\lceil K/4\rceil$ messages in the first time slot. This leaves us with  a $K_1$-unicast problem where $K_1=(K-\lceil K/4\rceil)$. Since the problem is still half-rate-feasible, an orthogonal scheme can now serve $\lceil K_1/4\rceil$ messages in the second time slot,  leaving us with  a $K_2$-unicast problem where $K_2=(K_1-\lceil K_1/4\rceil)$. Continuing this pattern for $n$ channel uses, we are left with a $K_n$-unicast problem where $K_n=(K_{n-1}-\lceil K_{n-1}/4\rceil)$. At this point, one can serve the remaining $K_n$ users one at a time, over $K_n$ channel uses. Define $K_0=K$. We can optimize over $n$. Thus, for half-rate feasible $K$-unicast topological interference management and index coding problems, orthogonal scheduling and partition multicast schemes, without fractionalization, can achieve at least a symmetric rate (DoF) value of 
 \begin{eqnarray*}
 DoF_{\mbox{\tiny sym}}&=& \sup_{n\in\mathbb{Z}^+}\frac{1}{n+K_n}\geq \sup_{n\in\mathbb{Z}^+}\frac{1}{n+\left(\frac{3}{4}\right)^nK}\geq\sup_{n\in\mathbb{R}^+}\frac{\left(\frac{4}{3}\right)^n}{n\left(\frac{4}{3}\right)^n+K}\geq\frac{1}{\log_{4/3}(K)+1}
 \end{eqnarray*}
 where in the last step, instead of the optimizing value $n=\log_{4/3}(K)+\log_{4/3}\log(4/3)$, we have chosen $n=\log_{4/3}(K)$ for a somewhat simpler expression.
 
\subsubsection{Multiple Unicasts: Half-rate-infeasible networks}

As we show next,  more significant improvements are possible when we we consider  networks where half-rate is not feasible, or we consider groupcast settings. The following theorem considers multiple unicast settings where 0.5 symmetric DoF is \emph{not} achievable.

\begin{theorem}\label{theorem:Hadamard}
There exists an explicit family of  multiple unicast topological interference management problems with $K$ messages where the  symmetric capacity (DoF) is at least $1/3$ per message, whereas  the highest achievable rate (DoF) of fractional orthogonal scheduling schemes as well as fractional partition multicast schemes is at most $(1+o(1))K^{-1/4}$. The best case improvement is therefore at least $(1/3+o(1))K^{1/4}$.
\end{theorem}
Theorem \ref{theorem:Hadamard} is a direct consequence of the equivalence results established in theorems \ref{theorem:one} and \ref{theorem:linear} and a corresponding result  for the index coding problem, established by Blasiak et al.  in Theorem 3 of \cite{Blasiak_Kleinberg_Lubetzky_2010}. Blasiak et al. present a symmetric instance of a $K$-unicast  index coding problem (where the side information graph is undirected) based on a  projective Hadamard graph, where a symmetric rate of $1/3$ per message is achievable, but the clique number, which bounds the sum-rate of fractional orthogonal schemes,  is  at most $(1+o(1))K^{3/4}$. Thus the symmetric rate of fractional orthogonal schemes cannot be higher than $(1+o(1))K^{-1/4}$ per message. The extension to fractional partition multicast follows from the observation that the projective Hadamard example is symmetric (side information graph is undirected), and therefore the subset of messages that achieves  the highest sum-rate with fractional partition multicast can be covered by disjoint cliques that achieve the same sum-rate \cite{Tehrani_Dimakis_Neely}. Thus, the same bounds hold for the highest sum-rate and the symmetric rate achievable by fractional partition multicast schemes. 

Since the result is only a lower bound on the best case improvement, a question of immediate interest is to determine if other topologies can be found for which the  improvement can be even higher, and conversely to develop non-trivial outer bounds on the best case improvement. An intriguing question to settle is whether improvements that scale linearly in $K$, such as the $K/2$ factor improvement with perfect CSIT,  are still possible with no CSIT beyond topology. 

\subsubsection{Multiple Groupcasts: Half-rate-feasible networks}
In this section we will study  the best case improvement in the groupcast setting for fractional orthogonal scheduling and fractional partition multicast. Let us start with fractional orthogonal scheduling schemes.

\begin{theorem}\label{theorem:groupcastorthogonal}
There exists an explicit family of  multiple groupcast topological interference management problems with $K$ messages where the  symmetric capacity (DoF) is 0.5 per message, whereas  the highest achievable rate (DoF) of fractional orthogonal  scheduling schemes is not more than $\frac{1}{K}$ per message. 
\end{theorem}
\addtocounter{theorem}{-1}
\begin{corollary}\label{corollary:groupcastorthogonal}
There exists an explicit family of  multiple groupcast index coding problems with $K$ messages where the  symmetric capacity  is 0.5 per message, whereas  the highest achievable rate  of  fractional hyperclique cover schemes is not more than $\frac{1}{K}$ per message. 
\end{corollary}
Note that a symmetric rate of $1/K$ is always achievable by a (non-fractional) orthogonal scheduling scheme in a network with $K$ messages. Therefore, Theorem \ref{theorem:groupcastorthogonal} characterizes the best case improvement over orthogonal and fractional orthogonal schemes for half-rate-feasible $K$-groupcast topological interference management problems as $K/2$. Remarkably, this is the best case improvement  over orthogonal (TDMA) schemes even with full CSIT.

Next let us bound the best case improvements over (fractional) partition multicast schemes. 

\begin{theorem}\label{theorem:groupcast}
There exists an explicit family of  multiple groupcast topological interference management problems with $K$ messages where the  symmetric capacity (DoF) is 0.5 per message, whereas  the highest achievable rate (DoF) of  fractional partition multicast schemes is not more than $\frac{1}{\sqrt{K}}$ per message. 
\end{theorem}
\addtocounter{theorem}{-1}
\begin{corollary}\label{corollary:groupcast}
There exists an explicit family of  multiple groupcast index coding problems with $K$ messages where the  symmetric capacity is 0.5 per message, whereas  the highest achievable rate of  fractional partition multicast schemes is not more than $\frac{1}{\sqrt{K}}$ per message. 
\end{corollary}

Now that we have a lower bound of $\sqrt{K}/2$ on the best case improvement over fractional partition multicast, let us find an upper bound. 
\begin{theorem}\label{theorem:groupcastmax}
For any multiple groupcast topological interference management problem with $K$ messages where the  symmetric rate (DoF) of 0.5 per message is feasible, a  partition multicast scheme can achieve a symmetric rate (DoF)  that is at least $\frac{1}{\lceil\sqrt{2K}\rceil}$ per message.
\end{theorem}
\addtocounter{theorem}{-1}
\begin{corollary}\label{theorem:groupcastmaxindex}
For any multiple groupcast index coding problem with $K$ messages where the  symmetric rate of 0.5 per message is feasible, a  partition multicast scheme can achieve a symmetric rate that is at least $\frac{1}{\lceil\sqrt{2K}\rceil}$ per message.
\end{corollary}

{\it Remark:} The proof presented in Section \ref{proof:groupcastmax} finds a tighter upper bound, $\frac{1}{\lceil \frac{\sqrt{8K+1}-1}{2}\rceil}$ instead of $\frac{1}{\lceil\sqrt{2K}\rceil}$  but the latter is reported in Theorem \ref{theorem:groupcastmax} for its relative simplicity.

Note that the outer bound is proved for fractional partition multicast while the inner bound is proved for partition multicast. Since fractional partition multicast includes partition multicast as a special case, the outer bound also applies to partition multicast and the inner bound also applies to fractional partition multicast.

In other words, the result says that there exist $K$-groupcast networks with no CSIT except topology, where everyone gets half the cake through interference alignment, but the best TDMA scheme cannot achieve more than the trivial value of $1/K$ of the cake per message, and where the best CDMA scheme cannot achieve more than $1/\sqrt{K}$ of the cake per message. For all such networks, the best CDMA scheme can always achieve at least $\frac{1}{\lceil\sqrt{2K}\rceil}$ of the cake per message. Thus, the best case improvement over TDMA (orthogonal schemes) is $K/2$, and the best case improvement over CDMA (partition multicast schemes) is between $\frac{\sqrt{K}}{2}$ and $\frac{\lceil\sqrt{2K}\rceil}{2}$.

These results are limited to half-rate-feasible networks. As shown in the multiple unicast case,  even stronger improvements over partition multicast schemes may be achievable in multiple groupcast  settings that are not half-rate-feasible. Since such settings are far from understood, we leave this as an open problem, and instead focus on exploring half-rate-infeasible networks in the next section.

\subsection{When is $1/K$ of the cake per message optimal?}\label{sec:1/K}

Here we are interested in $K$-groupcast settings where the symmetric capacity is $1/K$, i.e., the trivial solution that everyone gets $1/K$ of the cake is capacity optimal. In this section, we will obtain a complete characterization of such networks.

A sufficient condition for a $K$-unicast network to have symmetric capacity $1/K$ was obtained originally by  Bar-Yossef et al. in \cite{Yossef_Birk_Jayram_Kol_Trans} and generalized subsequently to the multiple groupcast setting by Neely et al. in \cite{Neely_Tehrani_Zhang}. A relevant notion here is the demand graph, defined for the multiple-groupcast index coding problem in \cite{Neely_Tehrani_Zhang}. Note that the definition applies also to multiple-unicast settings (as a special case of multiple-groupcast) and in both unicast and groupcast settings can be similarly applied to the topological interference management problem.

\noindent{\bf Demand Graph for Index Coding:} For a $K$-groupcast index coding problem, the demand graph is defined as the following directed bi-partite graph with the destination nodes on one side and message nodes on the other \cite{Neely_Tehrani_Zhang}. There is a directed edge from  message node $W_j$ to a destination node $D_i$ if the message $W_j$ is desired  by destination $D_i$, i.e., $W_j\in\mathcal{W}(D_i)$. There is a directed edge from a destination node $D_i$ to a message $W_j$ if the destination $D_i$ has message $W_j$ as antidote, i.e., $a_{ik}=1$, where $W_j\in\mathcal{W}(S_k)$.\\
{\bf Demand Graph for Topological Interference Management:} For a $K$-groupcast topological interference management problem, the demand graph is defined as the following directed bi-partite graph with the destination nodes on one side and message nodes on the other. There is a directed edge from  message node $W_j$ to a destination node $D_i$ if the message $W_j$ is desired  by destination $D_i$, i.e., $W_j\in\mathcal{W}(D_i)$. There is a directed edge from a destination node $D_i$ to a message $W_j$ if the destination $D_i$ cannot hear the source of the message $W_j$, i.e., $t_{ik}=0$, where $W_j\in\mathcal{W}(S_k)$.

A directed graph is acyclic if it contains no directed cycles, i.e., it  is not possible to return to any starting point while traversing a sequence of directed edges, while respecting the direction of the edges.

Bar-Yossef et al. showed in \cite{Yossef_Birk_Jayram_Kol_Trans}  that if the demand graph of a $K$-unicast index coding problem is acyclic then the symmetric capacity is $1/K$ per message. Neely et al. generalized the result in \cite{Neely_Tehrani_Zhang} to show that if the demand graph of a $K$-groupcast index coding problem is acyclic then the  symmetric capacity is $1/K$ per message. We state here a stronger form of these results and its extension to the \emph{exact} capacity of the original wireless network.

\begin{figure}[!h]
\begin{center}
\includegraphics[width=4.5in]{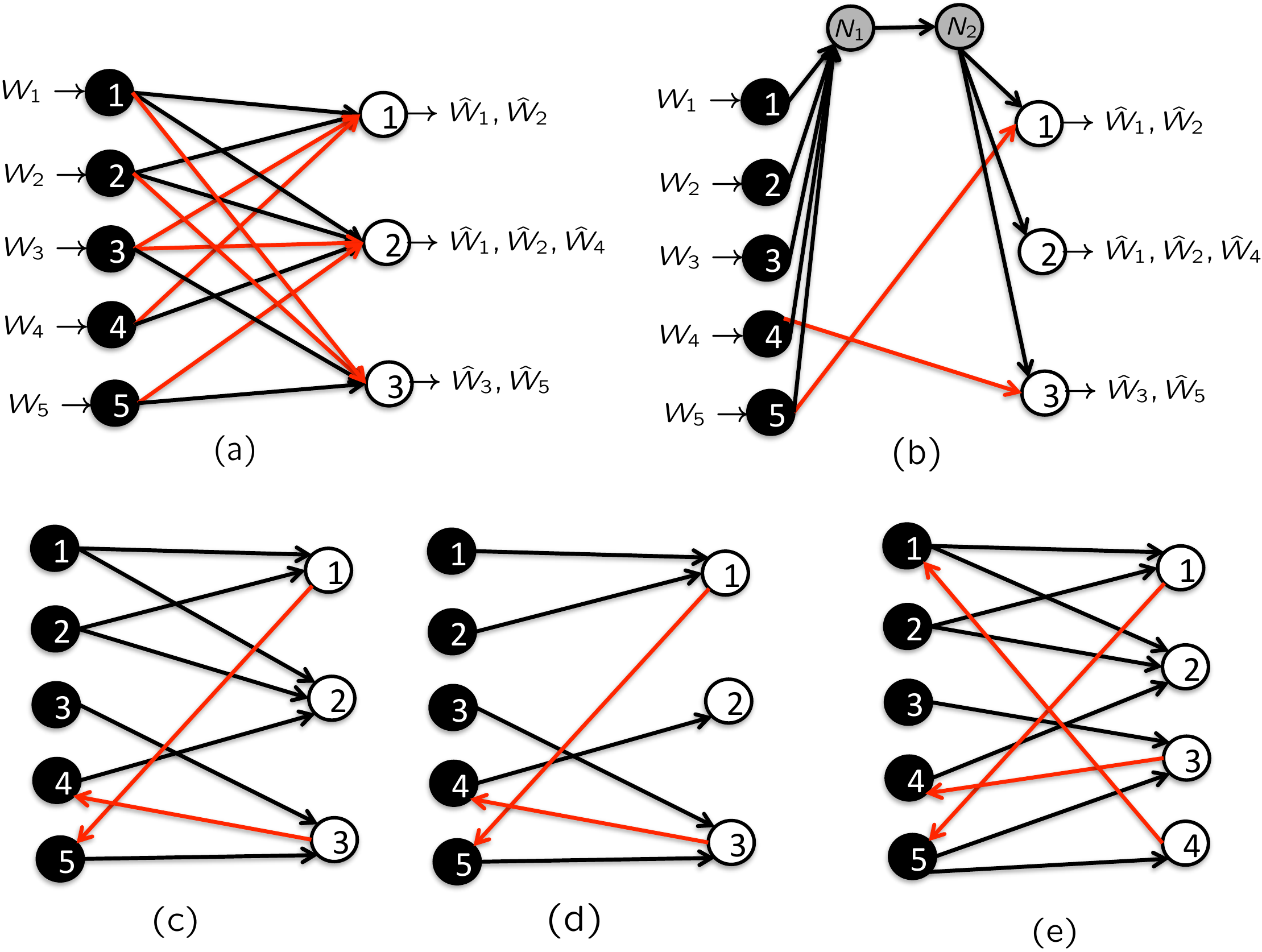}
\caption{\small \it (a) A $5$-groupcast topological interference management problem,  (b) corresponding index coding problem, (c) the demand graph for both (note that it is acyclic, so the symmetric capacity is $1/5$ for both networks), (d) demand graph for a $5$-unicast problem obtained by relaxing the 5-groupcast problem (e) a $5$-groupcast problem whose demand graph is not acyclic but its symmetric capacity is still $1/5$ (because it can be relaxed to obtain the same $5$-unicast network).}\label{fig:noalign}
\end{center}
\vspace{-0.5cm}
\end{figure}

\begin{theorem}\label{theorem:noalign}
For a $K$-unicast topological interference management problem to have  symmetric capacity (DoF) of $1/K$ per message it is necessary and sufficient that  its demand graph is acyclic. For a $K$-groupcast topological interference management problem to have symmetric capacity (DoF) of $1/K$ per message, it is sufficient but not necessary that its demand graph is acyclic. The necessary and sufficient condition for a $K$-groupcast topological interference management problem to have symmetric capacity (DoF) of $1/K$ per message, is that it  can be relaxed (by eliminating certain message demands) into a $K$-unicast topological interference management problem for which the demand graph is acyclic. Further, if each message originates at a distinct source node, then the exact symmetric capacity of the original wireless network is $\frac{1}{K}\log(1+K\mbox{SNR})$.
\end{theorem}
\addtocounter{theorem}{-1}
\begin{corollary}\label{corollary:noalignindex}
For a $K$-unicast index coding problem to have  symmetric capacity (DoF) of $1/K$ per message it is necessary and sufficient that  its demand graph is acyclic. For a $K$-groupcast index coding problem to have symmetric capacity (DoF) of $1/K$ per message, it is sufficient but not necessary that its demand graph is acyclic. The necessary and sufficient condition for a $K$-groupcast index coding problem to have symmetric capacity (DoF) of $1/K$ per message, is that it  can be relaxed (by eliminating certain message demands) into a $K$-unicast topological interference management problem for which the demand graph is acyclic.
\end{corollary}
Note that the relaxed problem must be a $K$-unicast problem, i.e., it must still have all $K$ messages, each of which is desired  by exactly one destination. We can eliminate (all but one) message demands only for those messages that are desired by more than one destination node. Since eliminating certain message demands cannot reduce the capacity region, the capacity region of the relaxed network must include the capacity region of the original network. Note that some destination nodes may be eliminated. For example,  Fig. \ref{fig:noalign}(d) is the demand graph of a  5-unicast network obtained by relaxing the network whose demand graph is shown in Fig. \ref{fig:noalign}(e), and destination $4$ is eliminated without reducing the number of messages.

The result in  Theorem \ref{theorem:noalign} and Corollary \ref{corollary:noalignindex} is stronger than the previous results in the following ways.  
\begin{enumerate}
\item For the multiple unicast index coding problem it is known that  the demand graph being acyclic is \emph{sufficient} to conclude that the symmetric capacity is $1/K$ \cite{Yossef_Birk_Jayram_Kol_Trans}. However, Corollary \ref{corollary:noalignindex} states that it is also necessary.
\item For the multiple groupcast index coding problem also it is known that the demand graph being acyclic is {sufficient} to conclude that the symmetric capacity is $1/K$ \cite{Neely_Tehrani_Zhang}. Interestingly, for multiple groupcast it is not necessary. This can be seen from the demand graph of Fig. \ref{fig:noalign}(e) where one more destination node (number 4) has been added, that desires message $5$ and has message $1$ available as antidote, thus introducing a cycle into the demand graph. However, since the additional destination cannot increase the capacity region, the symmetric capacity is still $1/5$ per message. Evidently the acyclic demand graph condition is not necessary for  $K$-groupcast index coding networks to have symmetric capacity $1/K$. What is necessary and sufficient, however, is that the $K$-groupcast network can be relaxed into a $K$-unicast network for which the demand graph is acyclic, as stated in Theorem  \ref{theorem:noalign} and Corollary \ref{corollary:noalignindex}. 
\item For this class of wireless networks we obtain not just a capacity approximation, but the capacity itself.
\end{enumerate}

Thus, we have a complete characterization of $K$-groupcast  networks (and $K$-unicast networks as a special case) that have symmetric capacity (DoF) of $1/K$ for all three cases: index coding, linear wired networks and partially connected wireless networks. 

Even for networks where the symmetric capacity is more than $1/K$, the acyclic demand graph property is useful to bound the rates of subsets of messages. This is a commonly used bound in index coding literature. Here we state the corresponding bound for topological interference management problem.

\begin{theorem}[\cite{Yossef_Birk_Jayram_Kol_Trans}]\label{theorem:acyclicbound}
The symmetric capacity (DoF)  of a topological interference management problem  is bounded above as
\begin{eqnarray}
C_{\mbox{\small sym}}&\leq&\frac{1}{\Psi}
\end{eqnarray}
where $\Psi$ is the maximum cardinality of an acyclic subset of messages.
\end{theorem}
\noindent{\bf Acyclic Subset of Messages:} A subset of messages $\mathcal{W}_o\in\mathcal{W}$ is said to be acyclic if the symmetric capacity (DoF) of this subset of messages is $\frac{1}{|\mathcal{W}_o|}$ per message. 

The definition applies to both index coding and topological interference management problems. Note that the necessary and sufficient condition for a subset of messages to be acyclic is easily obtained from Theorem \ref{theorem:noalign} for the topological interference management problem and from Corollary \ref{corollary:noalignindex} for the index coding problem by eliminating the rest of the messages.

\subsection{Between the extremes: When less than $1/2$ but more than $1/K$ of the cake is Optimal}
Beyond the non-trivial extremes of $1/2$ the cake or only $1/K$ of the cake per message, capacity characterizations are available for the index coding problem for certain special antidote graphs such as the Peterson graph, the Grotzsch graph, and the Chvatal graph, on 10, 11 and 12 vertices (messages), respectively,  for various Cayley graphs of cyclic groups \cite{Blasiak_Kleinberg_Lubetzky_2010}, for graphs associated with certain matroids \cite{Blasiak_Kleinberg_Lubetzky_2011},  and  cases where the trivial inner and outer bounds  (represented by the independence number and the clique cover numbers) coincide. Since the capacity achieving solutions in most of these cases are based on linear schemes the solutions can be translated directly into corresponding instances of the  topological interference management problem. However,  there are few exact capacity results or general principles available for large classes of problem instances. In this section we will use the interference alignment perspective to solve some interesting classes of index coding and topological interference management problems. 

\subsubsection{Internal Conflicts: An Interference Alignment Perspective}
Since we are dealing with half-rate-infeasible networks, these networks must have internal conflicts. Recall that an internal conflict is when two messages are in the same alignment set, and at least one of them causes interference to the other's desired destination. The ``conflict" is evident from an interference alignment perspective. Messages  in the same alignment set should be aligned as much as possible, but conflicting messages cannot be aligned. The tension created by this conflict is captured by the following outer bound on the symmetric capacity, derived previously by Blasiak et al. in  \cite{Blasiak_Kleinberg_Lubetzky_2010} from a graph theoretic perspective and by Maleki et al. in \cite{Maleki_Cadambe_Jafar} from an interference alignment perspective. As we have shown earlier in Theorem \ref{theorem:one},  the index coding problem provides an outer bound on the corresponding topological interference management problem, any  outer bound for the index coding problem is directly inherited by the corresponding topological interference management problem as well.

\begin{theorem}[\cite{Blasiak_Kleinberg_Lubetzky_2010,Maleki_Cadambe_Jafar}] \label{theorem:slide}
The symmetric capacity of a half-rate-infeasible multiple-groupcast index coding problem is bounded above as:
\begin{eqnarray}
C_{\mbox{\small sym}}&\leq&\frac{\Delta}{2\Delta+1}
\end{eqnarray}
where $\Delta$ is the minimum internal conflict distance.
\end{theorem}
\addtocounter{theorem}{-1}
\begin{corollary}\label{corollary:slide}
The symmetric capacity (DoF) of a half-rate-infeasible multiple-groupcast topological interference management problem is bounded above as:
\begin{eqnarray}
C_{\mbox{\small sym}}&\leq&\frac{\Delta}{2\Delta+1}
\end{eqnarray}
where $\Delta$ is the minimum internal conflict distance.
\end{corollary}

{\bf Conflict Distance:} For two nodes that have an internal conflict between them, the conflict distance is defined as the minimum number of alignment graph edges that need to be traversed to go from one node to the other. The minimum conflict distance of all internal conflicts is denoted as $\Delta$. 

{\it Remark:} Note that the alignment graph may have several connected components, each of which is an alignment set. For internal conflicts, the distance is over each set. Thus, we can find the minimum distance for each alignment set and then find the minimum over all alignment sets to get the conflict distance.

\begin{figure}[!h]
\begin{center}
\includegraphics[width=3.2in]{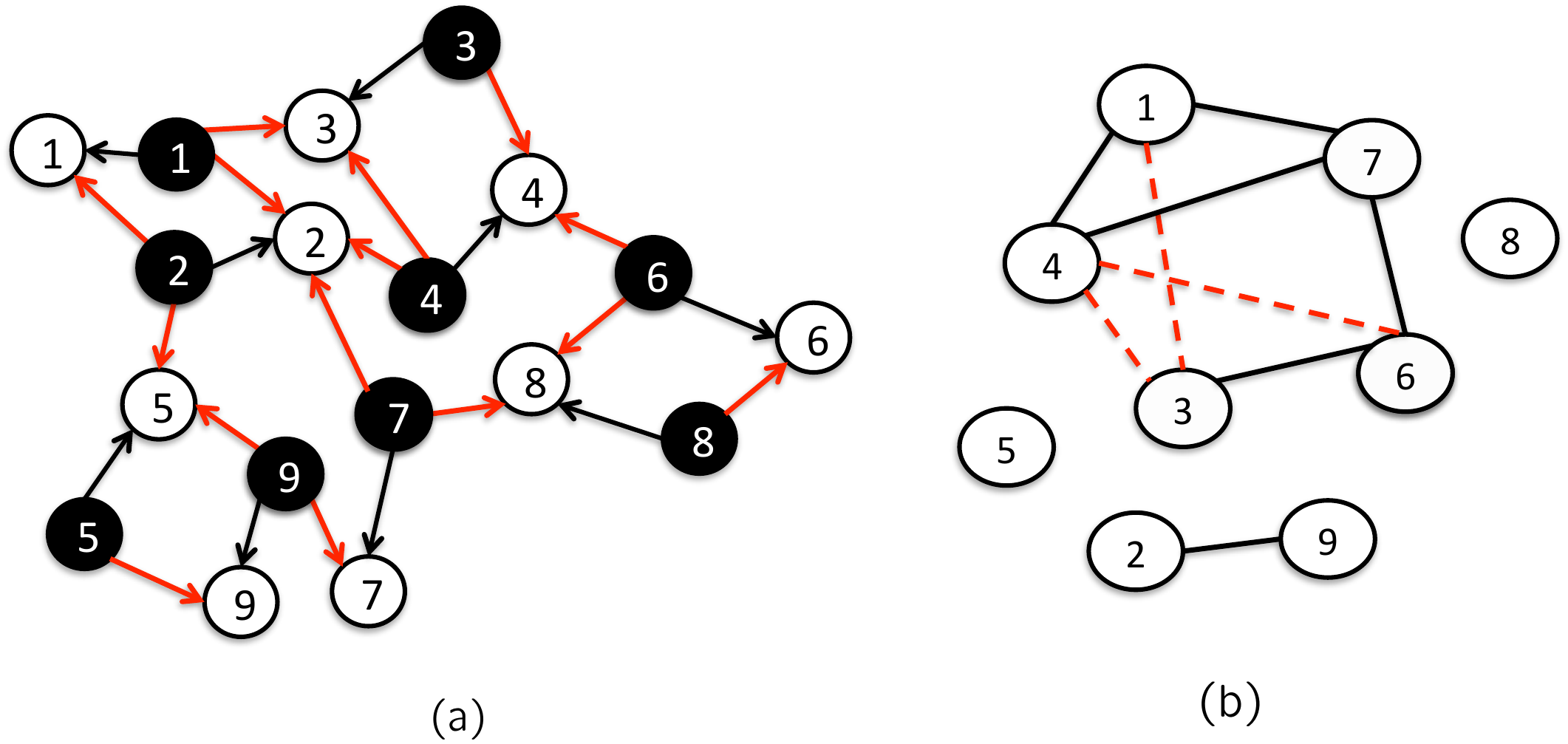}~~~~\includegraphics[width=2.1in]{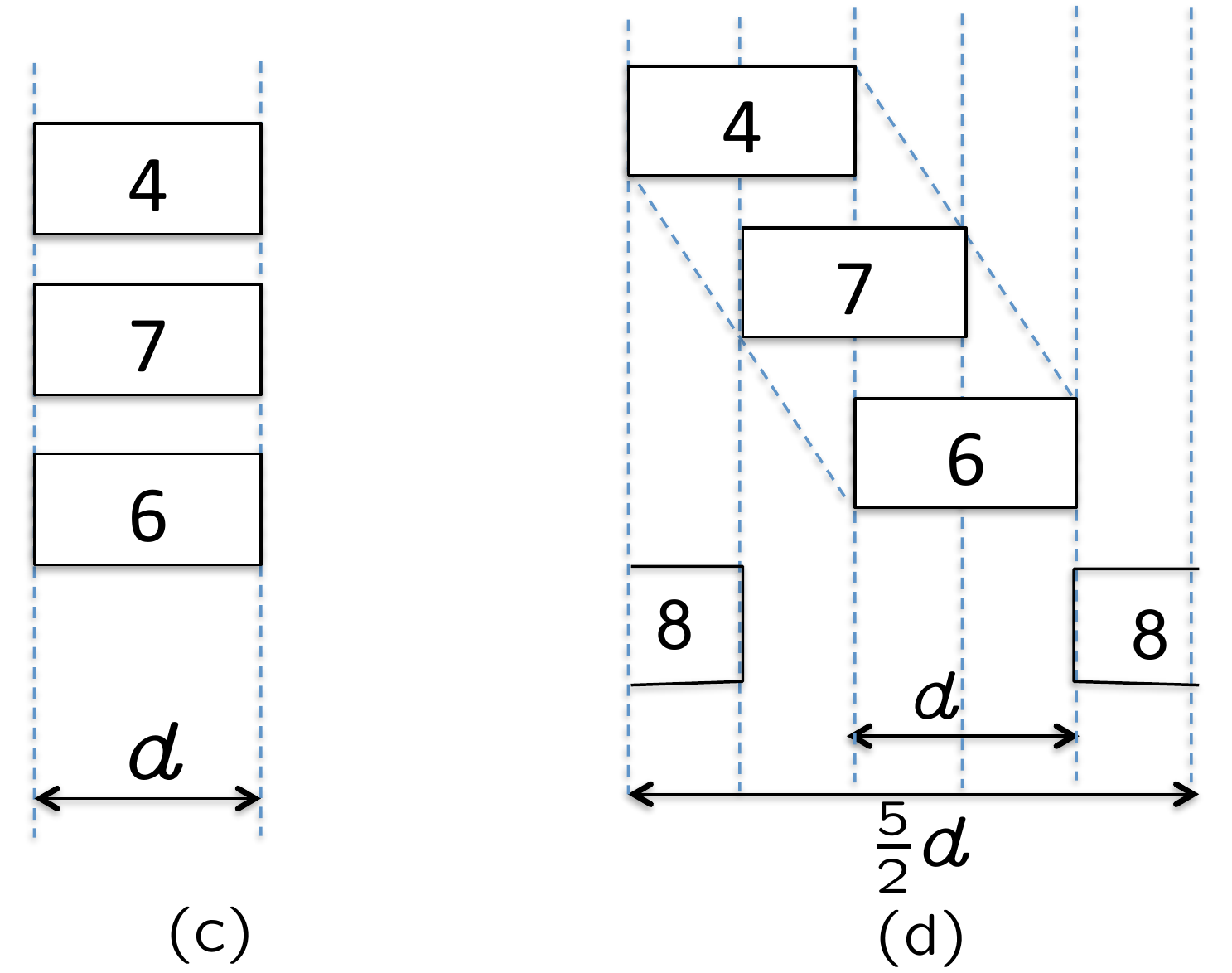}
\caption{\small \it (a) An instance of the topological interference management problem, and (b) its alignment graph (black edges) showing internal conflicts as dashed red edges. Conflicts that are not internal are not shown. (c) If there are no internal conflicts, nodes that belong to the same alignment set should align perfectly (d) With internal conflicts, conflicting nodes slide out of each other's way just enough to avoid overlaps while all nodes try to stay as aligned as possible with their adjacent nodes.}\label{fig:internal}
\end{center}
\end{figure}
As an example, consider the $9$-unicast topological interference management problem shown  in Fig.\ref{fig:internal}(a), whose alignment graph and internal conflicts are shown in Fig. \ref{fig:internal}(b) as black edges and dashed red-edges, respectively. Note that for clarity, conflicts that are not internal are not shown. For example, there is a conflict between nodes $6$ and $8$ that is not shown. There is an internal conflict  between nodes $3$ and $4$, and they are connected through the path 4 --- 1 --- 7 --- 6 --- 3 which is comprised of 4 edges. However, a shorter path exists, 4 --- 7 --- 6 --- 3, which is comprised of only 3 edges and which is also the shortest path between 3 and 4. So the conflict distance between 3 and 4 is 3.  However, this is not the minimum conflict distance. The conflict between 4 and 6 has conflict distance  2 because of the path 4 --- 7 --- 6, and  is the minimum conflict distance, so that $\Delta=2$ for this network.  Therefore, according to Corollary \ref{corollary:slide} the symmetric capacity (DoF) of this network cannot be more than $2/5$. Indeed, as we will see for this example, $2/5$ is the symmetric capacity.

Since the information theoretic proof of Theorem \ref{theorem:slide} is already available in \cite{Blasiak_Kleinberg_Lubetzky_2010,Maleki_Cadambe_Jafar}, here we will highlight only the intuition from an interference alignment perspective, using Fig. \ref{fig:internal}(c) and Fig. \ref{fig:internal}(d), which will naturally explain the achievable scheme as well.

Message $W_4$ should align as much as possible with message $W_7$ because they both cause interference to destination $D_2$. Similarly, message $W_7$ should align as much as possible with message $W_6$ because they both cause interference to destination $D_8$. If there were no internal conflicts, this would force $W_4, W_6, W_7$ to align perfectly (as done in half-rate-feasible networks). However, as it turns out in this network, $W_4$ cannot align with $W_6$ because they have an internal conflict ($W_6$ interferes with the desired destination of $W_4$). So, $W_4$ and $W_6$ must ``slide" out of each other's way, as shown in Fig. \ref{fig:internal}(d). Since both $W_4$ and $W_6$ want to align with $W_7$, the natural resolution is for half of $W_7$ to align with $W_4$ and the remaining half to align with $W_6$. If each message occupies $d$ signal dimensions, this means that there is a $d/2$ dimensional overlap between $W_4, W_7$ and a $d/2$ dimensional overlap between $W_6,W_7$ and no overlap between $W_4,W_7$. Next let us see how this leads us to the outer bound of $2/5$ on the symmetric capacity (DoF). 

Consider destination $D_8$ that desires message $W_8$, and sees interference from $W_6, W_7$.  Note that because of the partial overlap between them, $W_6, W_7$ together occupy $1.5d$ dimensions. The desired message $W_8$ cannot overlap with the interference space. So the total number of dimensions available to destination $D_8$ to resolve the desired signal from interference must be at least $d+1.5d=2.5d$ dimensions. Since each destination sees one normalized signal dimension per channel use, we have the outerbound:
\begin{eqnarray}
2.5d&\leq&1\\
\Rightarrow d&\leq&2/5.
\end{eqnarray}
That is, the symmetric capacity (DoF) cannot be more than $0.4$ for this network, as also stated in Corollary \ref{corollary:slide}.

The same intuitive explanation also leads us to the achievability scheme. To achieve  symmetric rate (DoF) of $2/5$ per user with a linear scheme, we will operate over $5$ channel uses and send 2 symbols from each message. The precoding vectors for $W_4,W_6,W_7$ are chosen as:
\begin{eqnarray}
W_4&:& {\bf v}_1, {\bf v}_2\\
W_7&:& {\bf v}_1, {\bf v}_4\\
W_6&:& {\bf v}_3, {\bf v}_4
\end{eqnarray}
where ${\bf v}_i$ are randomly generated vectors (in a sufficiently large field so that any $5$ of them are linearly independent). Note that $W_4, W_7$ align in half of their dimensions, as do $W_6,W_7$ but there is no overlap between $W_4,W_6$, consistent with the intuition highlighted in Fig. \ref{fig:internal}(d). Based on the alignment and conflict graphs, the remaining messages can be assigned precoding vectors as follows:
\begin{eqnarray}
\begin{array}{lcl}
W_1&:& {\bf v}_1, {\bf v}_2\\
W_3&:& {\bf v}_3, {\bf v}_5\\
W_2&:&{\bf v}_6, {\bf v}_7\\
\end{array}&&
\begin{array}{lcl}
W_9&:&{\bf v}_7, {\bf v}_8\\
W_5&:&{\bf v}_9, {\bf v}_{10}\\
W_8&:&{\bf v}_{11},{\bf v}_{12}
\end{array}
\end{eqnarray}
For simplicity one may use ${\bf v}_i=e_i$ for $i\in\{1,2,\cdots, 5\}$, where $e_i$ is the $i^{th}$ column of the $5\times 5$ identity matrix, and generate the remaining ${\bf v}_i$ as random linear combinations of the first 5.  Other optimizations are also possible. For example, we can assign the same vectors to $W_2, W_9$ since they have no internal conflict. Such optimizations of the ${\bf v}_i$ are not needed for the capacity of the wired case, and not necessary for the DoF, or even a constant gap approximation, of the wireless case. However, the ${\bf v}_i$ should be optimized if the gap to capacity in the wireless case is to be minimized, or linear solutions over smaller field sizes and few channel extensions are preferred in the wired case. As observed previously in the half-rate-feasibility study, such optimizations are computationally quite cumbersome (NP hard) but conceptually not particularly challenging. While an algorithmic study of these optimizations may be practically illuminating, in this work we are content with a guarantee of constant gap to capacity that does not depend on SNR.

\begin{figure}[!h]
\begin{center}
\includegraphics[width=3.2in]{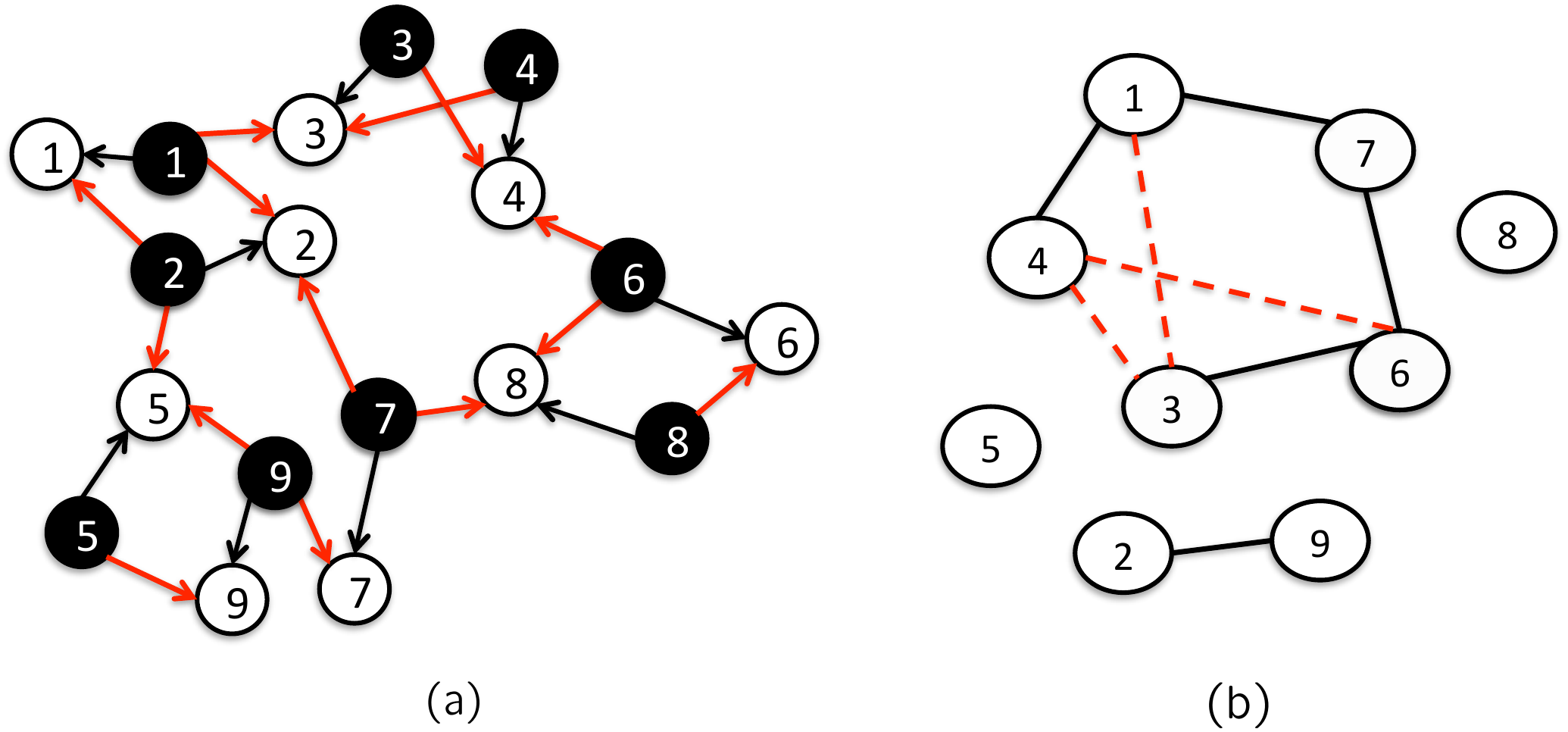}~~~~\includegraphics[width=2.3in]{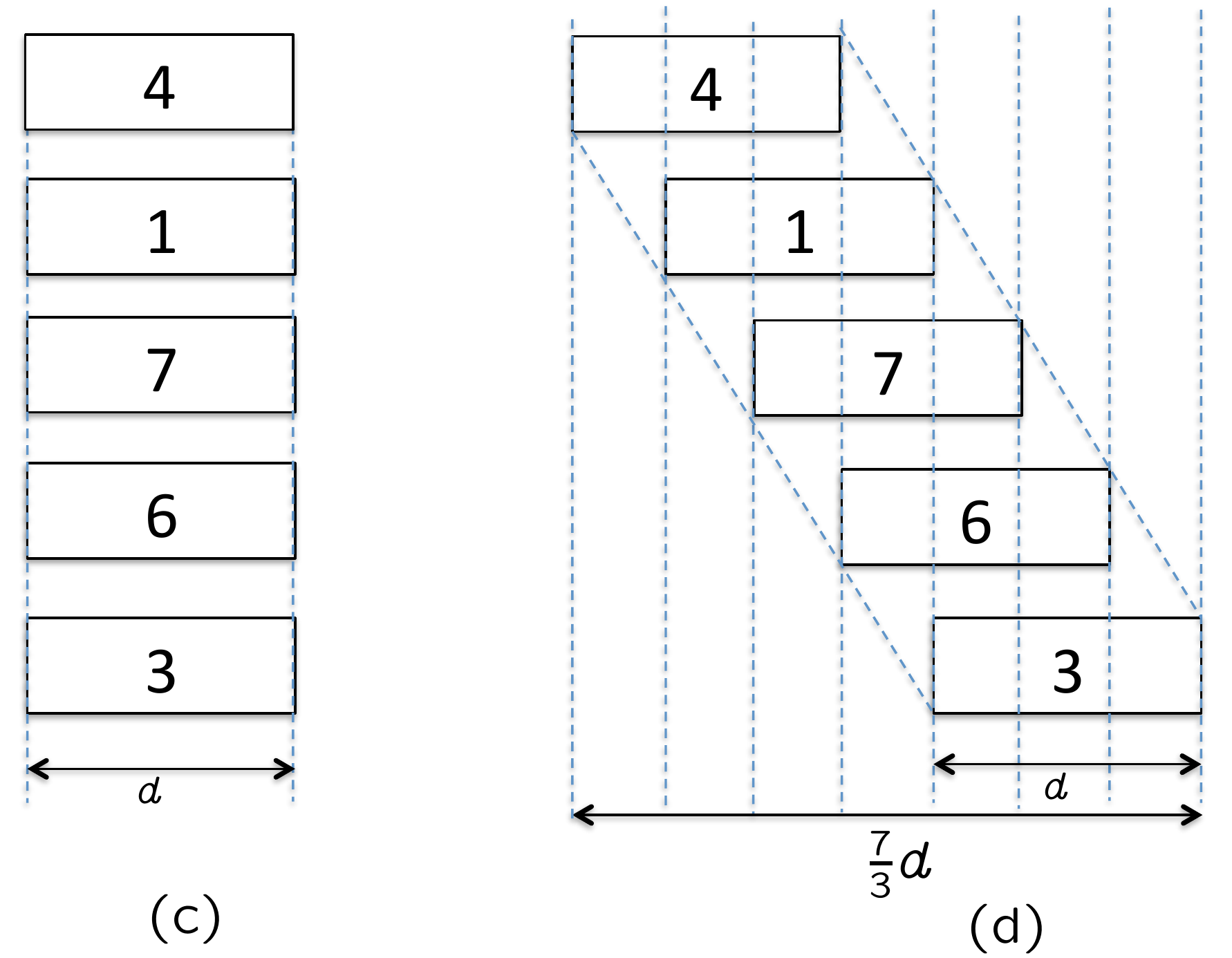}
\caption{\small \it (a) An instance of the topological interference management problem, and (b) its alignment graph (black edges) showing internal conflicts as dashed red edges. (c) Without internal conflicts, $W_4, W_1, W_7, W_6, W_3$ would align perfectly (d) With internal conflicts, conflicting nodes slide out of each other's way just enough to avoid overlaps while all nodes try to stay as aligned as possible with their adjacent nodes. This gives us the symmetric capacity (DoF) of the network as $3/7$.}\label{fig:internal2}
\end{center}
\end{figure}

To further illustrate the ideas, another example is provided in Fig. \ref{fig:internal2}, where the minimum conflict distance is $3$ and the symmetric capacity is $3/7$. Achievability follows by assigning 3 vectors to every message in a $7$ dimensional signal space (over 7 channel uses), such that adjacent messages overlap in $2/3$ dimensions (as shown in Fig. \ref{fig:internal2}(d)). For example, one could assign precoding vectors as, 
\begin{eqnarray}
W_4&:& {\bf v}_1,{\bf v}_2, {\bf v}_3\\
W_1&:& {\bf v}_2,{\bf v}_3, {\bf v}_4\\
W_7&:& {\bf v}_3,{\bf v}_4, {\bf v}_5\\
W_6&:& {\bf v}_4,{\bf v}_5, {\bf v}_6\\
W_3&:& {\bf v}_5,{\bf v}_6, {\bf v}_7\\
W_2&:&{\bf v}_8, {\bf v}_9, {\bf v}_{10}\\
W_9&:&{\bf v}_9, {\bf v}_{10}, {\bf v}_{11}\\
W_5&:&{\bf v}_{12}, {\bf v}_{13}, {\bf v}_{14}\\
W_8&:&{\bf v}_{15}, {\bf v}_{16}, {\bf v}_{17}
\end{eqnarray}
{\it Remark: } Note that generating the precoding vectors of each alignment set independently of the other alignment sets is sufficient to maintain the desired linear independence needed to resolve conflicts across alignment sets. This is a significant simplifying observation that allows us to decompose the problem and focus on each alignment set and its internal conflicts individually.

The examples of Fig. \ref{fig:internal} and Fig. \ref{fig:internal2} convey quite a few insights about interference alignment. A natural question now is whether these insights can be translated into systematic solutions for larger classes of  index coding and topological interference management problem instances. The following theorem presents such a result.

\begin{theorem}\label{theorem:tree}
If each alignment set of a half-rate-infeasible multiple-groupcast topological interference management problem has either no cycles, or  no forks\footnote{A fork is a vertex that is connected by three or more edges.}, then the symmetric capacity (DoF) of the network is $\min(\frac{\Delta}{2\Delta+1}, \frac{1}{\Psi})$, where $\Delta$ is the minimum internal conflict distance of the network and $\Psi$ is the maximum cardinality of an acyclic subset of messages.
\end{theorem}
\addtocounter{theorem}{-1}
\begin{corollary}\label{corollary:tree}
If each alignment set of a half-rate-infeasible multiple-groupcast index coding problem has either no cycles, or no forks, then the symmetric capacity of the network is $\min(\frac{\Delta}{2\Delta+1}, \frac{1}{\Psi})$, where $\Delta$ is the minimum internal conflict distance of the network and $\Psi$ is the maximum cardinality of an acyclic subset of messages.
\end{corollary}

\begin{figure}[!h]
\begin{center}
\includegraphics[width=6.5in]{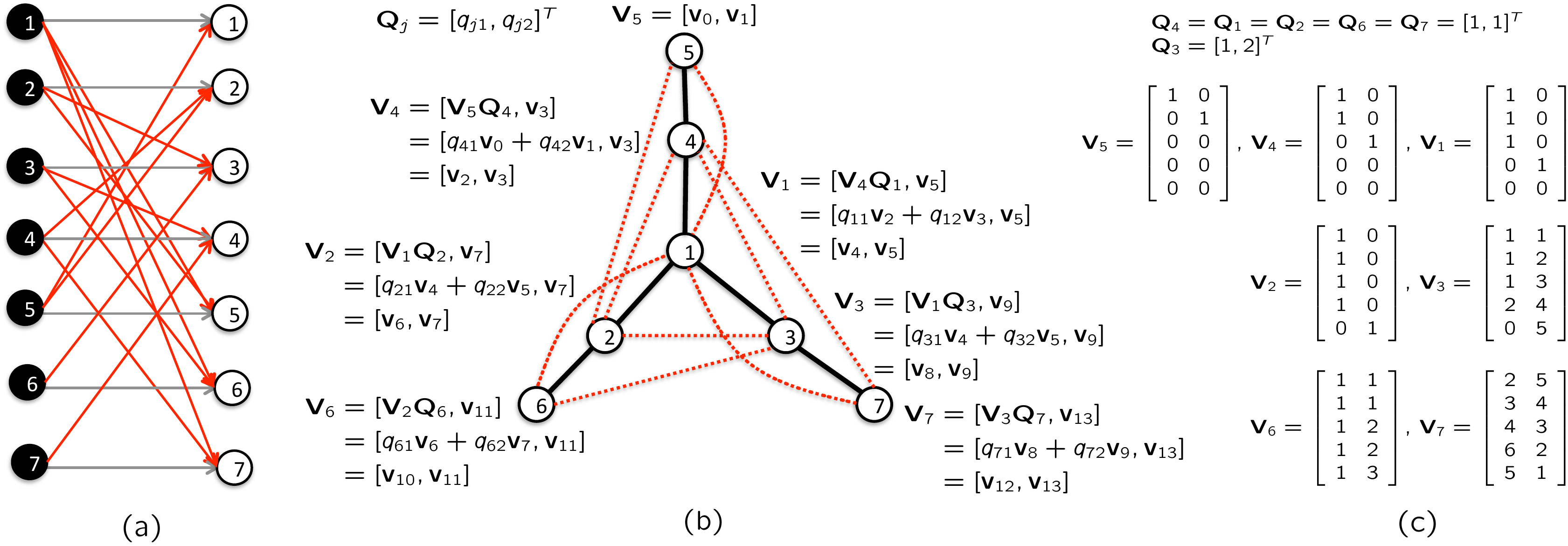}
\caption{\small \it (a) A $7$-unicast topological interference management problem for which the symmetric capacity (DoF) is $2/5$,  (b) its alignment graph (black edges), conflict graph (red edges), and generic  solution:   ${\bf v}_0, {\bf v}_1, {\bf v}_3,{\bf v}_5,{\bf v}_7,{\bf v}_9,{\bf v}_{11},{\bf v}_{13},$ are generic $5 \times 1$ vectors, $q_{ji}$ are generic scalars, (c) a particular  solution.}\label{fig:tree}
\end{center}
\end{figure}
 
 {\it Remark:} Note that the constraint that there are either no cycles, or no forks in each alignment set restricts the topology of the network such that there can be no more than three interferers at any receiver. Otherwise, the four or more interferers will form a clique in the alignment graph which will have both cycles and forks.  
 
With Theorem \ref{theorem:tree} we have the symmetric capacity (DoF) characterization of all index coding problems and all topological interference management problems for which the alignment graph does not contain cycles (such an example is presented in Fig. \ref{fig:tree}). We have a capacity (DoF) characterization of all index coding problems and all topological interference management problems for which the alignment graph has no forks. Theorem \ref{theorem:tree} also covers all cases where some alignment sets may have cycles (but no forks), some may have forks (but no cycles), some may have neither forks nor cycles (Fig. \ref{fig:internal2}), but none of the alignment sets have both forks and cycles (Fig. \ref{fig:internal}).  Since we do know the capacity of Fig. \ref{fig:internal} and it follows from the same principles as used in Theorem \ref{theorem:tree}, evidently there is room for further extension of the scope of the theorem, perhaps to the more general setting of all alignment sets that do not have overlapping cycles. 

\begin{figure}[!h]
\begin{center}
\includegraphics[width=6.1in]{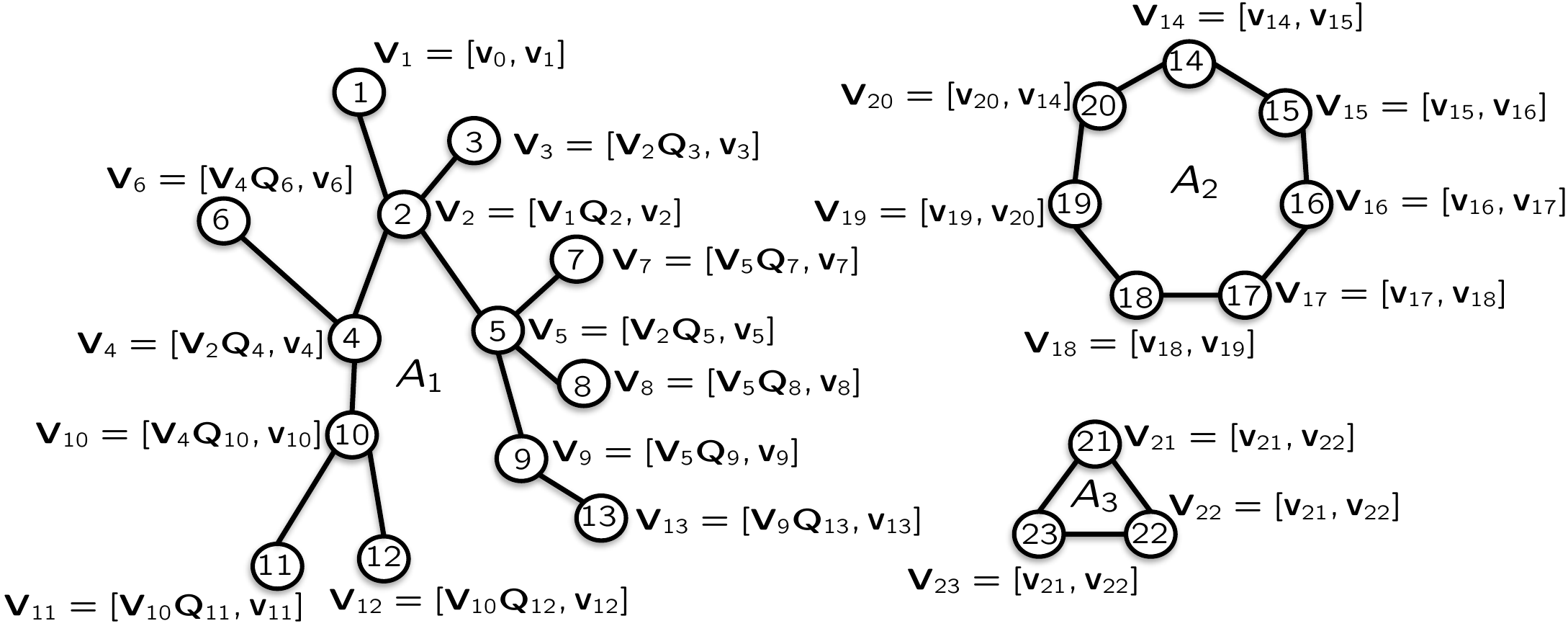}
\caption{\small \it An alignment graph with multiple alignment sets showing a generic solution that avoids all conflicts at distance 2 or more. Symmetric rate (DoF) achieved is $2/5$ per message.}\label{fig:nocyclenofork}
\end{center}
\vspace{-0.3cm}
\end{figure}

As another illustration of the result of Theorem \ref{theorem:tree} an alignment graph is shown in Fig. \ref{fig:nocyclenofork}. While conflicts are not shown, the minimum internal conflict distance is assumed to be 2. Note that the assignment of precoding vectors avoids all conflicts at distance 2 or more. Three alignment sets are shown, labeled as $A_1, A_2, A_3$. Alignment set $A_1$ has no cycles and $A_2$ has no forks. $A_3$ cannot have an internal conflict because the minimum conflict distance is 2. For this reason, it is possible to assign the same subspace to all messages that are a part of $A_3$. All ${\bf v}_i$ are randomly generated independent of each other as $5\times 1$ vectors. All ${\bf Q}_i$ are randomly generated $2\times 1$ matrices, meant to extract a random one dimensional subspace from the parent node's signal space.  Each message occupies $2/5$ dimensions and any two adjacent messages occupy no more than $3/5$ dimensions, leaving the remaining $2/5$ dimensions for the desired signal. Intuitively, the three main ideas involved in the assignment of precoding vectors are the following:
\begin{enumerate}
\item The precoding vectors for each alignment set are generated independently of other alignment sets. This avoids conflicts across alignment sets because independently generated subspaces are in general position over sufficiently large fields, and therefore will not overlap if the overall space is large enough to accommodate both of them.
\item To avoid conflicts across a cycle, the assignment of precoding vectors is also cyclic. See for example $A_2$ in Fig. \ref{fig:nocyclenofork}. The cyclic assignment of vectors allows the dependence between nodes to first diminish with increasing distance and then re-emerge as one returns to the beginning.
\item To avoid conflicts across forks, the assignment of precoding vectors is based on randomly chosen inherited subspaces from the parent node, complemented with independently generated vectors. The inherited space allows just the right amount of overlap between adjacent nodes and the addition of the independently generated vector allows the dependency to fade away at the distance where conflicts start to appear. See for example $A_1$ in Fig. \ref{fig:nocyclenofork}, where with high probability nodes $1, 3, 4, 5$ each have a one dimensional overlap with node $2$, but no overlap with each other or with any other node to which they are not directly connected. \footnote{This illustrates why (fractional) coloring solutions are not optimal. Fractional colorings correspond to one-to-one alignments, whereas forks require subspace alignments to avoid conflicts.}
\end{enumerate}

An interesting class of networks included in Theorem \ref{theorem:tree} is the partially connected $K$ user interference channel where $S=D=|\mathcal{W}|=K$,  each source sends one message to its corresponding destination, and the topology of the network is such that no source is heard by more than three destinations, and no destination can hear more than three sources. 
\begin{eqnarray}
\forall j\in\{1,2,\cdots, S\}, &&|\{i: t_{ij}=1\}|\leq 3\\
\forall i\in\{1,2,\cdots, D\}, &&|\{j: t_{ij}=1\}|\leq 3
\end{eqnarray}
and the corresponding index coding problem where again $S=D=|\mathcal{W}|=K$ and 
\begin{eqnarray}
\forall j\in\{1,2,\cdots, S\}, &&|\{i: a_{ij}=0\}|\leq 3\\
\forall i\in\{1,2,\cdots, D\}, &&|\{j: a_{ij}=0\}|\leq 3
\end{eqnarray}
This is  a class of sparsely connected networks. The number of sources that can be heard by a destination and the number of destinations that can hear a source, is indicative of the density of a wireless network. By virtue of Theorem \ref{theorem:tree} we have a symmetric capacity characterization for such networks.

\begin{figure}[!h]
\begin{center}
\includegraphics[width=4.3in]{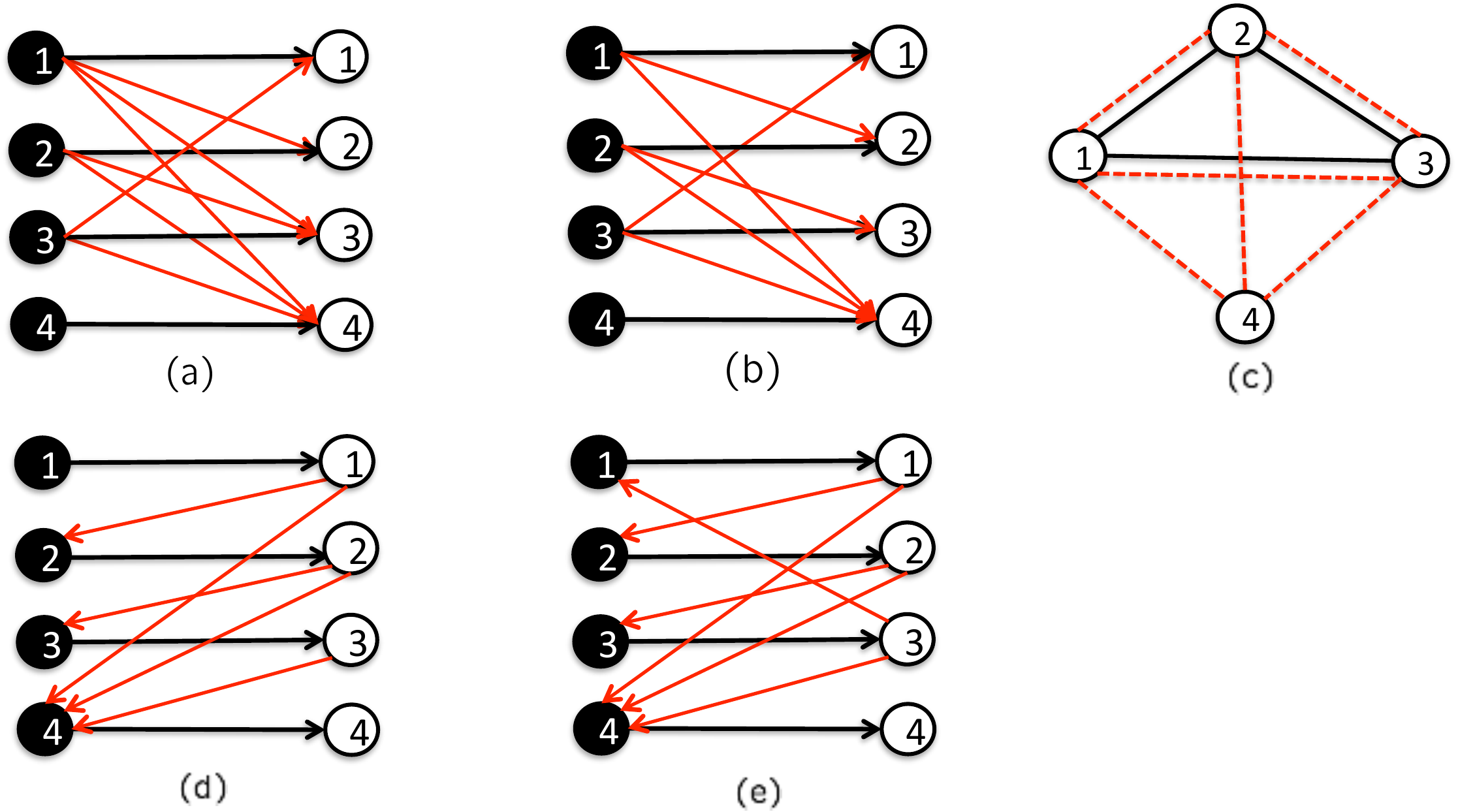}
\caption{\small \it (a) A 5-unicast topological interference problem, (b) A different instance of a 5-unicast topological interference problem, (c) Alignment and conflict graphs, identical for both problems, (d) Demand graph for the first problem is acyclic (e) Demand graph for the second problem is not acyclic. }\label{fig:samegraph}
\end{center}
\vspace{-0.3cm}
\end{figure}

We conclude this section with an interesting example illustrated in Fig. \ref{fig:samegraph}. Shown in Fig. \ref{fig:samegraph}(a) and Fig. \ref{fig:samegraph}(b) are two instances of a 5-unicast topological interference management problem. While both instances have the same alignment and conflict graphs, illustrated in Fig. \ref{fig:samegraph}(c), they have different symmetric capacity (DoF) values. The symmetric  capacity (DoF) of the first network is $\frac{1}{\Psi}=\frac{1}{4}$, because its demand graph, shown in Fig. \ref{fig:samegraph}(d) is acyclic. The symmetric capacity (DoF) of the second network is $\frac{\Delta}{2\Delta+1}=\frac{1}{3}$. One possible achievable scheme for  symmetric rate (DoF) of 1/3 per message for the second network is to use precoding vectors ${\bf V}_1=[1,0,0], {\bf V}_2=[0,1,0], {\bf V}_3=[1, 1, 0], {\bf V}_4=[0,0,1]$. 
Evidently, the alignment and conflict graphs do not capture all  the information needed to determine the optimal alignment of signals needed in general. This observation hints at the complexity of the general problem.



While we pursue information theoretic capacity characterizations, it is also worthwhile to consider the rates achievable by linear achievable schemes, especially since linear schemes, as we have seen for all the solutions presented in this work, are frequently capacity optimal. This is the issue explored in the next set of results.

\subsection{Linear Feasibility: Duality}
Duality of reciprocal channels, e.g., the duality of Gaussian multiple access (MAC) and broadcast channels (BC) established in \cite{Duality, Vishwanath_Jindal_Goldsmith, Viswanath_Tse_BC}, where the roles of transmitters and receivers are switched, can be a very useful property. When duality holds, the solution to a problem solves its dual problem as well. This is useful because often a problem that seems hard to solve, may have a dual that is much simpler. For example, the capacity region of the Gaussian MIMO broadcast channel may seem much more challenging to compute directly, but its dual problem the MIMO MAC is much simpler. The duality of linear interference alignment solutions for MIMO multiple-unicast networks (interference channels, $X$ networks), established by Gomadam et al. in \cite{Gomadam_Cadambe_Jafar} is useful to find iterative and distributed numerical interference alignment solutions. 

Note that the MAC-BC duality, as well as the DoF duality of linear solutions for MIMO multiple-unicast networks are shown for networks with full CSIT. This is a critical assumption. Duality relationships do not hold in general even at the DoF level with limited or no CSIT. For example, the DoF collapse for a fully connected MISO BC with no CSIT  but the fully connected SIMO MAC does not need CSIT. Since we are interested in interference networks with no CSIT beyond the topology knowledge, one might not expect a duality relationship to hold in this setting. However,  surprisingly, a duality relationship does hold for linear solutions, for both the topological interference management and index coding problems in the multiple unicast setting, as stated in the following theorem.

\begin{theorem}\label{theorem:duality}
Any rate tuple $R(W)$ achievable by a linear scheme in a  multiple unicast topological interference management problem ${\bf TIM}(\mathcal{T}, \mathcal{W}(S), \mathcal{W}(D), \mathbb{C})$ is also achievable by a linear scheme in the dual topological interference management problem, ${\bf TIM}(\mathcal{T}', \mathcal{W}'(S'),\mathcal{W}'(D'), \mathbb{C})$, where the roles of sources and destinations are reversed as follows,
\begin{eqnarray}
S'&=&D\\
D'&=&S\\
\mathcal{T}'&=&\mathcal{T}^T\\
\mathcal{W}'(S'_i)&=&\mathcal{W}(D_i), ~~\forall i\in\{1,2,\cdots, S'\}\\
\mathcal{W}'(D'_j)&=&\mathcal{W}(S_j), ~~\forall j\in\{1,2,\cdots, D'\}
\end{eqnarray}
\end{theorem}
\addtocounter{theorem}{-1}
\begin{corollary}\label{corollary:duality}
Any rate tuple $R(W)$ achievable by a linear scheme over a field $\mathbb{F}$ in a multiple unicast index coding problem ${\bf IC}(\mathcal{A}, \mathcal{W}(S), \mathcal{W}(D))$ is also achievable by a linear scheme over the same field $\mathbb{F}$ in the dual multiple unicast index coding problem, ${\bf IC}(\mathcal{A}', \mathcal{W}'(S'),\mathcal{W}'(D'))$,  where the roles of sources and destinations are reversed as follows,
\begin{eqnarray}
S'&=&D\\
D'&=&S\\
\mathcal{A}'&=&\mathcal{A}^T\\
\mathcal{W}'(S'_i)&=&\mathcal{W}(D_i), ~~\forall i\in\{1,2,\cdots, S'\}\\
\mathcal{W}'(D'_j)&=&\mathcal{W}(S_j), ~~\forall j\in\{1,2,\cdots, D'\}
\end{eqnarray}
\end{corollary}
As seen from the proof of Theorem \ref{theorem:duality} the duality relationship is quite explicit, when the transmitters and receivers switch roles, so do the precoding and receiver combining matrices, but otherwise the linear solution does not change, because the feasibility conditions for the original and dual networks are the same. The restriction to multiple-unicast networks is noteworthy. 

Duality relationships allow us to extend the scope of known results to their dual networks. As an example, the following theorem extends the scope of Theorem \ref{theorem:tree}. 
\begin{theorem}\label{theorem:dualtree}
If each alignment set of the dual of a half-rate-infeasible multiple-unicast topological interference management problem has either no cycles, or  no forks, then the symmetric capacity (DoF) of the network is $\min(\frac{\Delta}{2\Delta+1}, \frac{1}{\Psi})$, where $\Delta$ is the minimum internal conflict distance of the network and $\Psi$ is the maximum cardinality of an acyclic subset of messages.
\end{theorem}
\addtocounter{theorem}{-1}
\begin{corollary}\label{corollary:dualtree}
If each alignment set of the dual of a half-rate-infeasible multiple-unicast index coding problem has either no cycles, or no forks, then the symmetric capacity of the network is $\min(\frac{\Delta}{2\Delta+1}, \frac{1}{\Psi})$, where $\Delta$ is the minimum internal conflict distance of the network and $\Psi$ is the maximum cardinality of an acyclic subset of messages.
\end{corollary}
Remarkably, while the duality is based on linear achievable schemes, the extension of the result of Theorem \ref{theorem:tree} stated in Theorem \ref{theorem:dualtree} is still an information theoretic capacity result, not restricted to linear schemes. Note, however, the restriction to multiple-\emph{unicasts} in Theorem \ref{theorem:dualtree}, whereas Theorem \ref{theorem:tree} applies to groupcasts. This is because the duality relationship in Theorem \ref{theorem:duality} is established only for multiple unicast settings.

\begin{figure}[!h]
\begin{center}
\includegraphics[width=3in]{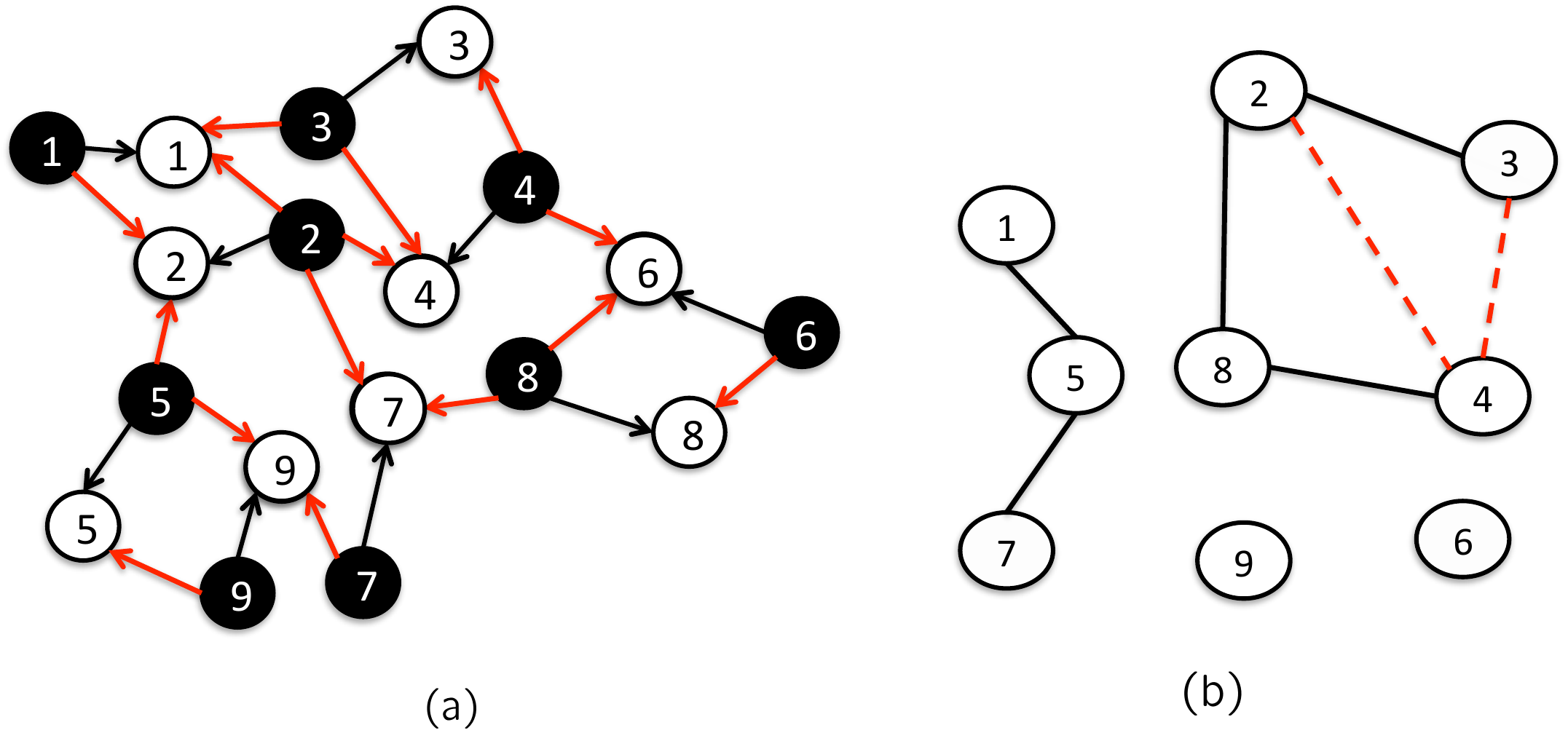}~~~~~~\includegraphics[width=3in]{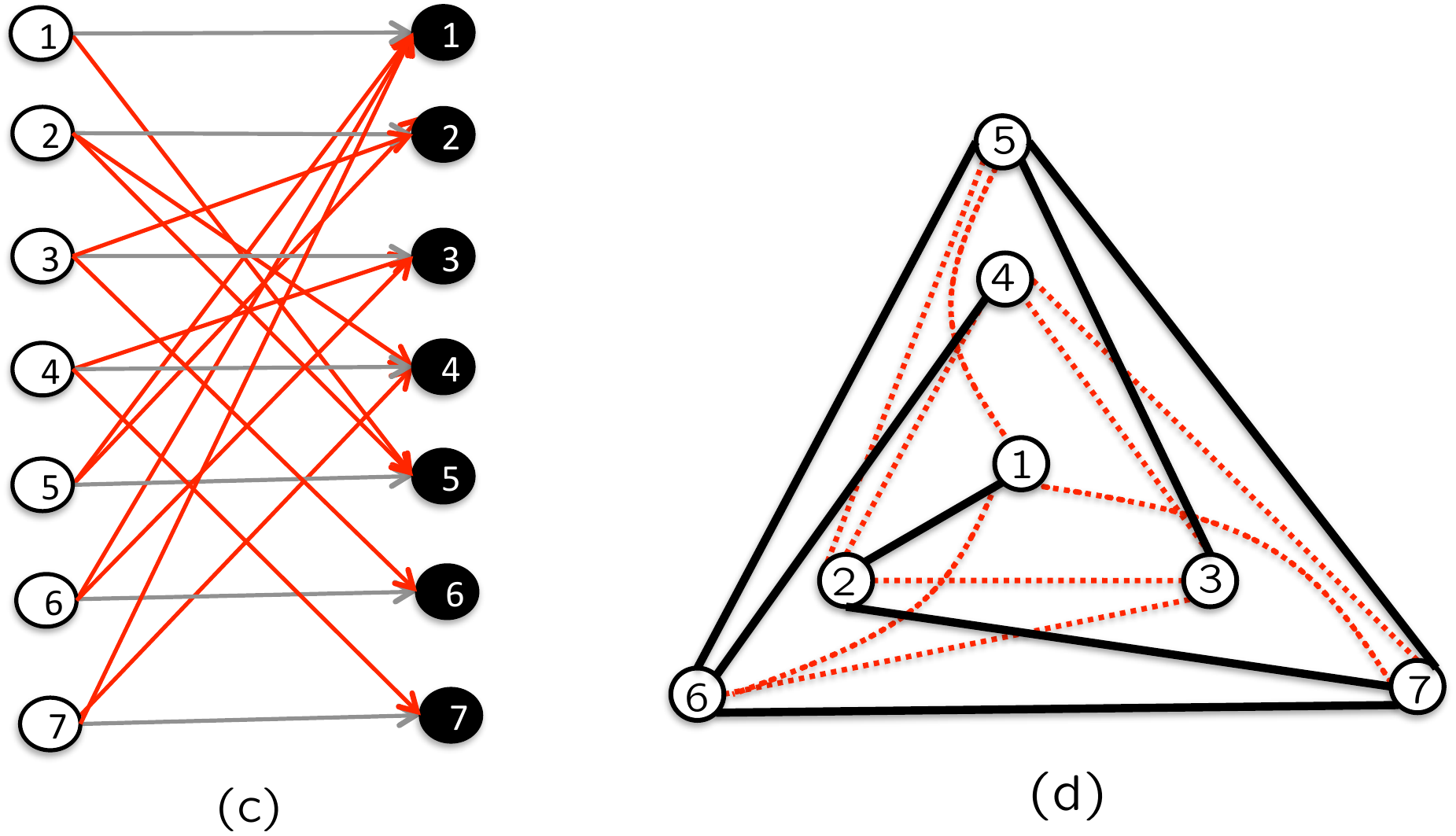}
\caption{\small \it (a) The dual 9-unicast topological interference management problem of the original problem shown in Fig. \ref{fig:internal}(a) and  (b) its alignment graph showing internal conflicts as dashed red edges. Conflicts that are not internal are not shown. (c) Dual of the network of Fig. \ref{fig:tree}(a) and, (d) its alignment and conflict graphs. }\label{fig:internaldual}
\end{center}
\end{figure}

As  examples of the application of duality, consider the network of Fig. \ref{fig:internal}(a) whose alignment graph had both forks and cycles in the same alignment set, so that the result of Theorem \ref{theorem:tree} did not apply to it. However, if we consider its dual network, shown in Fig. \ref{fig:internaldual}(a),  its alignment graph shown in Fig. \ref{fig:internaldual}(b) turns out to have neither cycles nor forks. Thus, Theorem \ref{theorem:dualtree} settles the symmetric capacity (DoF) of the network as $2/5$ per user. 
Similarly, consider the network shown in Fig. \ref{fig:internaldual}(c) whose alignment graph, shown in Fig. \ref{fig:internaldual}(d) has both cycles and forks in the same alignment set, so Theorem \ref{theorem:tree} does not apply to this network. Yet, the dual of this network is the network, shown in Fig. \ref{fig:tree}(a), has alignment graph, shown in Fig. \ref{fig:tree}(b) consisting of no cycles, so Theorem \ref{theorem:dualtree} establishes the symmetric capacity (DoF) as $2/5$ in this case  as well. 

As obvious from the examples above, the alignment graph of the original problem can have distinct properties from that of the dual problem. Remarkably, though, the conflict graphs are identical in both problems. Of course, because the alignment graph changes, an internal conflict in one problem may be only an external conflict (conflict between messages in different alignment sets) in its dual. This is quite intriguing because, as seen throughout this work, internal conflicts are quite significant whereas external conflicts seem to be irrelevant, at least from the perspective of linear schemes. Interestingly, as shown in the proof of Theorem \ref{theorem:dualtree}, the minimum internal conflict distance in the original problem is the same as that in its dual. Finally, we note that the duality of linear solutions, and the frequent optimality of linear schemes for topological interference management problems as well as index coding problems, raises the natural question whether an information theoretic version of this duality holds true as well. This is an interesting open problem for future work.

\subsection{Cellular Topologies}
In this section we start with an example of a small cluster of cells, regularly placed, and then consider infinite cellular arrays. 
\subsubsection{4-Cell Downlink Network}

\begin{figure}[!h]
\centering
\includegraphics[width=6.1in]{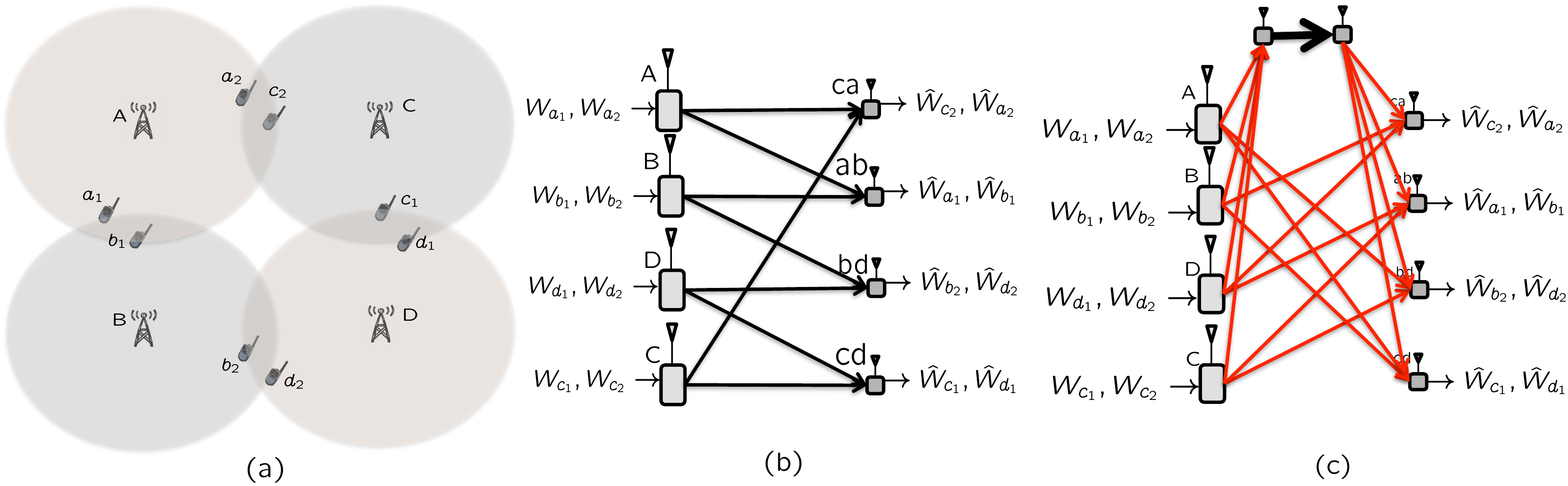}
\caption{\small\it (a) Locally connected 4-cell downlink network, (b) Equivalently, a partially connected X network, (c) Equivalent Index Coding Problem.}
\label{fig:intro}
\end{figure}

Consider the 4 cell downlink setting shown in Fig. \ref{fig:intro}(a). Each base station is heard at a significant SNR level only within the cell area shown as a shaded disk centered at that base station. This gives rise to interference for the users located in the overlapping cellular regions. In the figure we show two users in each boundary region, one for each of the two cells that overlap at that boundary. Instead of our regular notation which can be quite cumbersome for cellular networks, in this section we will adopt a more intuitive notation. For example, receivers $a_1, a_2$ want to receive independent messages $W_{a1}, W_{a2}$ from the base station $A$, but have to deal with interference from base stations $B, C$, respectively. 

Because the users inhabiting the same boundary are statistically indistinguishable from the transmitters' perspective, they have the same ability to decode messages and hence may be combined into one equivalent user without loss of generality, as shown in Fig. \ref{fig:intro}(b), producing a partially connected X network setting \cite{Cadambe_Jafar_X} with 8 independent messages, one from each transmitter to each connected receiver. So, e.g., the receiver labeled `$ab$' wants to decode  two independent messages sent from transmitters labeled $A, B$. This is the topological interference management problem that we want to solve. Fig. \ref{fig:intro}(c) also shows the corresponding index coding problem that will be automatically solved simultaneously.

A natural scheme would be  an orthogonal (frequency reuse) scheme which eliminates interference, e.g., cells $A, D$ are simultaneously active for half the time  (or over half the frequency band) and cells $B, C$ are simultaneously active for the remaining half of the time. This orthogonal scheme achieves 1/2 DoF per cell, i.e., 1/4 symmetric DoF per message, and it is easy to see that no orthogonal (or multicast) scheme can achieve higher DoF.  However, the optimal symmetric DoF value  \emph{with or without} CSIT, is  1/3 per message, i.e., an improvement by a factor of $4/3$. Note that $1/3$ per message is also the symmetric capacity of the equivalent index coding problem shown in Fig. \ref{fig:intro}(c).

\begin{figure}[!h]
\centering
\includegraphics[width=4.1in]{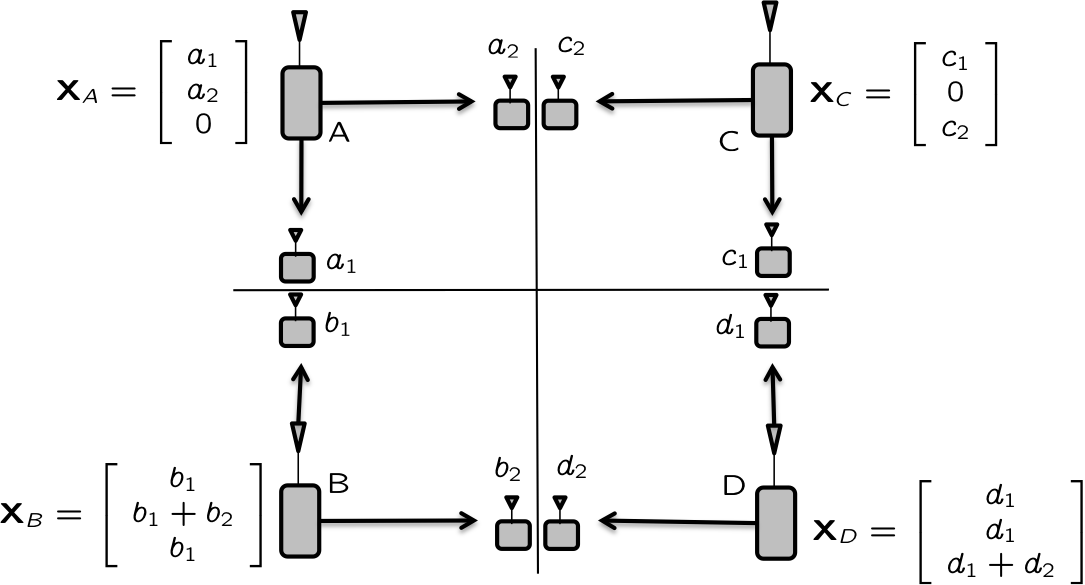}
\caption{\it \small Transmitted symbols over three time slots. 1/3 DoF per message are achieved. 1/3  is also the optimal symmetric DoF value even with  perfect CSIT. }
\label{fig:4cellsol}
\end{figure}

Let us understand the solution from an interference alignment perspective. It is easy to verify that the alignment graph for this network contains internal conflicts so it is not half-rate-feasible, it contains both cycles and forks so we cannot use Theorem \ref{theorem:tree}, the demand graph for this multiple unicast network is not acyclic, so we cannot use Theorem \ref{theorem:noalign}. Thus, this problem does not fall into any of the categories that have previously been solved. It is also clear that the alignment graph has minimum internal conflict distance of $\Delta=1$, e.g., $a_1, a_2$ cause interference to $c_2$ so they are connected by an edge in the alignment graph, but also $a_1$ causes interference to $a_2$, so they are also connected by an edge in the conflict graph. So, by Corollary \ref{corollary:slide} the symmetric capacity is bounded above by $1/3$. 

From achievability perspective, consider the receivers $a_2, c_2$.  The interfering messages seen by these receivers are $a_1, c_1$. So, let us align them along the same vector. Similarly, consider receivers $a_1, b_1$ which see interference from $a_2, b_2$, so let us align $a_2, b_2$ as well. Following this argument for receivers $c_1,d_1$ (align $c_2,d_2$) and $b_2,d_2$ (align $b_1,d_1$) we have the following assignments of signal vectors to messages.
\begin{eqnarray}
{a_1,c_1}&:&{\bf v}_1\\
{a_2,b_2}&:&{\bf v}_2\\
{c_2,d_2}&:&{\bf v}_3\\
{b_1,d_1}&:&{\bf v}_4
\end{eqnarray}
where ${\bf v}_1, {\bf v}_2, {\bf v}_3, {\bf v}_4$ are pairwise linearly independent $3\times 1$ vectors in a three dimensional space. If we choose 
${\bf v}_1=[1, 0, 0]^T, {\bf v}_2=[0, 1, 0]^T, {\bf v}_3=[0,0,1]^T, {\bf v}_4=[1,1,1]^T$ we obtain the solution illustrated in Fig. \ref{fig:4cellsol}. The same solution works for the wired (any finite field) and wireless case of the topological interference management problem, and for the index coding problem as well. Since the solution is linear and non-asymptotic, it also gives us a constant gap capacity approximation for the wireless case, although, as usual, the ${\bf v}_i$ can be optimized to make the gap as small as possible. Remarkably, 1/3  is the symmetric DoF value for this partially connected newtork even with perfect CSIT. 

Another interesting, and much more challenging example of a cluster of cells, comprised of 5 regularly placed cells, is presented in our follow up work in \cite{Maleki_Cadambe_Jafar}. 
\subsubsection{Infinite arrays of locally connected cells}

\begin{figure}[!h]
\centering
\includegraphics[width=6in]{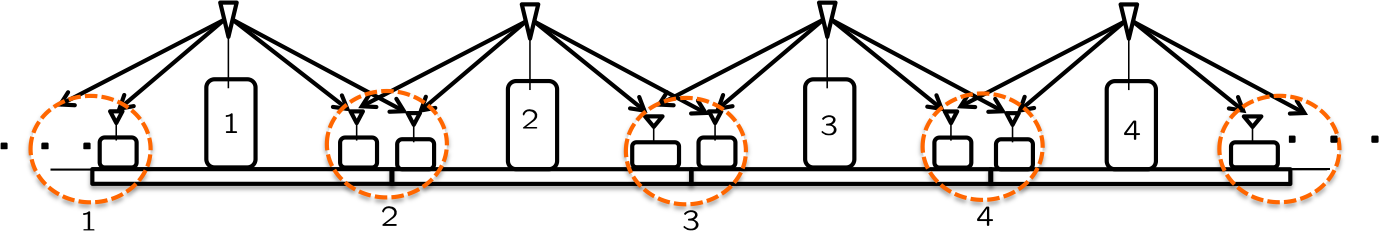}
\caption{\it Locally connected one-dimensional cellular array where boundary users hear both adjacent base stations.}
\label{fig:linear}
\end{figure}

Here we  consider  infinite arrays of uniformly placed cells that are locally connected. Three topologies are considered: 1) Linear Cellular Array, where cells are placed uniformly along a straight line, 2) Square Cellular Array, where the cell placements fall on a square grid, and 3) Hexagonal Cellular Array, where the cells are placed uniformly in a hexagonal grid pattern.  We are interested primarily in the users located at the boundaries between adjacent cells, since this is where interference is the most severe and where frequency planning is most needed.\footnote{This is consistent with the fractional frequency reuse principle used, e.g., in Mobile WiMAX,  which allows users near the cell center to follow a reuse factor of 1 and requires frequency planning only for cell edge users.}  A one-dimensional cellular model is shown in Fig. \ref{fig:linear}. Since in a linear model, each cell has 2 adjacent cells with which it shares boundaries, Fig. \ref{fig:linear} shows two representative users in each cell, one at each shared boundary.

\subsubsection*{Linear Cellular Array}
\begin{figure}[!h]
\centering
\includegraphics[width=6in]{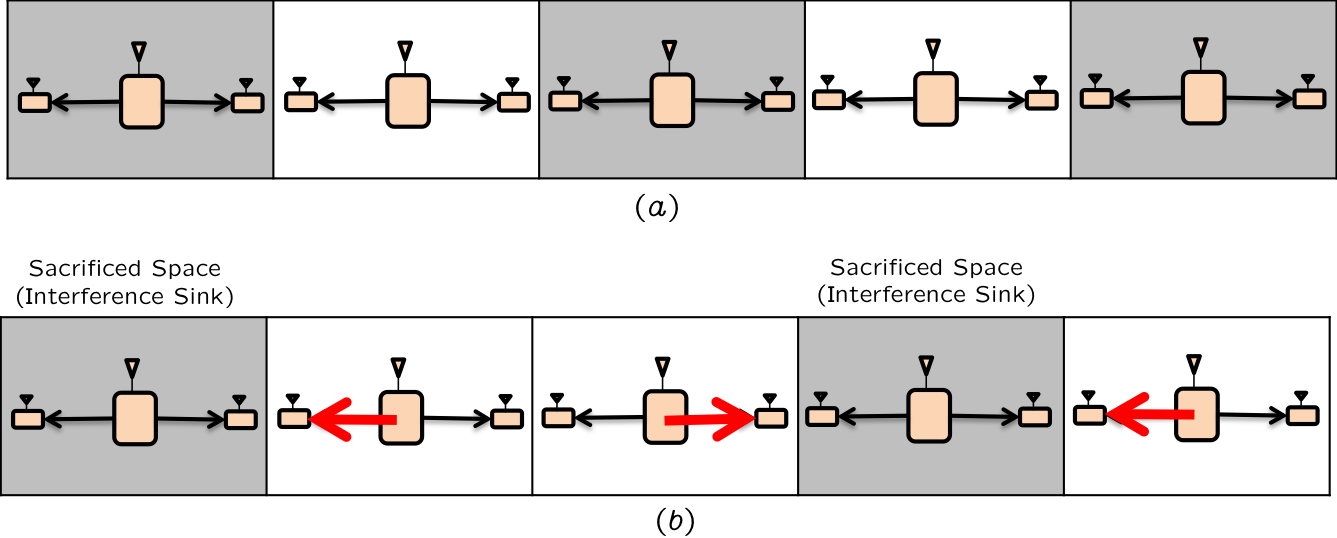}
\caption{\small\it (a) Conventional frequency reuse (periodic with period 2) allows 1/2 DoF per cell. (b) Aligned frequency reuse (periodic with period 3) allows 2/3 DoF per cell, an improvement of 33\% per cell. All transmissions are isotropic. Note that the red arrows do not indicate directional transmission (all transmission is isotropic), but rather the choice of the receiver to be served within each active cell. The remaining receivers are turned off.}
\label{fig:linearsol}
\end{figure}

As shown in Fig. \ref{fig:linear}, each base station is heard by all users around the cell boundary, both in its own cell as well as the immediately adjacent cells. However, due to path loss, the signals do not travel further beyond with a significant strength. Conventional spectral-reuse pattern used in this setting is shown in Fig. \ref{fig:linearsol}(a) where active cells that share a common cell edge are assigned different spectral bands, or equivalently, only alternating cells are activated, the even numbered cells for half the time, and the odd numbered cells for the remaining half of the time. Thus, each cell achieves 1/2 DoF and all inter-cell interference is eliminated. { The conventional frequency reuse solution, 1/2 DoF per cell, is the baseline} that we compare against, as we present next an alternative approach --- aligned frequency reuse (an orthogonal solution).

Aligned frequency reuse is illustrated in Fig. \ref{fig:linearsol}(b), where a periodic reuse pattern with period 3 cells is repeated along the infinite sequence of cells. Every third cell (shown as the grey shaded cells in Fig. \ref{fig:linearsol}(b)) is sacrificed, i.e., switched off. Since the transmitter of the sacrificed cell generates no signal, all the boundaries of sacrificed cell become interference free and therefore all the neighboring cells can serve users that are located on the boundary of the sacrificed cell. Clearly the throughput per cell and per user can be symmetrized by shifting the pattern so that each cell becomes the sink cell for 1/3 of the time. The resulting bandwidth allocation from this \emph{aligned} frequency reuse is 2/3 DoF per cell (1/3 per message) which corresponds to { an improvement of $33\%$ over the baseline}. Next we show that this orthogonal solution is also DoF optimal.

Let us consider any user in Fig. \ref{fig:linear}, e.g, the user in Cell 2 located at the boundary with Cell 3. Let us eliminate all messages except the desired message from BS 2 and the two undesired messages from BS 3. Clearly eliminating other messages cannot hurt the rates of the remaining messages. Now we argue that all three remaining messages are resolvable by this user. Since the desired message is decodable by design, the user can reliably reconstruct and subtract it from its received signal. This gives the user an invertible channel to BS 3 (within bounded variance noise distortion), from which it can reliably resolve both messages originating at BS 3. Thus, one user is able to resolve all 3 messages. Since the user has only 1 antenna, the total DoF of all three messages cannot be more than 1. This gives us an outer bound of $1/3$ DoF per message. Since each cell has two messages, it gives us an outer bound of $2/3$ DoF per cell.  Thus, the optimal solution is orthogonal. Remarkably, this is also the optimal symmetric DoF value even with perfect CSIT (because the outer bound argument presented above is not affected by the presence of CSIT).

\subsubsection*{Square Cellular Array}
The one-dimensional setting shows that the benefits of aligned frequency reuse can be significant, i.e., 33\% improvement over the  baseline corresponding to conventional frequency reuse. As we go from one dimensions to two, perhaps the most interesting question is to determine if the benefits diminish, increase, or remain the same.

\begin{figure}[!h]
\centering
\includegraphics[width=4in]{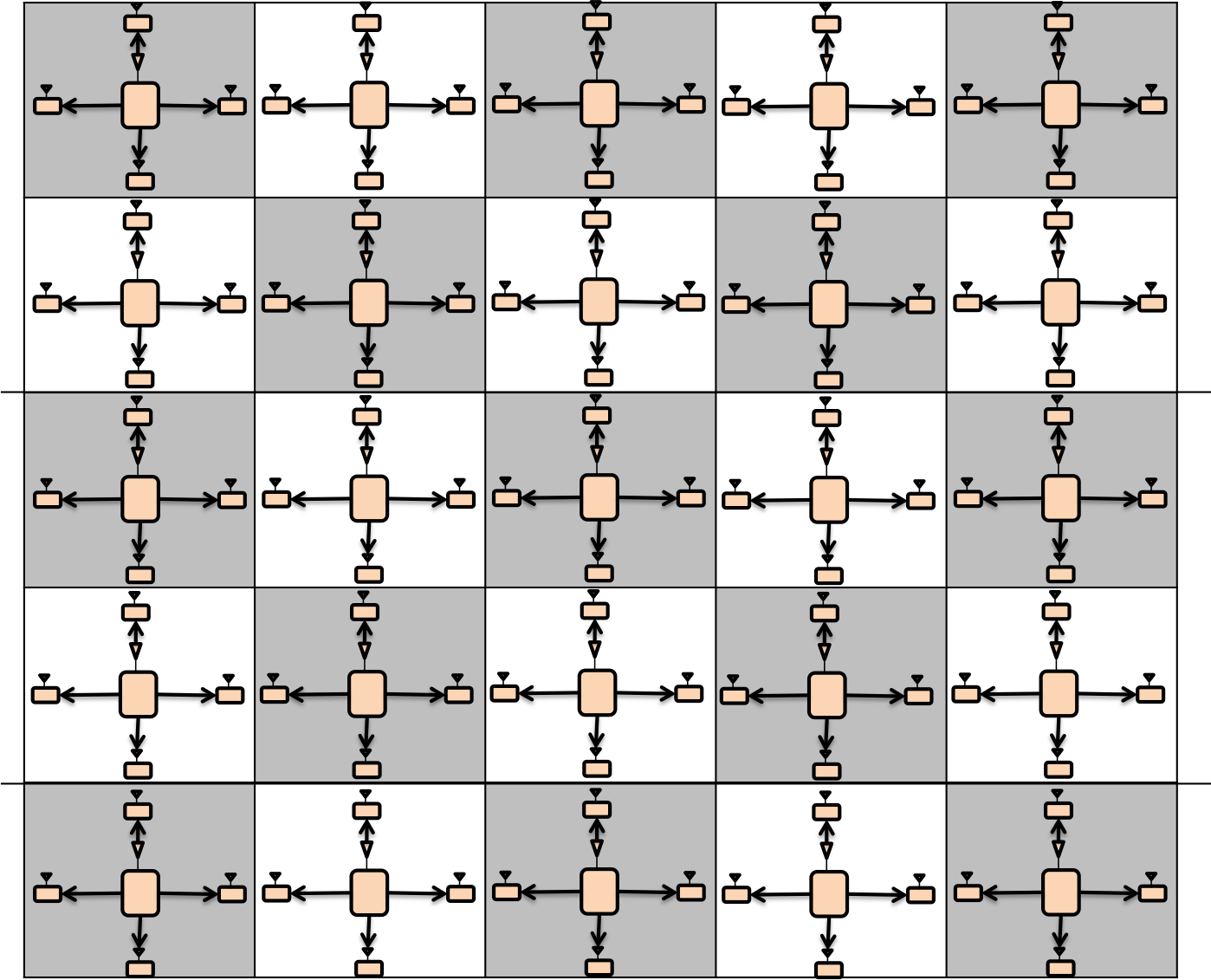}
\caption{\small\it Baseline frequency reuse pattern, achieves 1/2 DoF per cell.}
\label{fig:baselinearray}
\end{figure}

To answer this question, we consider a two dimensional square cellular grid with the same local connectivity assumptions as in the previous section. We assume, as before, that each user on the boundary between two cells, hears comparable signal strengths from both base stations, and each base station can be heard within its own cell and in the vicinity of its boundary with its adjacent cells. The conventional frequency reuse solution for this setting is shown in Fig. \ref{fig:baselinearray}, and corresponds to 1/2 DoF per cell as before.

\begin{figure}[!h]
\centering
\includegraphics[width=4in]{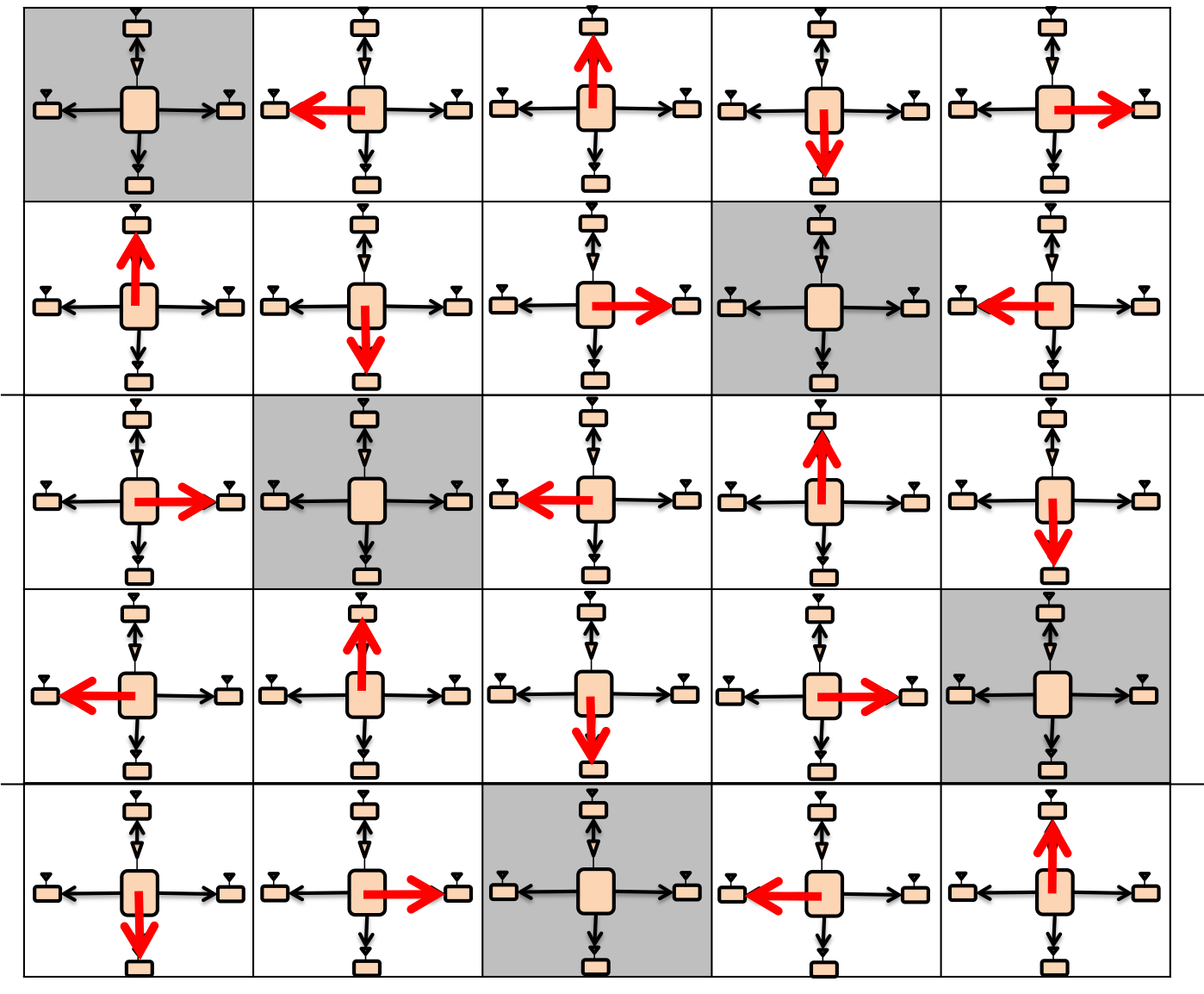}
\caption{\small\it Aligned frequency reuse, 60\% improvement over the baseline. All transmissions are isotropic. Red arrows only point out active receivers. All active receivers are adjacent to inactive (grey) cells.}
\label{fig:blindarray}
\end{figure}

The aligned reuse pattern is shown in Fig. \ref{fig:blindarray}, where again the interference sinks are shown in grey. The red arrows indicate the intended users whose messages are being transmitted. Recall that the BS antennas are isotropic, so the \emph{red arrows do not represent directional transmissions}. From Fig. \ref{fig:blindarray} it is clear that the achieved DoF value is 4/5 per cell, i.e., 1/5 per message. This corresponds to an { improvement of 60\% over the baseline}. Next we establish the optimality of this result. 

The proof of the outer bound is similar to the one-dimensional setting. Consider any user at the boundary between two cells. We claim that this user can resolve not only its desired message, which it must by design, but  also all 4 messages from its neighboring interfering BS. This is because if we eliminate all other messages, then the user can reconstruct and subtract the signal from its desired BS, and then invert the channel to its interfering BS, thus resolving all 5 messages. Since a user with only 1 antenna can resolve 5 messages, the DoF per message cannot be more than 1/5. 

Achievability of 4/5 DoF per cell is already shown in Fig. \ref{fig:linearsol}(b), and clearly requires no knowledge of channel realizations at the transmitters.  Once again it is remarkable that not only is the 4/5 DoF per cell the optimal DoF value in the absence of channel knowledge at the transmitters, but also that it is the optimal value \emph{even with perfect channel knowledge}. This is because the outer bound argument presented above is not affected by the presence or absence of CSIT.

\subsubsection*{Hexagonal Cellular Array}
\begin{figure}[!h]
\centering
\includegraphics[width=6in]{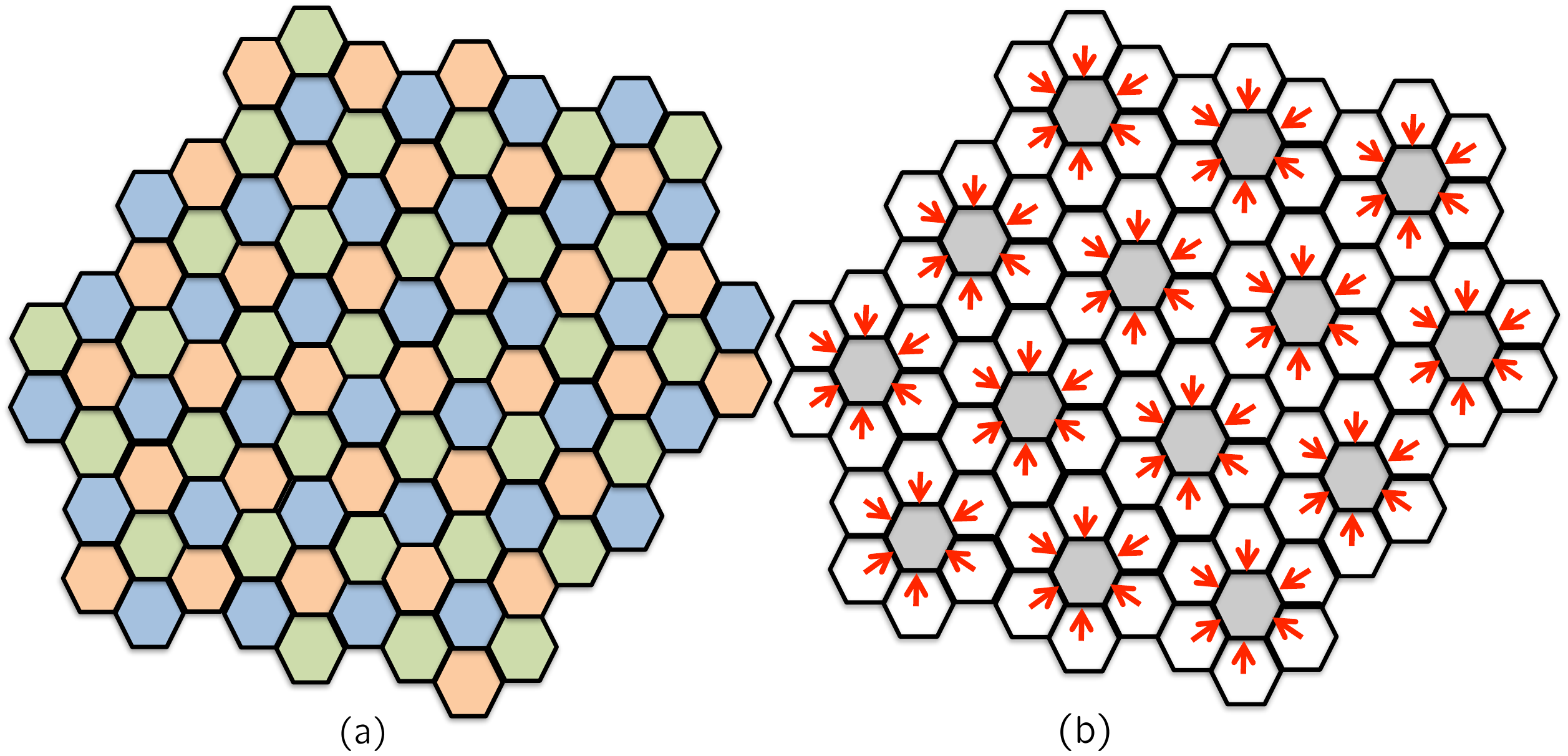}
\caption{\small\it Locally connected hexagonal array of cells. (a) Conventional frequency reuse (DoF=1/3 per cell). (b) Aligned frequency reuse (DoF=6/7 per cell) represents a 157\% improvement over the baseline. All active receivers are adjacent to inactive cells. While not shown explicitly, users are located on the cell boundaries, one on each side of the boundary. All transmissions are non-directional. Red arrows only indicate the locations of the active users. The base station transmitters in the grey cells are switched off.}
\label{fig:hexarray}
\end{figure}

Next, Fig. \ref{fig:hexarray}(a)  shows the  conventional frequency reuse baseline  and Fig. \ref{fig:hexarray}(b) shows the aligned frequency reuse pattern for a hexagonal cellular layout. While the conventional pattern achieves $1/3$ DoF per cell, the aligned frequency reuse pattern achieves $6/7$ DoF per cell, which represents an improvement of 157\% over the baseline.

The achieved value  is 6/7 DoF per cell, as obvious from the choice of active users illustrated in Fig. \ref{fig:hexarray}. The information theoretic optimality of 6/7 DoF per cell for this setting, both with or without CSIT, follows from similar arguments as the linear and square array examples. Considering any user let us eliminate all messages except the desired message and all the messages from the interfering base station. These 7 messages cannot have a total of more than 1 DoF because the single user with a single receive antenna can resolve all 7 messages. This follows from the logical argument that the user must be able to reliably decode his desired message by design, so he must be able to subtract the signal from the desired base station, which allows the user to re-construct the transmitted symbol from the interfering base station within bounded variance noise distortion, and thereby decode all 7 messages.

We note that for the linear and square grid models, the distance to the next strongest interferer is the same for both the conventional frequency re-use solution and the aligned frequency re-use solution. The aligned frequency reuse solution for the hexagonal setting on the other hand, is a bit more sensitive because the next-nearest interferer is  closer than with the conventional frequency re-use setting. While this is a sobering observation, we note that because the DoF improvement (157\%) is much stronger for the hexagonal setting than the linear or square cellular grids (33\% and 60\%, respectively), we do have additional room to accommodate interference.  Further, in going beyond DoF, what matters is not only  the strengths of the strongest interferers, but the number of such interferers as well. Since this is only an SNR offset effect it is not visible in DoF studies but is quite relevant for performance results at finite SNR. The aligned frequency reuse solution for even the linear and square cellular models will lose some of its projected multiplicative (DoF) gains over conventional frequency reuse at moderate SNR values due to this SNR offset. 
On the other hand, because of the overly pessimistic assumptions regarding CSIT, even modest  gains achieved in this setting bode well for the overall question of practical feasibility of  interference alignment schemes. 

It is quite remarkable that in a study of optimal scheduling patterns in cellular networks conducted by Bonald et al. in \cite{Bonald_Borst_Proutiere} under much more sophisticated and practical propagation models representative of 3G wireless networks, the optimal scheduling patterns that emerge from the numerical analysis are the same aligned frequency reuse patterns that we find from the information theoretic DoF analysis in this section.

\subsection{MIMO}
Multiple antennas have a special place in the DoF studies of wireless networks, as  the original setting that motivated the DoF metric, the idea of spatial multiplexing, which became a stepping stone to the study of multiuser networks. As such, the implications of multiple-antenna nodes in our setting, i.e., partially connected wireless networks with no CSIT beyond network topology, are worth studying.
\subsubsection{Symmetric MIMO}
As a first step, let us assume that all nodes are equipped with the same number  of antennas, $\Gamma$. We will call this the symmetric MIMO setting. Instead of (\ref{eq:channelmodel}) the channel model becomes:
\begin{eqnarray}
\left[\begin{array}{c}
Y_1(n)\\Y_2(n)\\ \vdots \\Y_D(n)
\end{array}
\right]
&=&\left[\begin{array}{cccc}
H_{11}&H_{12}&\cdots&H_{1S}\\
H_{21}&H_{22}&\cdots&H_{2S}\\
\vdots&\vdots&\vdots&\vdots\\
H_{D1}&H_{D2}&\cdots&H_{DS}\\
\end{array}\right]\left[\begin{array}{c}
X_1(n)\\X_2(n)\\ \vdots \\X_S(n)
\end{array}
\right]
+\left[\begin{array}{c} Z_1(n)\\Z_2(n)\\ \vdots \\Z_D(n) \end{array}\right]\label{eq:channelmodelMIMO}
\end{eqnarray}
where, over the $n^{th}$ channel use, $X_j(n)\in\mathbb{C}^{\Gamma\times 1}$ is the transmitted symbol from Source $S_j$, $Y_i(n)\in\mathbb{C}^{\Gamma\times 1}$ is the received symbol at Destination $D_i$, $Z_i(n)\in\mathbb{C}^{\Gamma\times 1}$ is the additive noise at Destination $D_i$, and $H_{ij}\in\mathbb{C}^{\Gamma\times \Gamma}$ is the constant channel coefficient matrix between Source $S_j$ and Destination $D_i$.  
The average transmit power constraint at  source $S_j$ is  set as $P_j$, i.e.,  $\mbox{E}\left[\sum_{n=1}^N\left||X_{j}(n)\right||^2\right]\leq P_j$, to ensure the following nominal interference-free rate guarantees for all desired links under i.i.d. Gaussian inputs:
\begin{eqnarray}
\log\left(1+\frac{P_j}{\Gamma N_o}\lambda^2_\gamma(H_{ij})\right)&\geq&\log(1+\mbox{SNR})~~\forall\gamma\in\{1,2,\cdots,\Gamma\},\nonumber\\
&&\forall i\in\mathcal\{1,2,\cdots,D\}, j\in\{1,2,\cdots,S\}, \mathcal{W}(D_i)\cap\mathcal{W}(S_j)\neq\phi.\label{eq:desiredpowersmimo}
\end{eqnarray}
where $\lambda_\gamma(H)$ is the $\gamma^{th}$ singular value of the channel matrix $H$. Thus, the power constraints are chosen such that, in the absence of all other messages, each message by itself can achieve a rate $\Gamma\log(1+\mbox{SNR})$. For a first order (DoF) analysis we study a partially connected network where the \emph{weak} channels are set to zero. 
\begin{eqnarray}
\mbox{Partially connected model:  } \forall i\in\{1,2,\cdots,D\}, j\in\{1,2,\cdots,S\}, \mbox{ if } t_{ij}=0 \mbox{ then } H_{ij}=0.\label{eq:partial2mimo}
\end{eqnarray}
In this partially connected model, we let SNR approach infinity (by increasing the transmit power for every source proportionately), and evaluate the achievable rates normalized by $\Gamma\log(SNR)$ to find the DoF value. 
\begin{eqnarray}
\mbox{DoF}(W)=\lim_{\mbox{\tiny SNR}\rightarrow\infty} \frac{R(W)}{\Gamma\log(SNR)}, \forall W\in\mathcal{W}
\end{eqnarray}
Note that we normalize by $\Gamma\log(SNR)$ because that is the (approximate) capacity of a MIMO link.

As it turns out, the symmetric MIMO problem is still essentially the index coding problem. All the DoF results found in this paper for partially connected SISO networks, apply to MIMO networks as well (with a corresponding normalization of the DoF). Since the outer bounds come from the index coding problem and achievability is linear for all results, it suffices to state the following two theorems for the symmetric MIMO case, which are the symmetric MIMO counterparts of the main results stated previously in Theorem \ref{theorem:one} and Theorem \ref{theorem:linear} for the SISO setting. 

\begin{theorem}\label{theorem:onemimo}
The  DoF region of the topological interference management problem \\{\bf TIM}$(\mathcal{T},\mathcal{W}(S),\mathcal{W}(D),\mathbb{C})$, for  wireless networks where each node is equipped with $\Gamma$ antennas, is bounded above by the  capacity region of the corresponding index coding problem {\bf IC}$(\mathcal{A},\mathcal{W}(S),\mathcal{W}(D))$, where  $\mathcal{A}=\overline{\mathcal{T}}$. 
\end{theorem}

\begin{theorem}\label{theorem:linearmimo}
The achievable DoF region through linear schemes, for {\bf TIM}$(\mathcal{T},\mathcal{W}(S),\mathcal{W}(D),\mathbb{C})$ where each node is equipped with $\Gamma>1$ antennas includes the DoF region of the SISO case $(\Gamma=1)$.
\end{theorem}

That the MIMO case should include the SISO case may seem trivial at first because one could simply discard all but one antenna. However, Theorem \ref{theorem:linearmimo} is not trivial because of the normalization of the DoF by the $1/\Gamma$ factor. What Theorem \ref{theorem:linearmimo} claims is that the DoF scale with the number of antennas, i.e., with $\Gamma$ antennas at each node, the prelog of the linear achievable rates is scaled by a factor of $\Gamma$. This is consistent with the spatial scale invariance property observed in MIMO interference networks with perfect CSIT \cite{Wang_Gou_Jafar_subspace, Sun_Geng_Gou_Jafar} which has been conjectured to hold in general for networks with perfect CSIT. Since we assume no CSIT beyond topology knowledge, it is remarkable that  spatial-scale invariance holds here as well.

\subsubsection{Asymmetric MIMO}
While the symmetric MIMO formulation of the topological interference management problem retains the essential character of an index coding problem, in this section we provide an example of an asymmetric MIMO topological interference management problem, which is fundamentally distinct from the index coding problem. Note that the asymmetric setting is richer than the index coding problem because of the varieties of antenna configurations one can have. For example, even the universal normalizing factor for the DoF metric, usually chosen as the DoF of one transmitter-receiver pair, is not applicable, since different links can have different DoF. Therefore the DoF values in this section will not be normalized by the antenna dimension.
\begin{eqnarray}
\mbox{DoF}(W)=\lim_{\mbox{\tiny SNR}\rightarrow\infty} \frac{R(W)}{\log(SNR)}, \forall W\in\mathcal{W}
\end{eqnarray}

The main idea illustrated here is \emph{interference diversity}. Interference diversity  refers to the observation that  each receiver experiences a different set of interferers, and therefore depending on the actions of its own set of interferers, the interference-free signal space at each receiver can be artificially made to fluctuate differently from other receivers. The knowledge of these pre-determined fluctuations, without requiring CSIT, creates  opportunities for blind interference alignment schemes  in the manner of  \cite{Jafar_corr, Wang_Gou_Jafar, Wang_Gou_Jafar_MIMO}, but without the caveats of these previous works. 

\begin{figure}[!h]
\centering
\includegraphics[width=4.5in]{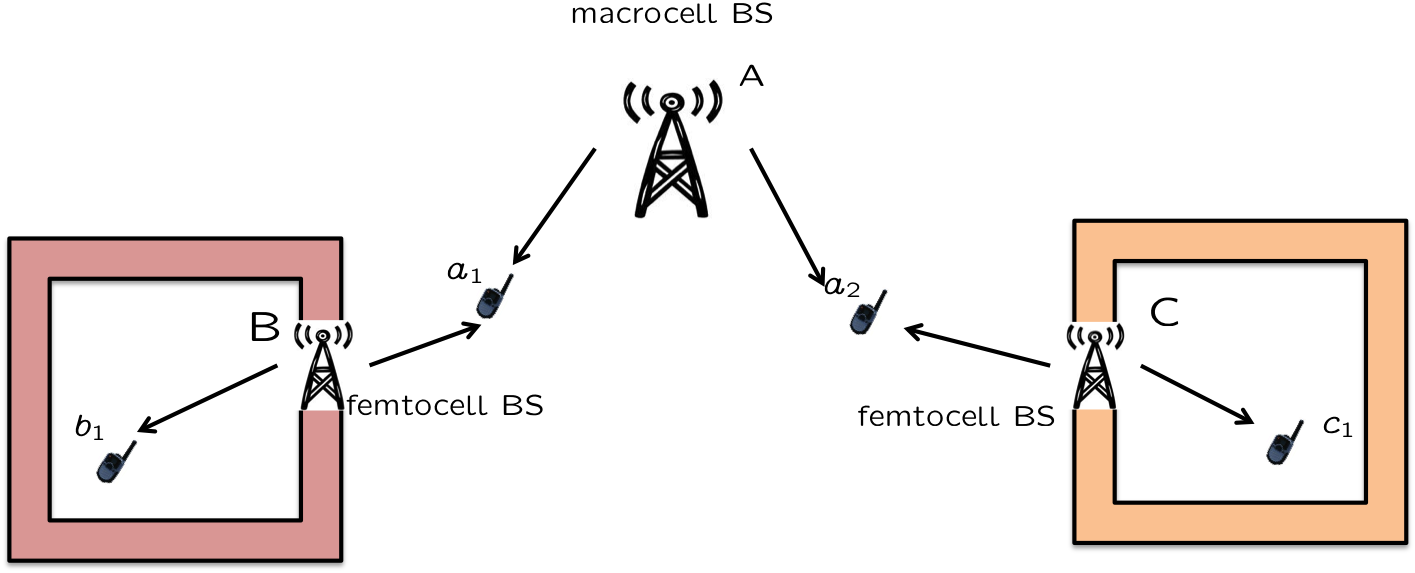}
\caption{\small\it Interference to macrocell users from in-band femtocell deployments}
\label{fig:macrofemto}
\end{figure}
We illustrate this concept through an example of a heterogeneous cellular downlink, shown in Fig. \ref{fig:macrofemto}. Such a situation would be typical for customer-deployed cells such as femtocells. For example, consider customers in areas around $b_1$ and $c_1$, who are located inside macrocell dead spots and set up their own femtocell base stations $B, C$ for wireless access, but then these femtocell base stations also interfere with neighboring macrocell users $a_1, a_2$, respectively. 

\begin{figure}[!h]
\centering
\includegraphics[width=4.5in]{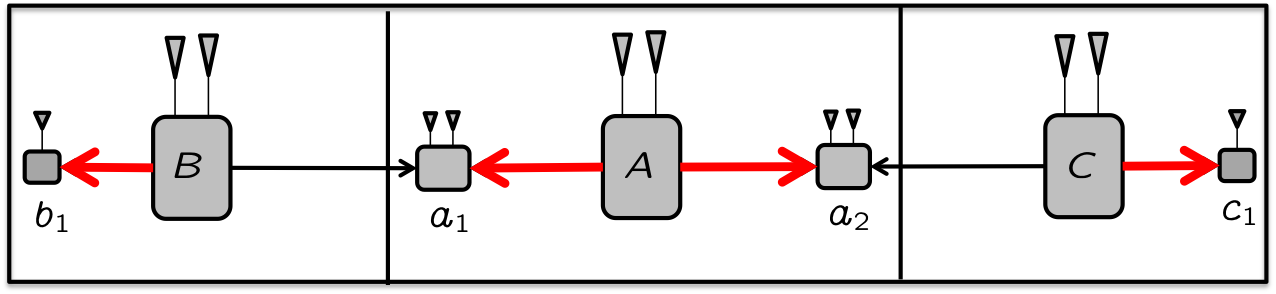}
\caption{\small\it An asymmetric MIMO instance of the topological interference management problem, exploiting interference diversity. Cell $A$ can achieve 4/3 DoF while Cells B and C achieve 1 DoF each.}
\label{fig:3cellMIMO}
\end{figure}

The resulting topological interference management problem is equivalently shown in Fig. \ref{fig:3cellMIMO} with asymmetric antenna configurations.   Note that while all transmitters (base stations) are equipped with two antennas, the receivers in cells $B, C$ are equipped with only single receive antennas.  The channels are constant for the duration of communication as assumed throughout this work and there is no CSIT beyond the topology of the network. Since the goal of this example is to highlight interference diversity, we will focus on cell $A$, and the two receivers $a_1, a_2$, who experience different interferers in transmitters $B, C$, respectively. {\it Suppose cells B, C achieve 1 DoF each. The question is to find out how many DoF  cell $A$ can achieve simultaneously.} A naive argument  might be as follows. Since cells $B, C$ achieve 1 DoF each and have only single antenna receivers, the interference from base stations $B, C$ must be consume one DoF  at each of receivers $a_1, a_2$, respectively. Essentially, this would mean that receivers $a_1, a_2$ would sacrifice one antenna to cancel the interference. That leaves cell A with a two-antenna transmitter, two single-antenna receivers and no CSIT which would suggest a collapse of DoF for cell $A$, i.e., cell $A$ should not be able to achieve any more than $1$ DoF. 

The naive reasoning is incorrect because it ignores the diversity of interference seen by the two receivers $a_1, a_2$. It turns out that cell $A$ can achieve $4/3$ DoF with no CSIT, no reconfigurable antennas and constant channels. The achievable scheme is simply the blind alignment scheme of \cite{Jafar_corr}, albeit without the restriction that nature must produce suitable distinct coherence patterns, and without the restriction to reconfigurable antennas as in \cite{Wang_Gou_Jafar}. Instead, the key here is interference diversity. We operate over three time slots. Transmitters $B, C$ send a new symbol in each time slot. However, transmitter $B$ uses his first transmit antenna for the first 2 time slots and then his second transmit antenna for the third time slot, whereas transmitter $C$ uses his first transmit antenna for the first time slot and his second transmit antenna for the last 2 time slots. In cell $A$, receivers $a_1, a_2$ simply discard the dimension along which interference is received. The net effect is that receiver $a_1$ becomes  a single antenna receiver, with a channel that remains the same for the first 2 time slots and then changes during the third time slot, whereas receiver $a_2$ becomes a single antenna receiver with a channel that changes after the first time slot and then stays constant across the last two time slots. In other words, we have created the staggered coherence blocks required for blind interference alignment \cite{Jafar_corr}, and thus, 4/3 DoF are easily achieved by cell $A$, but without requiring reconfigurable antennas, or relying on nature to create suitable statistical distinctions between the users. However, somewhat curiously, a tight DoF outer bound for this example has so far been elusive and the optimality of $4/3$ DoF for cell $A$ remains open. 

The interference diversity example suggests that the topological interference management problem, in spite of its similarity to the index coding problem in symmetric MIMO settings, can present a distinct set of challenges in asymmetric settings with arbitrary antenna configurations on top of arbitrary topology and arbitrary message sets.  

\section{Discussion}
The study of wireless networks with little to no CSIT offers a counterpoint to recent works on settings with abundant CSIT. As the abundant CSIT assumptions are incrementally relaxed and minimal CSIT assumptions incrementally enhanced, the hope is that the two perspectives will converge over practical regimes of interest, where CSIT is neither as abundant nor  as scant as often assumed in theoretical studies. The results presented here are more introductory   than conclusive, intended to lay down the groundwork for the complementary perspective, and to present supporting evidence of its promise.

Because of the coarse nature of the results several cautionary notes are in order, especially for wireless networks. The study undertaken here focuses on first order capacity characterizations (DoF) as a stepping stone to constant gap capacity approximations.  The quality of approximation improves with SNR, i.e., the difference of scale between desired channel strengths and the `weak' interferers whose net signal strength fall below the noise floor. Also, since the focus of the paper is not on minimizing the gaps, there is much room left for improving the bounds to make the gaps smaller which would be an interesting direction for future work. Another interesting issue here is the choice of the effective noise floor by the receivers. If a receiver chooses an effective noise floor too low, then most interferers would exceed the threshold, the network would become more  connected and there would be fewer opportunities to exploit topological interference management. On the other hand, if a receiver chooses the effective noise floor too high, then the network will be sparsely connected and there would be plenty of opportunities to exploit topological interference management, but the SNR itself will suffer. Therefore, the optimal choice of the effective noise floor threshold is another interesting research avenue for future work. 

It should also be pointed out that while the CSIT is minimally limited to topology, favoring robust results, there are still quite a few idealizations built into the model --- asynchronous networks, partial or mismatched topology knowledge and A/D saturation/non-linearities are some of the significant practical concerns that are ignored here.

From a theoretical standpoint, the unified framework of capacity of linear communication networks is intriguing and shows that DoF studies, often viewed as first order capacity approximations for wireless networks, are more generally useful as the exact capacity characterizations for the underlying linear communication network. Expanding this unified framework would be a worthwhile direction to pursue. Closely related to the study of linear networks is the optimality of linear solutions. While non-linear solutions are known to be necessary in some cases, it would be useful to understand whether such counter-examples are the exception or the norm. In particular, for the topological interference management problem, it is not clear whether linear solutions are sufficient.

The best case improvement of optimal solutions versus other conventional solutions such as weak fractional orthogonal scheduling, weak fractional partition multicast, local fractional coloring, for the $K$-unicast settings as well as the $K$-groupcast settings, remain open problems. Algorithm design for random instances of index coding and topological interference management problems, that incorporate the interference alignment principles, as used in the achievability proof of Theorem \ref{theorem:tree} is a promising direction. Duality of index coding and topological interference management problems, shown here from a linear coding perspective, remains an open problem from an information theoretic perspective.  The idea of splitting an index coding problem or a topological interference management problem by alignment sets, each of which can be solved as an independent problem, used to great advantage for linear schemes in this work, is also intriguing from a broader information theoretic perspective.

\bigskip
\section*{Appendix}
\appendix
\section{Index Coding Problem: {\bf IC}$(\mathcal{A}, \mathcal{W}(S),\mathcal{W}(D))$}\label{sec:indexcoding}
The following parameters  define an index coding problem.
\begin{enumerate}
\item An antidote matrix $\mathcal{A}\in\{0,1\}^{D\times S}$.
\item $S$ message sets $\mathcal{W}(S_j)$, $j\in\{1,2,\cdots, S\}$, collectively denoted as $\mathcal{W}(\mathcal{S})$.
\item $D$ message sets $\mathcal{W}(D_i)$, $i\in\{1,2,\cdots, D\}$, collectively denoted as $\mathcal{W}(\mathcal{D})$.
\end{enumerate}

These parameters define a  communication network with $S$ source nodes, labeled $S_1, S_2, \cdots, S_S$ and $D$ destination nodes, labeled $D_1, D_2, \cdots, D_D$, and two additional nodes, labeled $N_1$, $N_2$, that are connected by  a unit capacity edge going from $N_1$ to $N_2$, known as the bottleneck link. There is an infinite capacity link from every source to the node $N_1$, and an infinite capacity link from $N_2$ to every destination node. What it means is simply that $N_1$ knows all the messages, so all the coding is performed at $N_1$, and the output of the bottleneck link is available to all destination nodes.

\bigskip
\noindent {\bf Message Sets:} Source node $S_j$ has a set of independent messages, $\mathcal{W}({S_j})$, that it wants to send to their desired destinations. Destination node $D_i$ has a set of independent messages $\mathcal{W}(D_i)$ that it desires. The set of all messages is denoted as $\mathcal{W}$.
\begin{eqnarray}
\mathcal{W}&=&\bigcup_{i=1}^D\mathcal{W}(D_i)=\bigcup_{j=1}^S\mathcal{W}(S_j)
\end{eqnarray}
Each message has a unique source, i.e., $\mathcal{W}(S_j)\cap\mathcal{W}(S_{j'})=\phi$ if $j\neq j'$. If every message also has a unique destination node, it is called the multiple unicast setting. However, in general a message may be desired by multiple destinations. To distinguish it from the unicast setting defined earlier and the multiple multicast setting where every message is desired by every destination, we call this the multiple groupcast setting \cite{Maleki_Cadambe_Jafar}.

\bigskip
\noindent{\bf Antidotes:} The antidotes are defined by the matrix $\mathcal{A}=[a_{ij}]_{D\times S}$ of zeros and ones, with
\begin{eqnarray}
a_{ij}&=&\left\{
\begin{array}{ll}
0,&\mbox{no path exists from $S_j$ to $D_i$ except through the bottleneck link,}\\
1,&\mbox{a direct  link of infinite capacity exists from $S_j$ to $D_i$.}
\end{array}
\right.
\end{eqnarray}
The presence of an antidote link, i.e., an infinite capacity link between $S_j$ and $D_i$, simply means that the messages $\mathcal{W}(S_j)$ are already available to $D_i$ for free.  To avoid degenerate scenarios, we will assume throughout that $a_{ij}=0$ whenever $\mathcal{W}(D_i)\cap\mathcal{W}(S_j)\neq\phi$, i.e., if destination node $D_i$ desires a message that originates at source node $S_j$, then there is no infinite capacity link between them. The desired messages must, therefore, pass through the bottleneck link, and their  rate can at most be one.

Coding schemes, probability of error, achievable rates and capacity region are defined in the standard information theoretic sense of vanishing probability of error, albeit it is noteworthy that Langberg et al. have shown in \cite{Langberg_Effros} that the vanishing-error capacity of the index coding problem is the same as the zero-error capacity. 

\bigskip
\noindent{\bf Field:} While the field is irrelevant for the information theoretic capacity of the index coding problem, in order to identify linear schemes we will associate an auxiliary field $\mathbb{F}$ with the bottleneck link. If $\mathbb{F}$ is a finite field, i.e., the bottleneck link can transmit one symbol from $\mathbb{F}$ each channel use, then we will express the capacity of the index coding problem, normalized by $\log(|\mathbb{F}|)$. If $\mathbb{F}$ is the field of complex numbers, then we will interpret the bottleneck link as an AWGN channel  with transmit power $S^2P$, AWGN power $N_o$, and capacity $\log(1+S^2\mbox{SNR})$, where SNR = $P/N_o$, and express the capacity of the index coding problem normalized by $\log(1+S^2\mbox{SNR})$. Note that, because of network equivalence theorem of \cite{Koetter_Effros_Medard}, the choice of field is irrelevant to the normalized capacity of the index coding problem. However, the field specification will be useful to deal with linear solutions. 

\bigskip
\noindent{\bf Linear Scheme:} {\it A linear scheme over $N$ channel uses, when $\mathbb{F}$ is a finite field, achieving the rates 
\begin{eqnarray}
R(W)=\frac{L(W)}{N}, \forall W\in\mathcal{W}
\end{eqnarray}
and when $\mathbb{F}$ is the field of complex numbers, achieving the DoF, 
\begin{eqnarray}
\mbox{DoF}(W)=\frac{L(W)}{N}, \forall W\in\mathcal{W}
\end{eqnarray}
where $L(W)$ are non-negative integer values, consists of 
\begin{enumerate}
\item precoding matrices ${\bf V}(W)\in\mathbb{F}^{N\times L(W)}$, $\forall W\in\mathcal{W}$,
\item receiver combining matrices ${\bf U}_i(W)\in\mathbb{F}^{L(W)\times N}$, $\forall  W\in \mathcal{W}(D_i), i \in\{1,2,\cdots, D\}$
\end{enumerate}
such that the following properties are satisfied
\begin{eqnarray}
\mbox{Property 1:} &&{\bf U}_i(W){\bf V}(\tilde W)=0,\\
&&\forall i\in\{1,2,\cdots, D\}, j\in\{1,2,\cdots, S\},W\in\mathcal{W}(D_i), \tilde W\in\mathcal{W}(S_j),\nonumber\\
&& \mbox{ such that } W\neq \tilde W \mbox{ and }  a_{ij}=0.\nonumber\\
\mbox{Property 2:}&& det({\bf U}_i(W){\bf V}(W))\neq 0, ~\forall W\in\mathcal{W}(D_i).
\end{eqnarray}
}
So, each message $W$ is split into $L(W)$ independent scalar streams, collectively represented by the column vector ${\bf X}(W)=(x_1(W), x_2(W), \cdots,x_{L(W)}(W))^T\in\mathbb{F}^{L(W)\times 1}$, each of which carries one symbol from $\mathbb{F}$, and is transmitted along the corresponding column vectors (the ``beamforming" vectors) of the precoding matrix  ${\bf V}(W)$. In the finite field case, the symbols $x_l(W)$ are uniformly distributed over the finite field $\mathbb{F}$, each carrying $\log|\mathbb{F}|$ bits of information. When $\mathbb{F}=\mathbb{C}$, the field of complex numbers, the $x_l(W)$ are independent Gaussian codebooks, each with power $S^2P/(\sum_W L(W))$, and the columns of ${\bf V}(W)$ are scaled to have unit norm (which does not affect Property 1 or 2), so that the power constraints are satisfied.

\noindent Over the $N$ channel uses, node $N_1$ sends on the bottleneck link to $N_2$ the $N\times 1$ vector,
\begin{eqnarray}
{\bf X}&=&\sum_{W\in\mathcal{W}}{\bf V}(W){\bf X}(W).
\end{eqnarray}
$N_2$ receives the $N\times 1$ vector ${\bf Y}={\bf X}+{\bf Z}$ and passes it  to all destinations through the infinite capacity links that connect $N_2$ to each destination. Note that ${\bf Z}=0$ in the finite field case.

Destination $D_i$ removes the contribution from known antidotes and then for each desired message $W\in\mathcal{W}(D_i)\cap\mathcal{W}(S_j)$, projects the remaining signal into the ${\bf U}_i(W)$ space to obtain

\begin{eqnarray}
\overline{{\bf Y}}_i(W)&=&{\bf U}_i(W)\left({\bf Y}-\sum_{j: a_{ij}=1}\left(\sum_{W'\in\mathcal{W}(S_j)}{\bf V}(W'){\bf X}(W')\right)\right)\\
&=&{\bf U}_i(W){\bf V}(W){\bf X}(W)+{\bf U}_i(W){\bf Z},
\end{eqnarray}
where the contributions from all other messages are eliminated  due to Property 1. Now,  according to Property 2, ${\bf U}_i(W){\bf V}(W)$ is an invertible matrix, so that the following non-interfering channels are obtained for each desired symbol stream. 
\begin{eqnarray}
\overline{\overline{{\bf Y}}}_i=\left[{\bf U}_i(W){\bf V}(W)\right]^{-1}\overline{{\bf Y}}_i(W)&=&{\bf X}(W)+\underbrace{\left[{\bf U}_i(W){\bf V}(W)\right]^{-1}{\bf U}_i(W){\bf Z}_i}_{\overline{\overline{{\bf Z}}}_i}\\
\Rightarrow \overline{\overline{y}}_{i,l}(W)&=&x_l(W)+\overline{\overline{z}}_{i,l}, ~~~~l\in\{1,2,\cdots,L(W)\}.
\end{eqnarray}
Over $\mathbb{F}=\mathbb{C}$, each non-interfering channel with AWGN contributes $1/N$ DoF (it contributes 1 DoF, but because $N$ channel uses are required by the linear coding scheme, the normalized value is $1/N$ per channel use), so that DoF of $L(W)/N$ is achieved for each message $W$.  In the finite field case, there is no noise, and a rate of $L(W)/N$ is achieved for each message $W$. 

\bigskip
\noindent{\bf DoF Optimal Non-Asymptotic Linear Scheme:} Suppose an index coding problem has symmetric capacity $C_{\mbox{\small sym}}$ per message. If a non-asymptotic (i.e., precoding over finite number, $N$, of channel uses) linear precoding scheme can achieve symmetric DoF $= C_{\mbox{\small sym}}$ per message for this index coding problem over $\mathbb{C}$, then the linear scheme is said to be DoF optimal. Clearly, $C_{\mbox{\small sym}}$ must be a fraction $L/N$ in this case.

\section{Conventional Access: Orthogonal (TDMA) and Multicast (CDMA)}\label{sec:conventionalaccess}
The essence of orthogonal and multicast  schemes is defined as follows.\\
\noindent {\bf Orthogonal Transmission:} An orthogonal transmission  serves, over a single channel use, a subset of messages $\mathcal{W}_o\subset\mathcal{W}$  such that no message in $\mathcal{W}_o$ sees interference from any other message in $\mathcal{W}_o$. In terms of the conflict graph, the set of  messages served simultaneously over a channel use by an orthogonal scheme, $\mathcal{W}_o$, is an independent set of the conflict graph. Since the transmission is interference-free, the rate (DoF) allocated to each message in $\mathcal{W}_o$ over this one channel-use is unity, and the sum-rate (sum-DoF) value over this one channel use is $|\mathcal{W}_o|$. Orthogonal schemes may be seen as  TDMA (time division).

\noindent{\bf Multicast Transmission:} A multicast transmission serves, over $k$ channel uses, a subset of messages $\mathcal{W}_{m}\subset\mathcal{W}$, such that every destination that desires a message from $\mathcal{W}_m$ sees interference from less than $k$ other messages within $\mathcal{W}_m$. Since, no destination sees more than $k-1$ interferers in addition to its desired message, a total of $k$ linearly independent equations delivered over $k$ channel uses suffices for every destination to resolve one symbol from each of  the interfering messages and one symbol from the desired message. This is accomplished by  `spreading' each symbol by a different $k\times 1$ pseudo-random  code vector. Since only one symbol is sent from each message in $\mathcal{W}_m$ over $k$ channel uses, the rate (DoF) allocated to each message in $\mathcal{W}_m$ over these $k$ channel uses, is $\frac{1}{k}$ and the sum-rate (sum-DoF) value over these $k$ channel uses is $\frac{|\mathcal{W}_m|}{k}$. Multicast schemes may  be  seen as CDMA (code division).

If the goal is to achieve only the best sum-rate (sum-DoF) possible through orthogonal or multicast schemes, then it suffices to serve the best (corresponding to the highest sum-rate) subset of messages $\mathcal{W}_o$ or $\mathcal{W}_m$, and  the remaining messages are ignored. For orthogonal schemes, the highest sum-rate (sum-DoF) is the independence number of the conflict graph. Note that an orthogonal scheme is a special case of multicast. This is because if we  choose $\mathcal{W}_m=\mathcal{W}_o$ then multicast  achieves the same sum-rate (DoF) as an orthogonal scheme. Therefore, for multicast schemes, the highest sum-rate (DoF) is at least as high as that for orthogonal schemes.

Beyond sum-rates, e.g., if the goal is to optimize the  symmetric rate achieved by all messages, then both orthogonal and multicast schemes are naturally extended to  all messages $\mathcal{W}$ by appropriately partitioning the set of messages.

\noindent{\bf Symmetric Rate (DoF) -- Orthogonal Scheduling and Partition Multicast:}
 For orthogonal scheduling, the conflict graph is partitioned into independent sets, i.e., the symmetric rate (DoF) $\frac{1}{\alpha}$ per message is achieved  if the vertices of the conflict graph can be covered by $\alpha$  independent sets that are mutually disjoint. This is done by scheduling the independent sets one after another through an orthogonal scheme. 
 
For multicast, the set of messages is partitioned into mutually disjoint subsets and each subset is served by a multicast scheme \cite{Tehrani_Dimakis_Neely}. If $\mathcal{W}_1, \mathcal{W}_2, \cdots, \mathcal{W}_p$ are  $p$ mutually-disjoint and collectively exhaustive subsets of  the set of all messages, $\mathcal{W}$, and if each destination that desires a message from $\mathcal{W}_i$ sees less than $m_i$ interfering messages from within $\mathcal{W}_i$, then by serving the partitions one after another through a multicast scheme, the symmetric rate (DoF) value achieved is $\frac{1}{m_1+m_2+\cdots+m_p}$ per message.

\begin{figure}[!h]
\begin{center}
\includegraphics[width=1.5in]{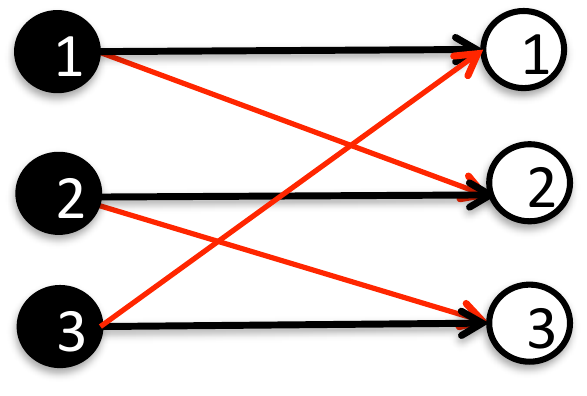}
\caption{\small \it A 3-unicast setting where the best orthogonal scheme can only achieve symmetric DoF $1/3$ per message but a multicast scheme achieves symmetric DoF $1/2$ per message (information theoretically optimal).}\label{fig:321}
\end{center}
\vspace{-0.5cm}
\end{figure}

Once again, note that orthogonal cover is a special case of partition multicast. Thus, the symmetric rate of partition multicast is at least as high as that of  orthogonal scheduling. In general symmetric rates achieved by partition multicast can be strictly higher, as illustrated in Fig. \ref{fig:321}.

It is also worthwhile to mention that orthogonal scheduling and partition multicast are equivalent for the class of symmetric unicast networks (represented by undirected side-information graphs in \cite{Tehrani_Dimakis_Neely}) where if message $W_i$ interferes with the desired destination of message $W_j$ then message $W_j$ interferes with the desired destination of (message $W_i$, as shown by Tehrani et al. in \cite{Tehrani_Dimakis_Neely}.

\noindent{\bf Fractional Orthogonal Scheduling and Fractional Partition Multicast:}
While the subsets of messages as defined above, are disjoint, the achievable rate (DoF) regions of both orthogonal cover and partition multicast schemes can be enlarged by allowing each message to be a part of multiple subsets  and considering only the average rate (DoF) of each message in the long term. This gives us the fractional version  \cite{fractionalgraphtheory} of orthogonal cover  and partition multicast schemes, and the resulting schemes are therefore called \emph{fractional} orthogonal scheduling and \emph{fractional} partition multicast.

\begin{figure}[!h]
\begin{center}
\includegraphics[width=3.5in]{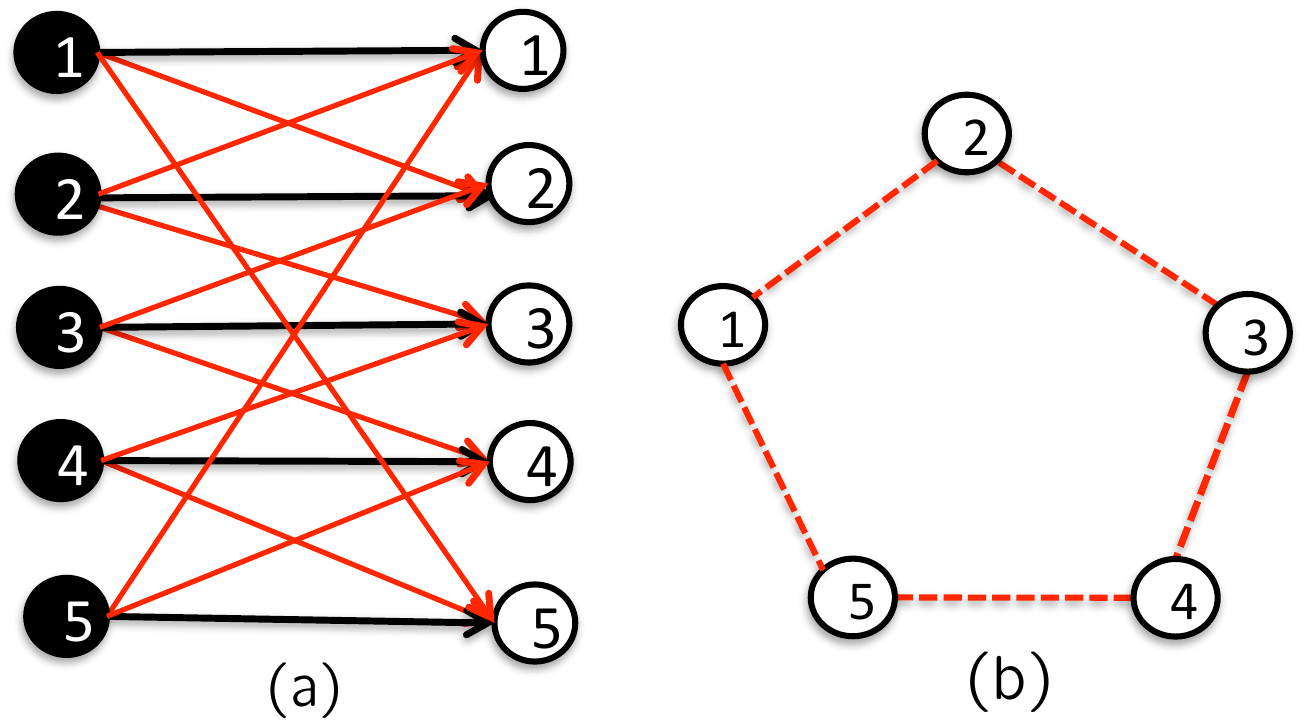}
\caption{\small \it (a) A 5-unicast topological interference management problem, (b) its conflict graph, (c) A 5-groupcast topological interference management problem that has the same conflict graph}\label{fig:5cycle}
\end{center}
\vspace{-0.5cm}
\end{figure}

The benefits of fractional schemes are evident through the example illustrated in Fig. \ref{fig:5cycle}(a). It is easy to verify that no orthogonal scheduling or partition multicast scheme can achieve symmetric DoF higher than $1/3$ in this network, which may be achieved by successively serving orthogonal subsets of messages such as $\{W_1, W_3\}, \{W_2, W_4\}, \{W_5\}$. However, a symmetric rate (DoF) of $2/5$ per message is achievable in Fig. \ref{fig:5cycle}(a) by the fractional orthogonal scheduling: $\{W_1, W_3\}$, $\{W_3, W_5\}$, $\{W_5, W_2\}$, $\{W_2,W_4\}$,$\{W_4,W_1\}$. Note that each message is a part of two subsets. As usual, fractional orthogonal scheduling schemes are a special case of fractional partition multicast. Incidentally, for this example the best fractional orthogonal scheduling scheme is also the best fractional partition multicast scheme, because it achieves the information theoretically optimal symmetric rate (DoF) value for the network of Fig. \ref{fig:5cycle}(a), as shown in \cite{Blasiak_Kleinberg_Lubetzky_2010, Maleki_Cadambe_Jafar}. 

\noindent{\bf Groupcast: Strong (messages) and Weak (destinations) Partitioning:}
In this work we use only the definitions provided above, where the multicast partitioning or orthogonal scheduling is based on messages. It is possible, however, to weaken the definitions by allowing partitioning or scheduling based on destinations rather than messages. 
For multiple unicast, where each message is desired by exactly one destination and without loss of generality one can assume that each destination desires exactly one message, there is a 1 to 1  association of messages with destinations. Thus, partitioning/scheduling messages is equivalent to partitioning/scheduling destinations. However, for groupcast settings, where each message may be desired by multiple destinations, partitioning/scheduling of messages is not the same as the partitioning/scheduling of destinations. The distinction is noted by Blasiak et al.  in Definition 2.5 of \cite{Blasiak_Kleinberg_Lubetzky_2010} where based on a side-information hypergraph representation, the orthogonal scheduling of messages is called a strong hyperclique  cover and the orthogonal scheduling of destinations is called a weak hyperclique cover. Note that an orthogonal scheduling of destinations means that the \emph{destinations} being scheduled simultaneously must not see interference from each others' desired messages. The strong hyperclique is a special case of a weak hyperclique, i.e., orthogonal scheduling of messages is a special case of orthogonal scheduling of destinations. The  weak orthogonal  scheduling schemes in general can achieve \emph{higher} rates (DoF) than  strong orthogonal  scheduling schemes. The same applies to partition multicast and the fractional versions of both orthogonal scheduling and partition multicast as well, where we can similarly define weak partition multicast,  fractional weak partition multicast, and fractional weak orthogonal scheduling. While the class of weak (fractional) schemes is the most powerful class for both orthogonal scheduling and partition multicast, in this work we  only consider \emph{strong} (fractional) schemes, and the qualifier ``strong" will be omitted for compact notation.

\section{Proofs}\label{sec:proofs}
\subsection{Proof of Theorem \ref{theorem:one}}\label{proof:theorem1}
 Starting from a reliable encoding/decoding scheme for the partially connected wireless network setting, we will go through a series of steps, each of which cannot reduce capacity, to arrive at the corresponding index coding problem. 
\begin{enumerate}
\item $\forall i\in\{1,2,\cdots,D\}, j\in\{1,2,\cdots,S\}$, if $t_{ij}=1$, then set the value of the channel coefficient as follows
\begin{eqnarray}
h_{ij}&=&\sqrt{\mbox{SNR}\times\frac{N_o}{P_j}}
\end{eqnarray}
Since this is one of the values that these channels can take, any reliable coding scheme must work for this choice of channel coefficient values as well. 
\item $\forall i\in\{1,2,\cdots,D\}, j\in\{1,2,\cdots,S\}$, if $t_{ij}=0$, then let a Genie provide the messages $\mathcal{W}(S_j)$ to destination node $D_i$. Note that none of these are desired messages. 
\item $\forall i\in\{1,2,\cdots,D\}, j\in\{1,2,\cdots,S\}$, if $t_{ij}=0$, then replace the (previously zero) channel coefficient values with new non-zero values $h_{ij}$ such that 
\begin{eqnarray}
h_{ij}&=&\sqrt{\mbox{SNR}\times\frac{N_o}{P_j}}
\end{eqnarray}
and allow full CSIR. This cannot hurt because of the previous step which provided all the messages from the ``formerly disconnected" transmitters to each receiver. Knowing all the messages from a transmitter allows the receiver to construct the transmitted codewords. Since the receiver has full CSIR, it can simply subtract the new interferers.
\item Allow full cooperation between all sources and allow full CSIT.
\item At this point, note that all channels have strength SNR, the $D\times S$ channel coefficient matrix is rank 1,  and all destinations see statistically equivalent signals. Since capacity depends only on marginals, let us assume, without loss of generality, that all destinations see the same received signal. Since all destinations see the same output signal, we have an $1\times S$ multiple input single output (MISO) channel whose output is available to all destinations. With the transmit signal power from each source and the additive noise power at the destination all normalized to unity, the channel vector is $\sqrt{SNR}\underbrace{[1, 1, \cdots, 1]}_{1\times S}$, and the total transmit power is $S$. This MISO channel has capacity $\log(1+S^2\mbox{SNR})$ \cite{Goldsmith_Jafar_Jindal_Vishwanath}. By the network equivalence theorem of \cite{Koetter_Effros_Medard} we replace this MISO channel with a noise-free link of capacity $\log(1+S^2\mbox{SNR})$. This becomes the bottleneck link and the transformation to the index coding problem is complete.
\end{enumerate}
Since at each step the capacity region is not reduced, the capacity region of the resulting index coding problem is an outer bound on the capacity region of the original wireless network. Note that in the index coding problem the bottleneck link capacity is normalized to unity. This means that if the rate allocation $R(W)$ is on the boundary of the capacity region of the index coding problem, then the rate allocation $R(W)\log(1+S^2SNR)$ must be either on the boundary or outside the capacity region of the partially connected wireless network. Dividing the rates by $\log(SNR)$ and taking the limit $\mbox{SNR}\rightarrow\infty$, we find that $R(W)$ is on the boundary or outside the DoF region of the partially connected wireless network.

A similar transformation is used in the wired case as well.
 Starting from a reliable encoding/decoding scheme for the partially connected wired network setting, we will go through a series of steps, each of which cannot reduce capacity, to arrive at the corresponding index coding problem. 
\begin{enumerate}
\item $\forall i\in\{1,2,\cdots,D\}, j\in\{1,2,\cdots,S\}$, if $t_{ij}=1$, then set the value of the channel coefficient as follows
\begin{eqnarray}
h_{ij}&=&1
\end{eqnarray}
Since this is one of the values that these channels can take, any reliable coding scheme must work for this choice of channel coefficient values as well. 
\item $\forall i\in\{1,2,\cdots,D\}, j\in\{1,2,\cdots,S\}$, if $t_{ij}=0$, then let a Genie provide the messages $\mathcal{W}(S_j)$ to destination node $D_i$. Note that none of these are desired messages. Equivalently, instead of a Genie, we can   include an infinite capacity antidote link between $S_j$ and $D_i$ whenever $t_{ij}=0$. Such a link allows the messages $\mathcal{W}(S_j)$ to be made available to $D_i$. 
\item $\forall i\in\{1,2,\cdots,D\}, j\in\{1,2,\cdots,S\}$, if $t_{ij}=0$, then replace the (previously zero) channel coefficient values with new non-zero values $h_{ij}$ such that 
\begin{eqnarray}
h_{ij}&=&1
\end{eqnarray}
and allow full CSIR. This cannot hurt because of the previous step which provided all the messages from the ``formerly disconnected" transmitters to each receiver. Knowing all the messages from a transmitter allows the receiver to construct the transmitted codewords. Since the receiver has full CSIR, it can simply subtract the new interferers.
\item Allow full cooperation between all sources and allow full CSIT.
\item At this point, note that all channel coefficient values are identical and all destinations see the same signal. Since all destinations see the same output signal, we have an $1 \times  S$ multiple input single output (MISO) channel of capacity $\log|\mathbb{GF}|$ whose output is available to all destinations. This becomes the bottleneck link. The Genie signals are made available through antidote links, and the transformation to the index coding problem is complete.
\end{enumerate}
\hfill\QED
\subsection{Proof of Theorem \ref{theorem:linear}}\label{proof:linear}
 The proof of Theorem \ref{theorem:linear} is evident from the description of linear schemes for the topological interference management problem, and the index coding problem. In particular, note that  replacing $\mathcal{A}$ with $\overline{\mathcal{T}}$, the feasibility conditions  on ${\bf U}, {\bf V}$ matrices are identical in both settings. Hence, the linear achievable rates are identical in both settings as well.\hfill\QED

\subsection{Proof of Theorem \ref{theorem:gap}}\label{proof:gap}
Suppose the index coding problem has symmetric capacity $C_{\mbox{\small sym}}$. By Theorem \ref{theorem:one}  the symmetric capacity of the wireless topological interference management problem is bounded above by $C_{\mbox{\small sym}}\log(1+S^2\mbox{SNR})$ per message. This is also an outer bound for the original wireless network because removing interference cannot hurt. Now consider achievability. We are given that a linear scheme is DoF optimal for the index coding problem over $\mathbb{C}$. So we must have $C_{\mbox{\small sym}}=\frac{L}{N}$ for some finite, non-negative integer values $L, N$, corresponding to the number of symbols sent per message and the number of channel uses, respectively. By Theorem \ref{theorem:linear} a linear scheme can achieve the same symmetric DoF for the topological interference management problem, with the same $L, N$. Let ${\bf V}(W), {\bf U}_i(W), \forall W\in\mathcal{W}, i\in\{1,2,\cdots, D\}$ be the precoding and receiver combining matrices for the DoF optimal linear scheme. The signal to noise ratio for the interference-free channel (\ref{eq:interferencefree}) is 
\begin{eqnarray}
\frac{|h_{ij}|^2P_j}{N_o}\frac{1}{|\mathcal{W}(S_j)|L\delta_{i,l}}&\geq& \mbox{SNR}\frac{1}{|\mathcal{W}(S_j)|L\delta_{i,l}}\geq\mbox{SNR}\frac{1}{KL\delta_{\max}}
\end{eqnarray}
where $\delta_{i,l}$ is the $(l,l)^{th}$ term of $[{\bf U}_i(W){\bf V}(W)]^{-1}{\bf U}_i(W){\bf U}_i^\dagger(W)\left([{\bf U}_i(W){\bf V}(W)]^{-1}\right)^\dagger$ and $\delta_{\max}$ is the maximum values it takes across $i,l$. Therefore an achievable rate guarantee for each message $W$ is 
\begin{eqnarray}
R(W)&=&C_{\mbox{\small sym}}\log\left(1+\mbox{SNR}\frac{1}{KL\delta_{\max}}\right), ~~\forall W\in\mathcal{W}
\end{eqnarray}
The gap between this achievable rate and the outer bound of $C_{\mbox{\small sym}}\log(1+S^2\mbox{SNR})$ is no more than $C_{\mbox{\small sym}}\log(S^2KL\delta_{\max})$. Clearly it can be made smaller, but since this is already an SNR independent gap, it suffices for the present purpose. \hfill\QED

\subsection{Proof of Theorem \ref{theorem:feasible}}\label{proof:feasible}
Since the outer bound  is shown for the index coding problem in \cite{Blasiak_Kleinberg_Lubetzky_2010, Maleki_Cadambe_Jafar} and the achievable scheme is linear over any field,  Theorem \ref{theorem:one} and Theorem \ref{theorem:linear} directly translate the result into the topological interference management setting. \hfill\QED

\subsection{Proof of Theorem \ref{theorem:hrfcompare}}\label{proof:hrfcompare}
Here we will prove the achievability result for fractional orthogonal scheduling schemes. Since fractional orthogonal scheduling is a special case of fractional partition multicast, the achievability result for the latter is automatically implied.

Let us assume that each destination desires only one message. There is no loss of generality in this assumption because if multiple messages are desired by a destination, then one can replace it with multiple destinations with identical received signals, each interested in only one of the originally desired messages. Because this is a unicast setting, now we have $D$ messages and $\mathcal{W}(D_j)=W_j$. Let us construct the alignment graph and conflict graph for the network and let the resulting alignment sets be $A_1, A_2, \cdots, A_m$, where $m$ is the number of alignment sets. 

Consider an $m\times m$ matrix $M$,  whose elements are disjoint and collectively exhaustive sets of messages. Let $M(i,j)$ be the element in the $i^{th}$ row and $j^{th}$ column of the matrix $M$. Then we have: 
$$M(i,j)\subset\mathcal{W}$$ 
$$M(i,j)\cap M(i',j')=\phi \mbox{ if } (i,j)\neq (i',j')$$
$$\bigcup_{i,j\in\{1,2,\cdots,m\}}M(i,j)=\mathcal{W}$$
 A message $W_i$ belongs to  $M(j,k)$ if $W_i$ is in the alignment set  $A_j$ and  its desired destination $D_i$ sees interference from a message in alignment set $A_k$. Note that all the interfering messages seen by a destination  belong to the same alignment set (by definition of alignment sets), so there is no ambiguity in this assignment of $W_i$ to $M(j,k)$.  
If the destination $D_i$ sees no interference, and $W_i$ belongs to the alignment set $A_j$, then we let $W_i\in M(j,j+1)$. The indices $i,j$ in $M(i,j)$ are interpreted modulo $m$, so $m+1$ is the same as $1$. Note that the diagonal elements of $M$ are empty sets.

We will now identify a set of non-interfering messages that can be simultaneously transmitted by an orthogonal scheme. From the integers $1,2,\cdots, m$, choose  $\lfloor m/2\rfloor$ distinct  values, $r_1, r_2, \cdots, r_{\lfloor m/2 \rfloor}$, that will serve as  indices for extracting a submatrix of $M$. Specifically from the $m\times m$ matrix $M$, let us eliminate columns $r_1, r_2, \cdots, r_{\lfloor m/2 \rfloor}$ and eliminate all rows \emph{except} rows $r_1, r_2, \cdots, r_{\lfloor m/2 \rfloor}$.  Let the set of all messages  in the surviving submatrix be denoted as $M_{[r_1, r_2, \cdots, r_{\lfloor m/2 \rfloor}]}$. Then these messages cause no interference to each other and can be simultaneously transmitted by an orthogonal scheme. This is because, the way $M$ is constructed, a message in column $r$ of $M$ can only see interference from messages in row $r$ of $M$. Since the surviving row indices are exclusive of the surviving column indices, there can be no interference among surviving messages.

Now let repeat this orthogonal scheduling process for every possible choice of the $\lfloor m/2\rfloor$ distinct  values, $r_1, r_2, \cdots, r_{\lfloor m/2 \rfloor}$. There are a total of $\binom{m}{\lfloor m/2\rfloor}$ choices of indices. The messages in $M(i,j), i\neq j$ survive only if the chosen indices contain $i$ and do not contain $j$. This can happen in $\binom{m-2}{\lfloor m/2\rfloor -1}$ ways. Thus, over $\binom{m}{\lfloor m/2\rfloor}$ channel uses, every message gets scheduled a total of $\binom{m-2}{\lfloor m/2\rfloor -1}$ times, giving us an achieved DoF value 

\begin{eqnarray}
\mbox{DoF}(W)&=&\frac{\binom{m-2}{\lfloor m/2\rfloor-1}}{\binom{m}{\lfloor m/2\rfloor}}, \mbox{ for all } W\in\mathcal{W}\\
&=& \frac{\lfloor m/2\rfloor (m-\lfloor m/2\rfloor)}{m(m-1)}\\
&=&\left\{
\begin{array}{ll}
0.25+\frac{1}{4(m-1)}&\mbox{ if $m$ is even,}\\
0.25+\frac{1}{4m}&\mbox{ if $m$ is odd.}
\end{array}
\right.
\end{eqnarray}

Thus, whenever symmetric DoF of 0.5 per message is achievable through any means, fractional orthogonal scheduling (and therefore, also fractional partition multicast) schemes can  achieve at least symmetric DoF of $0.25$ per message. We conclude with an illustration of the details of the proof for the example presented in Fig. \ref{fig:feasible}.

\begin{eqnarray*}
\begin{array}{rcl}
A_1&=&\{W_1,W_5, W_{10}\}\\
A_2&=&\{W_2,W_7\}\\
A_3&=&\{W_3\}\\
A_4&=&\{W_4,W_6,W_8\}\\
A_5&=&\{W_9\}
\end{array}
&&
M=\begin{array}{|c|c|c|c|c|}\hline
\{\}&\{W_{10}\}&\{W_1\}&\{W_5\}&\{\}\\\hline
\{W_2\}&\{\}&\{W_7\}&\{\}&\{\}\\\hline
\{\}&\{W_3\}&\{\}&\{\}&\{\}\\\hline
\{W_4, W_8\}&\{\}&\{W_6\}&\{\}&\{\}\\\hline
\{\}&\{\}&\{\}&\{W_9\}&\{\}\\\hline
\end{array}\\
\end{eqnarray*}
\begin{align*}
&M_{[1,2]}=\{W_1,W_5,W_7\},~~M_{[1,3]}=\{W_3, W_5,W_{10}\},~~M_{[1,4]}=\{W_1,W_6,W_{10}\}\\
&M_{[1,5]}=\{W_1,W_5,W_9, W_{10}\}, ~~M_{[2,3]}=\{W_2\},~~M_{[2,4]}=\{W_2, W_4, W_6, W_7, W_8\},~~M_{[2,5]}=\{W_2, W_7, W_9\}\\
&M_{[3,4]}=\{W_3, W_4, W_8\},~~M_{[3,5]}=\{W_3, W_9\}, ~~M_{[4,5]}=\{W_4, W_6, W_8\}
\end{align*}
Note that over 10 channel uses every message is scheduled a total of 3 times, thus achieving a symmetric DoF value of $0.25+\frac{1}{4m}=0.3$ per message.

\hfill\QED

\subsection{Proof of Theorem \ref{theorem:hrfoptimal}}\label{proof:hrfoptimal}
We will construct a network where 0.5 symmetric DoF is achievable and find an outer bound on the best \emph{sum} rate achievable by  partition multicast. This is also the best sum rate achievable by fractional partition multicast. Since the \emph{sum}-rate achieved by (fractional) partition multicast cannot be smaller than the \emph{sum}-rate achieved by (fractional) orthogonal scheduling schemes, the outer bound will also apply to (fractional) orthogonal scheduling schemes. To bound the symmetric rate, then we will  use the property that the symmetric rate cannot be larger than the sum-rate divided by the number of messages.

Let us construct a network with $m(m-1)$ messages (choose $m$ as an even number), where 0.5 symmetric DoF is achievable, and for which the $M$ matrix, constructed as in the proof of Theorem \ref{theorem:hrfcompare}, contains exactly one message in every off-diagonal element. The network has $m(m-1)$ sources and $m(m-1)$ destinations, i.e., $S=D=m(m-1)$, and source $S_i$ wants to send the message $W_i$ to destination $D_i$. The alignment graph for this network has $m$ alignment sets, each with $m-1$ elements. The message in $M(i,j), i\neq j$ is in alignment set $A_i$, and its desired destination sees interference from all the $m-1$ messages in the alignment set $A_j$. Since all interferers come from other alignment sets, there are no internal conflicts, and 0.5 symmetric DoF is feasible. 

For ease of exposition let us illustrate the construction through an example where $m=4$. The alignment sets and the matrix $M$ for this example are shown below.
\begin{eqnarray*}
\begin{array}{rcl}
A_1&=&\{W_1,W_2, W_{3}\}\\
A_2&=&\{W_4,W_5, W_6\}\\
A_3&=&\{W_7, W_8, W_9\}\\
A_4&=&\{W_{10},W_{11},W_{12}\}
\end{array}
&&M=
\begin{array}{|c|c|c|c|}\hline
\{\}&\{W_1\}&\{W_2\}&\{W_3\}\\\hline
\{W_4\}&\{\}&\{W_5\}&\{W_6\}\\\hline
\{W_7\}&\{W_8\}&\{\}&\{W_9\}\\\hline
\{W_{10}\}&\{W_{11}\}&\{W_{12}\}&\{\}\\\hline
\end{array}
\end{eqnarray*}
 If a message appears in the $j^{th}$ column of the matrix $M$, then the destination that desires that message,  sees interference from precisely the $3$ messages that are in the $j^{th}$ row of the matrix $M$. For example, since the message $W_1$ is in the second column of $M$, the destination that desires $W_1$ sees interference from the three messages $W_4, W_5, W_6$ that appear in the second row. The destination that desires $W_8$ also sees interference from $W_4, W_5, W_6$ whereas the destination that desires $W_4$ sees interference from $W_1, W_2, W_3$, and so on.

Leaving the example, now let us return to the general construction where $m$ is a large even number. For this network, consider any partition multicast (or orthogonal scheduling) scheme. Specifically, consider the partition that achieves the highest sum-DoF among all partitions. This is a group of messages that are multicast over the network. Let us call these the \emph{active} messages.  Suppose the maximum number of interferers seen by a destination node within the group of active messages is $l$ (for orthogonal schemes,  set $l=0$ throughout this proof). So the multicast takes place over $l+1$ channel uses and the DoF achieved per active message  is $\frac{1}{l+1}$. To find the sum-DoF we need to bound the number of active messages. Suppose the active messages comes from $n_1$ distinct columns of $M$:  specifically from columns $j_1, j_2, \cdots, j_{n_1}$. Now, a message in column $j_1$ of $M$ sees interference from   all $m-1$ messages in the alignment set $A_{j_1}$. Since no more than $l$ interferers are seen within the active set of messages, no more than $l$ messages in $A_{j_1}$ can be active. By the same logic, each of the $n_1$ alignment sets $A_{j_1}, A_{j_2}, \cdots, A_{j_{n_1}}$ cannot contain more than $l$ active messages. Therefore, the corresponding $n_1$ rows of $M$ cannot contain a total of more than $n_1l$ active messages. Since all active messages come only from $n_1$ columns, the remaining rows of $M$ cannot contain more than $(m-n_1)n_1$ active messages. Thus, the total number of active messages cannot be more than $(m-n_1)n_1+n_1l=n_1(m-n_1+l)$. The sum-DoF of this partition therefore is bounded above by $n_1(m-n_1+l)/(l+1)$, and the symmetric DoF achieved by this partition multicast scheme are bounded above by:
\begin{eqnarray}
\mbox{DoF}_{\mbox{\small sym}}(\mbox{Partition Multicast})&\leq&\frac{n_1(m-n_1+l)}{l+1}\times\frac{1}{m(m-1)}
\end{eqnarray}

If $n_1\leq m-1$,  then the term $\frac{m-n_1+l}{l+1}$ is no larger than $m-n_1$ and  the product $n_1(m-n_1)$ is no larger than $m^2/4$. Therefore the achieved symmetric DoF is no larger than $\frac{1}{4}+\frac{1}{4(m-1)}$. For any $\epsilon>0$, we can make $\frac{1}{4(m-1)}\leq\epsilon$ by choosing $m\geq\frac{1}{4\epsilon}+1$.

If $n_1=m$, then $ml/(l+1)$ is no larger than $m-1$ (because $l\leq m-1$), and the achieved symmetric DoF is no larger than $\frac{1}{m}$. This is no larger than $0.25+\epsilon$ if $m\geq 4$. 

Therefore, choosing $m\geq \max\left(4,\frac{1}{4\epsilon}+1\right)$, we have the desired statement for the network thus created --- symmetric DoF of 0.5 per message is feasible, but partition multicast or orthogonal schemes cannot achieve a symmetric DoF higher than 0.25+$\epsilon$.

\hfill\QED

\subsection{Proof of Theorem \ref{theorem:groupcastorthogonal}}\label{proof:groupcastorthogonal}
Let us construct a half-rate-feasible $K$-groupcast network with $K$ sources, each with one message. We have $K-1$ desired destinations for each message,  each destination desiring only one message, so that the total number of destinations is $K(K-1)$. In addition to its desired message, each destination sees  only one interferer. Each of the desired destinations of a particular message sees a different interferer. In this network, because every message causes interference to every other message at some destination, an orthogonal scheme cannot schedule more than one message simultaneously. Thus, the highest symmetric DoF achievable by a fractional  orthogonal scheduling scheme is $1/K$ for this network. Note that because each destination sees only one interferer, a multicast approach easily achieves symmetric DoF  of 0.5 in this network, so it is indeed half-rate-feasible.\hfill\QED

\subsection{Proof of Theorem \ref{theorem:groupcast}}\label{proof:groupcast}
Consider a half-rate feasible network  with $m$ alignment sets, each containing $m-1$ messages, so that the total number of messages is $K=m(m-1)$.  Each message originates at a distinct source node. There are a total of $m-1$ destinations interested in each message. Each destination desires only one message. So the total number of destinations is $m(m-1)^2$. Of the $m-1$ destinations that desire each message, each is uniquely associated with one of the $m-1$ alignment sets that do not contain the desired message, in the sense that it sees interference from all $m-1$ messages in its associated alignment set. For example, suppose message $W_1$ is in alignment set $A_1$ and is desired by destinations $D_1, D_2, \cdots, D_{m-1}$. Then $D_1$ sees interference from all the $m-1$ messages in the alignment set $A_2$, $D_2$ sees interference from all the $m-1$ messages in the alignment set $A_3$, $\cdots$, and $D_{m-1}$ sees interference from all the $m-1$ messages in the alignment set $A_m$. Since there are no internal conflicts, this network has symmetric DoF 0.5 per message.

Once again, let us illustrate the construction through an example where $m=4$. The alignment sets and the matrix $M$ for this example are shown below.
\begin{eqnarray*}
\begin{array}{rcl}
A_1&=&\{W_1,W_2, W_{3}\}\\
A_2&=&\{W_4,W_5, W_6\}\\
A_3&=&\{W_7, W_8, W_9\}\\
A_4&=&\{W_{10},W_{11},W_{12}\}
\end{array}
&&M=
\begin{array}{|c|c|c|c|}\hline
\{\}&\{W_1\}&\{W_2\}&\{W_3\}\\\hline
\{W_4\}&\{\}&\{W_5\}&\{W_6\}\\\hline
\{W_7\}&\{W_8\}&\{\}&\{W_9\}\\\hline
\{W_{10}\}&\{W_{11}\}&\{W_{12}\}&\{\}\\\hline
\end{array}
\end{eqnarray*}
In this example, for every message there are 3 destination nodes that desire that message, and there are 3 alignment sets that do not contain this message. Each of these three destinations that desire the given message, experiences interference from all 3 messages in one of the 3 alignment sets that do not contain the given message. So, for example, message $W_1$ is desired by three destinations, one of which sees interference from $W_4, W_5, W_6$, one that sees interference from $W_7, W_8, W_9$ and one that sees interference from $W_{10}, W_{11}, W_{12}$. Note that messages in the same row belong to an alignment set and do not interfere with each other, so that symmetric DoF value is 0.5 per message.

Leaving the example and returning to the general $m$ construction,  now consider a partition multicast scheme for the network. Specifically, consider the partition that achieves the highest sum-DoF among all partitions. This is a group of messages that are multicast over the network. Let us call these the \emph{active} messages.  Suppose the maximum number of interferers seen by a destination node within the group of active messages is $l$. So the multicast takes place over $l+1$ channel uses and the DoF achieved per message  is $\frac{1}{l+1}$. To find the sum-DoF we need to bound the number of active messages. We will consider two possibilities.

First, suppose all active messages come from the same alignment set. In this case, there is no interference between them, $l=0$, the DoF per active message is $1$,  the sum-DoF value is at most $m-1$, and the symmetric DoF achieved by this partition multicast scheme can be no more than $(m-1)/(m(m-1))=1/m$.

Now, suppose the active messages come from more than one alignment set. Let the maximum number of active messages that come from the same alignment set be $n_o$. Then, $l=n_o$, and the DoF achieved per active message is $1/(n_o+1)$.  Since there are $m$ alignment sets, and each can contribute no more than $n_o$ active messages, the total number of active messages cannot be more than $mn_o$. The sum-DoF value achievable through partition multicast is no more than $mn_o/(n_o+1)$. The symmetric DoF achieved through fractional  partition multicast scheme is no more than the highest possible sum-DoF value achieved by partition multicast divided by the total number of messages. So the  symmetric DoF are bounded above as:
\begin{eqnarray}
\mbox{DoF}_{\mbox{\small sym}}(\mbox{Fractional Partition Multicast})&\leq&\frac{mn_o}{n_o+1}\times\frac{1}{m(m-1)}\\
&\leq&\frac{1}{m}
\end{eqnarray}
This is because $n_o/(n_o+1)$ is an increasing function of $n_o$ and the maximum possible value of $n_o$ is $m-1$. 

Therefore in every case the symmetric DoF achieved by the fractional partition multicast scheme cannot be more than $1/m$. Since the number of messages, $K=m(m-1)$, we have $m>\sqrt{K}$. The symmetric DoF achieved by  fractional partition multicast schemes therefore cannot be more than $1/\sqrt{K}$.\hfill\QED

\subsection{Proof of  Theorem \ref{theorem:groupcastmax}}\label{proof:groupcastmax}
Consider a multiple groupcast network with $K$ messages that is half-rate-feasible. Assume, without loss of generality, that each destination desires exactly one message. Suppose there are $m$ alignment sets, $A_i$,  labeled in order of decreasing cardinality, i.e., $|A_1|\geq |A_2|\geq\cdots\geq|A_m|$. Let $T$ be an integer value to be defined later. Consider the following algorithm that sends one symbol from each message over $T$ time slots.

\noindent{\bf begin}\\
\begin{algorithm}[H]
t=1\;
\While{$|A_t|\geq T-t+1$}{
transmit all messages from $A_t$ in time slot $t$\;
$t\rightarrow t+1$\;
}
Over the remaining $T-t+1$ time slots, multicast the
remaining messages\;
\end{algorithm}
\noindent{\bf end}

The algorithm has two phases. The first part is an orthogonal transmission phase and the second part is a multicast phase. Note that a half-rate-feasible network does not have internal conflicts, so the messages that belong to the same alignment set do not interfere with each other. It is therefore possible to multicast all messages from the same alignment set over one time slot. This part is an orthogonal scheme. So the algorithm uses orthogonal transmission for the first $t-1$ time slots to multicast messages from $A_1, A_2, \cdots, A_{t-1}$, one alignment set per time slot. The orthogonal transmission phase stops when $|A_t| \leq T-t$. At this point there are $T-t+1$ time slots left, and all remaining alignment sets, $A_t, A_{t+1}, \cdots, A_m$ have no more than $T-t$ messages each. Since all interfering messages seen by a destination have to be in the same alignment set, evidently no destination sees more than $T-t$ interfering messages in addition to its own desired message. Therefore, $T-t+1$ time slots are sufficient to multicast all remaining messages.

The algorithm can only fail if the orthogonal phase continues until the end and even after $T$ time slots not all messages have been transmitted. Mathematically, the algorithm can fail only if both of the following conditions are true.
\begin{eqnarray}
|A_t|&\geq&T-t+1, ~~\forall t\in\{1,2,\cdots, T\}\\
\mbox{and }K&>&\sum_{t=1}^T|A_t|.
\end{eqnarray}
So the algorithm is guaranteed to succeed if
\begin{eqnarray}
K&\leq&\sum_{t=1}^T (T-t+1)\\
&=&\frac{T(T+1)}{2}
\end{eqnarray}
Or, equivalently, the algorithm is guaranteed to succeed if $T^2+T-2K\geq0$. This condition is true if 
\begin{eqnarray}
T\geq \frac{\sqrt{8K+1}-1}{2}
\end{eqnarray}
So, let us set $T=\lceil \frac{\sqrt{8K+1}-1}{2}\rceil$. Thus, a symmetric DoF of $\frac{1}{T} = \frac{1}{\lceil \frac{\sqrt{8K+1}-1}{2}\rceil}$ per message is always achievable by a partition multicast scheme. 

A slightly weaker result, but with a simpler expression, is obtained by recognizing that $ \frac{\sqrt{8K+1}-1}{2}\leq \sqrt{2K}$. So by setting $T=\lceil \sqrt{2K}\rceil$, a symmetric DoF of $\frac{1}{\lceil\sqrt{2K}\rceil}$ is always achievable by a partition multicast scheme.
 \hfill\QED

\subsection{Proof of Theorem \ref{theorem:noalign}}\label{proof:noalign}
Bar-Yossef et al. \cite{Yossef_Birk_Jayram_Kol_Trans} have shown that for a $K$-unicast index coding problem, that the demand graph is acyclic, is sufficient to conclude that the network has  symmetric capacity  $1/K$. In fact, their proof also implies the stronger statement:  an acyclic demand graph implies that the \emph{sum} capacity is $1$ (because there exists a node with no antidotes that is able to decode all messages). To show that the acyclic demand graph condition is also necessary, we construct a proof by contradiction. Consider a $K$-unicast index coding problem where the demand graph is not acyclic, i.e., it contains one or more cycles, but still has symmetric capacity is $1/K$.  Consider any such cycle.  Let the set of message nodes involved in this cycle be denoted by $\mathcal{W}(\mbox{cycle})$.  Since there is only one outgoing edge from any message node,  going to the only destination node that desires the message (unicast),  the cycle only includes the desired destinations of the messages in $\mathcal{W}(\mbox{cycle})$. Since they are a part of a cycle, each of these destinations must have at least one outgoing edge to a message in $\mathcal{W}(\mbox{cycle})$, i.e., each destination has at least one antidote from the messages within the cycle. Therefore, by themselves all the messages in $\mathcal{W}(\mbox{cycle})$  can be simultaneously multicast to achieve a symmetric rate $1/(|\mathcal{W}(\mbox{cycle})|-1)$, i.e., a sum-capacity of $\frac{|\mathcal{W}(\mbox{cycle})|}{(|\mathcal{W}(\mbox{cycle})|-1)}$ which is greater than 1. But this is a contradiction because the the sum-capacity of the original network is only 1. Thus, it is established that the acyclic demand graph condition is both necessary and sufficient for a $K$-unicast index coding problem to have symmetric capacity $1/K$ per message. 

Since the capacity region of the index coding problem is an outer bound on the corresponding topological interference management problem and symmetric rate (DoF) of $1/K$ is always achievable, the necessary and sufficient condition for a $K$-unicast topological interference management problem is the same as that for the corresponding index coding problem. Thus, it is established that the acyclic demand graph condition is both necessary and sufficient for a $K$-unicast topological interference management problem to have symmetric capacity (DoF) $1/K$ per message. 

Next, let us consider a $K$-groupcast index coding problem and show that it has symmetric capacity $1/K$ per message if and only if it can be relaxed into a $K$-unicast setting with an acyclic demand graph. The ``if" part  is trivial because the relaxing operation (eliminating demands) cannot reduce the capacity region of the network, and the resulting $K$-unicast network has symmetric capacity $1/K$ per message (because it has an acyclic demand graph). It remains to show the ``only if" part.

We start with the observation that a $K$-groupcast index coding problem can have symmetric capacity $1/K$ only if at least one destination node has no antidotes. This is because if every destination has at least one antidote, then a multicast (CDMA) approach can achieve a symmetric rate $1/(K-1)$ per message, so $1/K$ cannot be the symmetric capacity. This observation is essential to the proof.

Consider any instance of a $K$-groupcast index coding problem that has symmetric capacity $1/K$ per message. We will relax this problem into a $K$-unicast index coding problem that has an acyclic demand graph. From the groupcast problem, select a destination node, $D_i$, that has no antidotes. As just proved, such a node must exist. For all the messages desired by this node,  $\mathcal{W}(D_i)$, eliminate all other demands, i.e., relax the network so that any $W\in\mathcal{W}(D_i)$ is only desired by $D_i$. 
Clearly, after this relaxation $D_i$ and $\mathcal{W}(D_i)$ cannot be a part of any cycle in the demand graph, because the outgoing edges from message nodes $W\in\mathcal{W}(D_i)$ only go to $D_i$ and $D_i$ has no outgoing edges (because it has no antidotes). Include the node $D_i$ and the messages $\mathcal{W}(D_i)$ into our unicast construction and eliminate them from the original groupcast problem. After this elimination, also remove any  destination nodes that are left with no desired messages (because all their desired messages were in $\mathcal{W}(D_i))$. At this point the unicast problem has only $|\mathcal{W}(D_i)|$ messages. Our goal is to build it up to $K$ messages.

Consider the remaining  $K_1$-groupcast index coding problem, where $K_1=K-|\mathcal{W}(D_i)|$. Since the demand graph is still acyclic,  this network must have symmetric capacity $\frac{1}{K_1}$. Therefore, once again we can find a destination $D_j$ that has no antidotes among these $K_1$ messages, eliminate all other demands for the messages desired by $D_j$, then  include $D_j$ and $\mathcal{W}(D_j)$ into the unicast construction and eliminate them from the original groupcast problem. Note that $D_j$ may have antidotes from $\mathcal{W}(D_i)$ which will be preserved, but as explained earlier, these cannot be a part of a cycle in the relaxed unicast construction. 

The procedure described above can be repeated until we have transferred all $K$ messages into  the unicast network. The construction also guarantees that the demand graph is acyclic, completing the proof of the ``only if" part. Once again, because the index coding problem capacity provides an outer bound for the topological interference management problem, and the $1/K$ rate (DoF) is always achievable, the necessary and sufficient condition for symmetric capacity (DoF) of $1/K$ in the $K$-groupcast topological interference management problem is inherited directly from the corresponding index coding problem.

Lastly, let us derive the exact capacity result for the original wireless network.  First, the outer bound: Set all weak interference channels to zero, and the significant interference channels to be the same as desired channels, i.e., all with strength SNR and all with zero phase. Any achievable scheme must work for all possible channel realizations and this is one possible channel realization, so it can be used for the outer bound argument. Now, because the demand graph is acyclic  the argument presented in \cite{Neely_Tehrani_Zhang} leads us to the conclusion that there is a destination node in the network that can decode all messages.  For example, in the example Fig \ref{fig:noalign} it is Destination $2$. After decoding its desired signals $W_1, W_2, W_4$, it can subtract out the signal from Source $4$ from its received signal to obtain a signal statistically equivalent to the signal received by Destination $3$, from which it can then decode all the messages desired by Destination $3$, i.e., $W_3, W_5$. Thus, Destination $2$ can decode all the messages in the network. Using the multiple-access-channel capacity bound for this node gives us a sum-capacity outer bound $\log(1+K \mbox{SNR})$. This automatically implies a symmetric capacity outer bound of $\frac{1}{K}\log(1+K\mbox{SNR})$ per message. 

Achievability of this symmetric rate is described as follows. The transmit power of each source, $P_i$, is such that each desired message is capable of the same average SNR guarantee (\ref{eq:desiredpowers}). There are $K$ sources, one for each message, so let each source transmit for only $1/K$ fraction of the time according to a TDMA scheme, at a power level $KP_i$, so that the average power constraint is still satisfied, and the rate achieved is $\frac{1}{K}\log(1+K\mbox{SNR})$ per message, which matches the outer bound. Thus, we have an exact characterization of the symmetric capacity for this class of networks.\hfill\QED

\subsection{Proof of Theorem \ref{theorem:tree}}\label{proof:tree}
Since the outer bound is already available from Corollary \ref{corollary:slide}, we need to prove only the achievability. 

Note that if each alignment set contains either no cycles, or no forks, then the maximum cardinality of an acyclic subset of messages is $4$. This is because in an acyclic subset there must be a message whose desired destination sees interference from all other messages. Any destination that sees interference from more than 3 messages will introduce a clique of size 4 or higher into the alignment set, which has both cycles and forks. 

So, let us first consider the case where an acyclic subset of cardinality $4$ is present. The only way this can happen if there exist $4$ messages $W_{j_1}, W_{j_2}, W_{j_3}, W_{j_4}$ such that messages $W_{j_1}, W_{j_2}, W_{j_3}$ interfere with  a destination that desires $W_{j_4}$;  $W_{j_1}, W_{j_2}$ interfere with a destination that desires $W_{j_3}$; and $W_{j_1}$ interferes with a destination that desires $W_{j_2}$.  In this case, an independently generated $4\times 1$ vector is assigned to every message $W\in\mathcal{W}$.   Since no destination can see more than 3 interferers in a network where each alignment set has either no cycles or no forks, a multicast (CDMA) approach can always achieve a rate  (DoF) value of $1/4$ per message. Since this is also a capacity (DoF) outer bound according to Theorem \ref{theorem:acyclicbound}, the capacity (DoF) characterization is complete for this case. Incidentally,  this is the only case where the symmetric capacity is not $\frac{\Delta}{2\Delta+1}$. So in the remainder of the proof we will prove the achievability of $\frac{\Delta}{2\Delta+1}$ for all other cases.

The goal is to operate over $2\Delta+1$ channel uses and choose $\Delta$ precoding vectors for each message, along which $\Delta$ symbols for that message will be sent.  A key idea here is that the precoding for each alignment set is designed independently. So we will describe the precoding vector design for each \emph{type} of alignment set.

\bigskip
{\bf Alignment sets with no internal conflicts:} \\
For each alignment set $A_i$ that has no internal conflicts, we randomly generate a $(2\Delta+1)\times \Delta$ matrix ${\bf V}(A_i)$. 
\begin{eqnarray}
{\bf V}(A_i)&=&\mbox{rand}(2\Delta+1, \Delta)
\end{eqnarray}
where rand($a,b$) is a function that returns a randomly generated $a\times b$ matrix. The same precoding matrix ${\bf V}(A_i)$ will  be used by every message node in $A_i$.  That is, $\Delta$ symbols for each message $W\in A_i$ will be sent along the  $\Delta$ columns (normalized, in the wireless case, to satisfy power constraints) of ${\bf V}(A_i)$. As an example, note that this is the case for alignment set $A_3$ in Fig. \ref{fig:nocyclenofork}.

Next we describe precoder design for the  alignment sets that have internal conflicts. These are further classified as follows. 

\bigskip
{\bf Alignment sets with no cycles:} \\
\begin{enumerate}
\item From each alignment set $A_i$ with no cycles (may have multiple forks), arbitrarily choose one message node to be the root node of that alignment set. Without loss of generality, let the chosen root node of alignment set $A_i$ be the message $W_1(A_i)$. 
\item For each root node $W_1(A_i)$, randomly and independently generate a $(2\Delta+1)\times \Delta$ matrix ${\bf V}_1(A_i)$. 
\begin{eqnarray}
{\bf V}_1(A_i)&=&\mbox{rand}(2\Delta+1, \Delta),
\end{eqnarray}
 This is the precoding matrix to be used by the root node message. 
 \item Since this alignment set is acyclic and undirected and each connected component has a designated root, each node that is not a root node has a unique parent node. For each non-root message node $W_j(A_i), j\neq 1$, let its parent node be denoted as $W_{\pi(j)(A_i)}$. 
\item The precoding matrix for every non-root node $W_j(A_i), j\neq 1$  is generated as:
\begin{eqnarray}
{\bf Q}_j(A_i)&=&\mbox{rand}(\Delta, \Delta-1)\\
{\bf V}_j(A_i)&=&[{\bf V}_{\pi(j)}(A_i){\bf Q}_j(A_i)~~~~ \mbox{rand}(2\Delta+1,1)]
\end{eqnarray}
\end{enumerate}
The random matrix ${\bf Q}_j(A_i)$ is simply meant to choose a generic $\Delta-1$ dimensional subspace from the signal space of the parent node. This is appended with an independently generated vector that will  (with high probability over a sufficiently large field) be in general position with respect to all previously generated vectors, i.e., linearly independent of any $2\Delta$ of  previously generated vectors. Thus, any two messages that are connected by an edge in the alignment graph will have a $\Delta-1$ dimensional overlap between their signal spaces. Because the construction always includes an independent random vector in addition to the space inherited from the parent, message nodes that are connected by a path of two edges in the alignment graph,  have an overlap of $\Delta-2$ dimensions, message nodes that are connected by a path of three edges  have an overlap of $\Delta-3$ dimensions, and so on, so that messages that are connected by a path of $\Delta$ edges (or more) have no overlap. Thus, all conflicts are avoided. 
An example of the construction is provided in Fig. \ref{fig:tree}(c) where the alignment set has no cycles, $\Delta=2$ and message $W_5$ is chosen as the root node. Another example is alignment set $A_1$ in Fig. \ref{fig:nocyclenofork} where node $1$, corresponding to $W_1$ is chosen as the root node.

\bigskip
{\bf Alignment sets with no forks :--- Cycles of length greater than 3:}\\
An alignment set $A_i$ with no forks, that contains a cycle, can only be a cycle itself.  Let the length of the cycle be  $|A_i|=l$. Note that since the cycle contains internal conflicts, the length of the cycle, $l\geq 2\Delta$. Label the messages in the cycle (in order) as $W_1(A_i), W_2(A_i), \cdots, W_l(A_i)$. Randomly generate $l$ vectors, each $(2\Delta+1)\times 1$, and call them ${\bf v}_1(A_i), {\bf v}_2(A_i), \cdots, {\bf v}_l(A_i)$. Now assign the vectors cyclically (subscripts modulo $l$) as follows:
\begin{eqnarray}
W_1(A_i)&:& {\bf v}_1(A_i), {\bf v}_2(A_i), \cdots, {\bf v}_{\Delta}(A_i)\\
W_2(A_i)&:& {\bf v}_2(A_i), {\bf v}_3(A_i), \cdots, {\bf v}_{\Delta+1}(A_i)\\
W_3(A_i)&:& {\bf v}_3(A_i), {\bf v}_4(A_i), \cdots, {\bf v}_{\Delta+2}(A_i)\\
\vdots &:&\vdots\\
W_l(A_i)&:& {\bf v}_l(A_i), {\bf v}_1(A_i), \cdots, {\bf v}_{\Delta-1}(A_i)
\end{eqnarray}
Note that  this construction ensures that there are no overlaps between conflicting nodes. As an example, note that this is the case for alignment set $A_2$ in Fig. \ref{fig:nocyclenofork}.

\bigskip
{\bf Alignment sets with no forks :--- Cycle of length 3:}\\
Consider an alignment set $A_i$ that is a cycle of length 3 (i.e., a triangle) with internal conflicts $(\Delta=1)$, and comprised of messages $\{W_{j_1}, W_{j_2}, W_{j_3}\}$.  If no destination sees all three messages as interference, then  our regular construction works, i.e., simply assign a randomly generated $3\times 1$ vector to each of the three message nodes that form the triangle. 

\begin{figure}[h]
\begin{center}
\includegraphics[width=3.2in]{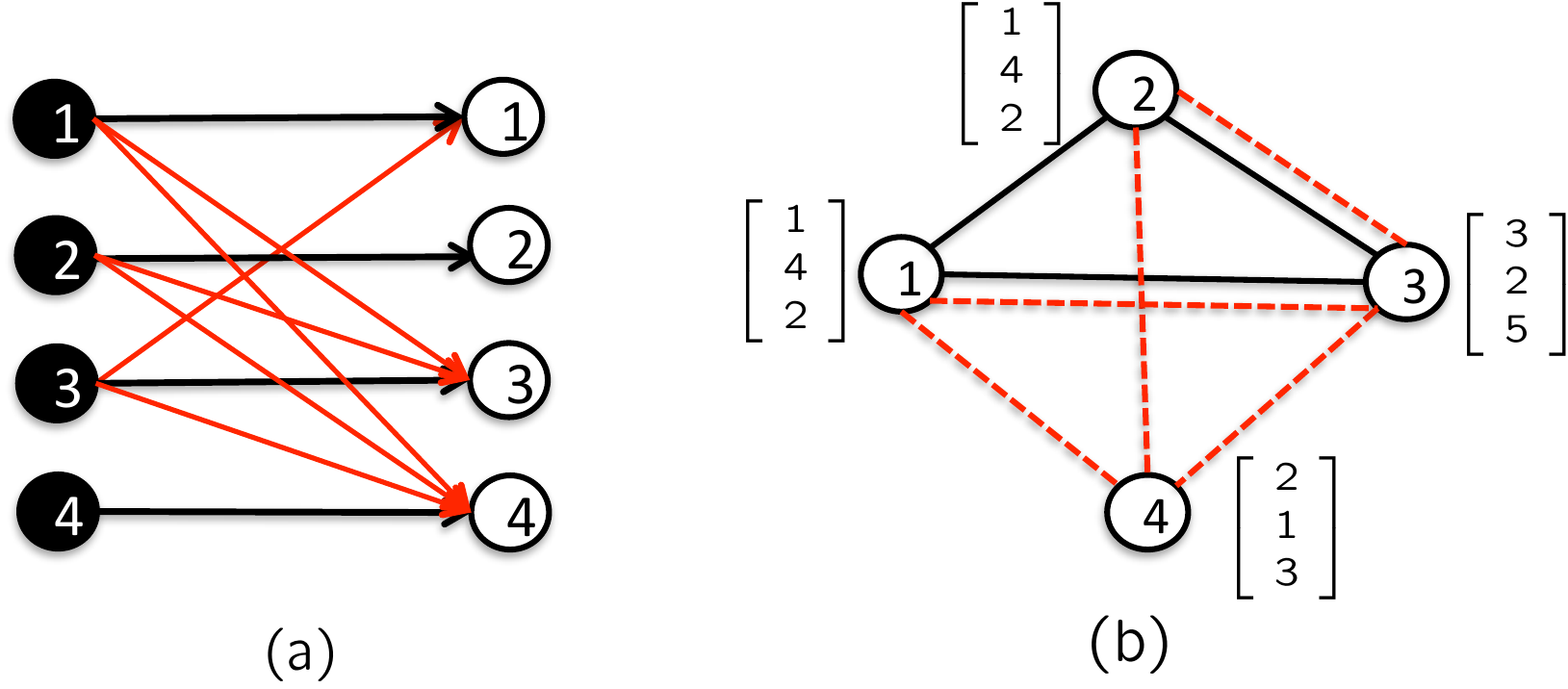}
\caption{\small \it (a) Example of a  TIM problem (b) Signal vector assignments}\label{fig:3cyclecase1}
\end{center}
\end{figure}

Now consider the case where a destination that desires message $W_{j_4}$ sees  interference from all three messages $W_{j_1}, W_{j_2}, W_{j_3}$.  Here also there are two cases. First, the case where $\{W_{j_1}, W_{j_2}\}$ interfere with a destination that desires $W_{j_3}$. Note that there cannot be conflict between $W_{j_1}, W_{j_2}$ otherwise we would have an acyclic subset of messages of cardinality 4 (a case that we already considered at the beginning of this proof). In this case, the precoding vectors are assigned as follows.
\begin{eqnarray}
W_{j_1}(A_i), W_{j_2}(A_i)&:&\mbox{rand}(3,1)\\
W_{j_3}(A_i)&:&\mbox{rand}(3,1)
\end{eqnarray}
Thus, the same random vector is assigned to messages $W_{j_1}, W_{j_2}$ and an independently generated vector is assigned to $W_{j_3}$. Thus, together $W_{j_1}, W_{j_2}, W_{j_3}$ span a two dimensional subspace, leaving  one dimension for $W_{j_4}$, so that the rate $1/3$ per message can be achieved. An example  is illustrated in Fig. \ref{fig:3cyclecase1}. 

\begin{figure}[h]
\begin{center}
\includegraphics[width=3.2in]{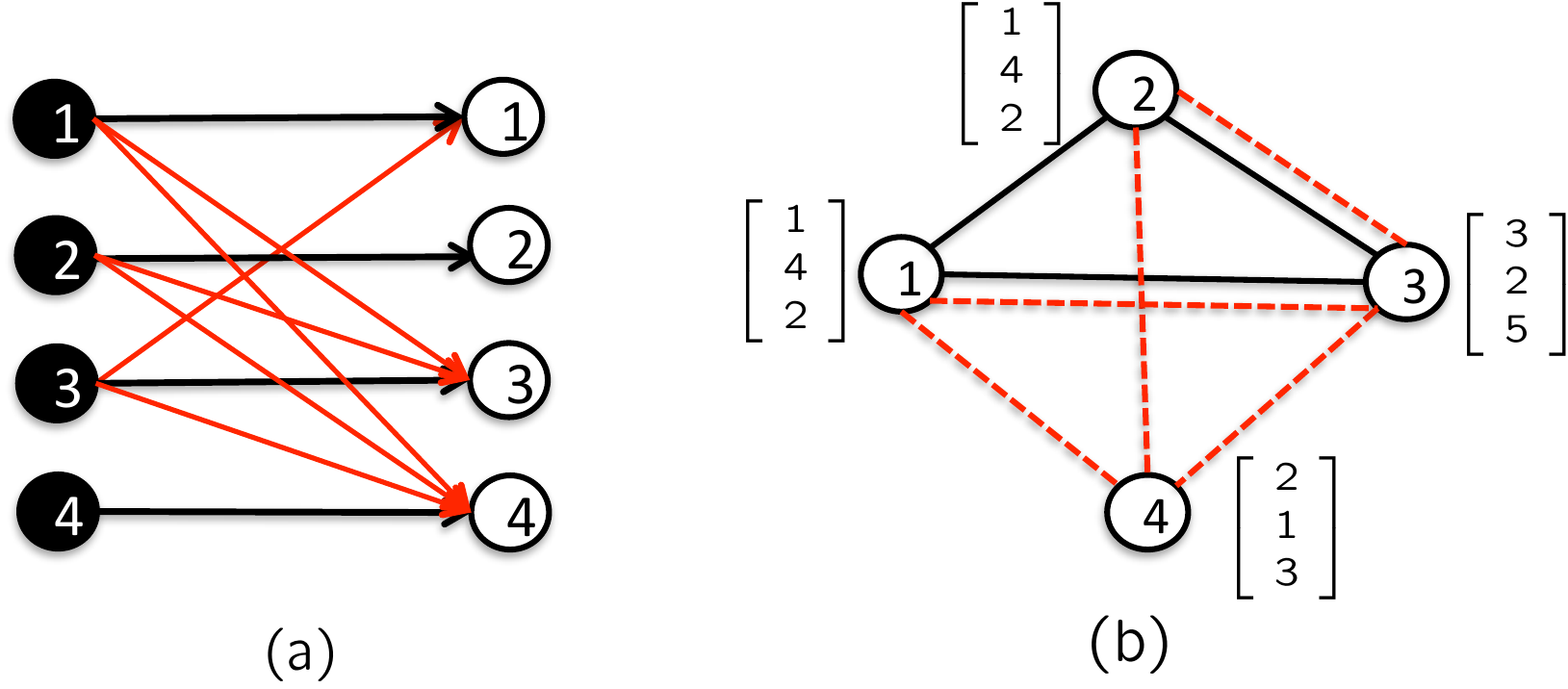}
\caption{\small \it (a) Example of a TIM problem (b) Signal vector assignments. }\label{fig:3cyclecase2}
\end{center}
\end{figure}

The last remaining case is where no two of $W_{j_1}, W_{j_2}, W_{j_3}$ cause interference to a destination that desires the third message from this group. Such an example is illustrated in Fig. \ref{fig:samegraph}(b). In this case if a destination that desires one of these three messages sees interference from another one of these messages (internal conflict) then it must see only one interferer. To generate the precoding vectors we will generate three random vectors in a two dimensional space. The vectors are assigned as follows:
\begin{eqnarray}
{\bf Q}(A_i)&=&\mbox{rand}(3,2)\\
W_{j_1}(A_i)&:&{\bf Q}(A_i)\mbox{rand}(2,1)\\
W_{j_2}(A_i)&:&{\bf Q}(A_i)\mbox{rand}(2,1)\\
W_{j_3}(A_i)&:&{\bf Q}(A_i)\mbox{rand}(2,1)
\end{eqnarray}
Again, together $W_{j_1}, W_{j_2}, W_{j_3}$ span a two dimensional subspace, leaving one dimension for $W_{j_4}$. An example  is illustrated in Fig. \ref{fig:3cyclecase2}. Here, we have only one alignment set $A$,  ${\bf Q}(A)=[1,3;4,2;2,5]$, and the $2\times 1$ projection vectors are $[1;0]$, $[0;1]$ and $[1;1]$.

This completes the construction. 
What remains is to show that with high probability  (i.e., probability 1 in the wireless case, and a probability that can be as close to 1 as needed over a sufficiently large field in the wired case)  the desired signals at each destination have no overlap with the interference. Without loss of generality we will assume each destination desires one message. Because the proof for the case of acyclic subset of cardinality 4 has already been completed at the beginning of this proof, here we will assume that no such acyclic subset exists.

Consider a destination whose interfering messages come from an alignment set that has no internal conflicts. Thus, all interfering messages span the same $\Delta$ dimensional space, and the desired signal (because it belongs to a different alignment set) spans an independently generated $\Delta$ dimensional space. Since the overall number of dimensions is $2\Delta+1$, with high probability these two spaces have no overlap. 

Henceforth we consider only destinations whose interfering messages come from an alignment set that has  internal conflicts.

Consider  a destination that sees three interfering messages. The three interfering messages form an alignment set that is a cycle of length 3, and the minimum conflict distance is $1$. It is easy to verify that the achievable scheme allocates only a 2 dimensional space to the interferers, independently generated from the precoding vector for the desired message. Since the overall space is 3 dimensional (3 channel uses), the desired signal is separable from interference.

Henceforth we consider only destinations that see no more than two interfering messages.

Consider a destination whose interfering messages come from an alignment set that has no cycles (but may have multiple forks).  Suppose the desired message, say $W_i$, sees two interferers $W_j, W_k$. Then $W_j, W_k$ must be connected by an edge in the alignment graph. Therefore, they must have a $\Delta-1$ dimensional overlap, so that together they must span $\Delta+\Delta-(\Delta-1)=\Delta+1$ dimensions. Further, if $W_i$ is in the same alignment set, then  $W_j, W_k$ must be at least $\Delta$ edges away from $W_i$, so that with high probability the union of the spans of ${\bf V}_j, {\bf V}_k$ is in general position with respect to ${\bf V}_i$. Since the total space is $2\Delta+1$ dimensional, it is big enough to accommodate the interference and the desired signal without forcing them to overlap. Thus, the desired signal does not overlap with interference with high probability.  If the message $W_i$ is in a different alignment set then again its signal space is independently generated and with high probability has no overlap with the space spanned by the interference. If the message $W_i$ sees only one interferer, $W_j$, then once again because $W_i, W_j$ are at least $\Delta$ edges apart (or belong to different alignment sets), the signal spaces ${\bf V}_i, {\bf V}_j$ have no overlap with high probability.

Lastly, consider a destination whose two interfering messages come from an alignment set that is a  cycle. Because of the construction, these two interfering messages span a $\Delta+1$ dimensional space. If the desired signal is in the same alignment set as the interfering messages, then it is separated by at least $\Delta$ edges, and therefore with high probability has no overlap with interference. If the desired signal is in a different alignment set, then its signal space is independently generated from the interference and again has no overlap with interference with high probability.

{\it Remark:} Note that the ``high probability" condition is meant in the sense of the standard random coding argument, as a proof of existence, and does not imply that the codebooks are actually random (unknown a-priori). Over a sufficiently large field, all the generic properties (linear independence conditions) are true with high probability for randomly generated signal spaces. This means that there must exist a choice of signal spaces for which all linear independence conditions hold. This is the \emph{deterministic} choice of signal spaces that is actually used by the linear scheme.\hfill\QED

\subsection{Proof of Theorem \ref{theorem:duality}}\label{proof:duality}
The proof follows from a direct mapping of precoding and combining matrices from the original problem to the dual problem. If the rate tuple $R(W)=\frac{L(W)}{N}$ is achievable through linear schemes over $N$ channel uses, then according to Section \ref{sec:linear}, there must exist precoding matrices ${\bf V}(W)\in\mathbb{F}^{N\times L(W)}, \forall W\in \mathcal{W}$ and receiver combining matrices ${\bf U}_i(W)\in \mathbb{F}^{L(W)\times N}, \forall W\in\mathcal{W}(D_i), i\in\{1,2,\cdots, D\}.$

Let us define ${\bf U}(W)\in\mathbb{F}^{L(W)\times N}$ as follows:
\begin{eqnarray}
\forall W\in\mathcal{W}, {\bf U}(W) & = & {\bf U}_i(W), \mbox{ such that } W\in\mathcal{W}(D_i)
\end{eqnarray}
Note that this definition is unambiguous only in the multiple unicast setting, where each message has a unique desired destination. With this definition, Property 1 and Property 2 can be written as:
\begin{eqnarray}
\mbox{Property 1:} &&{\bf U}(W){\bf V}(\tilde W)=0,\\
&&\forall i\in\{1,2,\cdots, D\}, j\in\{1,2,\cdots, S\},W\in\mathcal{W}(D_i), \tilde W\in\mathcal{W}(S_j),\nonumber\\
&& \mbox{ such that } W\neq \tilde W \mbox{ and }  t_{ij}=1.\nonumber\\
\mbox{Property 2:}&& det({\bf U}(W){\bf V}(W))\neq 0, ~\forall W\in\mathcal{W}.
\end{eqnarray}
which can further be re-written, equivalently, as:
\begin{eqnarray}
\mbox{Property 1:} &&({\bf V}(W))^T({\bf U}(\tilde W))^T=0,\label{eq:prop1primal}\\
&&\forall i\in\{1,2,\cdots, S\}, j\in\{1,2,\cdots, D\},W\in\mathcal{W}(S_i), \tilde W\in\mathcal{W}(D_j), \nonumber\\
&& \mbox{ such that } W\neq \tilde W \mbox{ and }  t_{ji}=1.\nonumber\\
\mbox{Property 2:}&& det(({\bf V}(W))^T({\bf U}( W))^T)\neq 0, ~\forall W\in\mathcal{W}.\label{eq:prop2primal}
\end{eqnarray}

In the dual problem, for the same rate tuple $R(W)=\frac{L(W)}{N}$ to be achievable through linear schemes over $N$ channel uses,  there must exist precoding matrices ${\bf V}'(W)\in\mathbb{F}^{N\times L(W)}, \forall W\in \mathcal{W}$ and receiver combining matrices ${\bf U}'(W)\in \mathbb{F}^{L(W)\times N}, \forall W\in\mathcal{W}$, such that:
\begin{eqnarray}
\mbox{Property 1:} &&{\bf U}'(W){\bf V}'(\tilde W)=0,\label{eq:prop1dual}\\ 
&&\forall i\in\{1,2,\cdots, D'\}, j\in\{1,2,\cdots, S'\},W\in\mathcal{W}'(D_i'), \tilde W\in\mathcal{W}'(S_j'),\nonumber\\
&&\mbox{ such that } W\neq \tilde W \mbox{ and }  t'_{ij}=1.\\
\mbox{Property 2:}&& det({\bf U}'(W){\bf V}'(W))\neq 0, ~\forall W\in\mathcal{W}.\label{eq:prop2dual}
\end{eqnarray}

\noindent Let us set the values of the dual precoding and combining matrices as follows.
\begin{eqnarray}
{\bf U}'(W)&=&\left({\bf V}(W)\right)^T\\ 
{\bf V}'(W)&=&\left({\bf U}(W)\right)^T
\end{eqnarray}
In the dual problem, since $S'=D, D'=S, \mathcal{W}'(D_i')=\mathcal{W}(S_i), \mathcal{W}'(S_j')=\mathcal{W}(D_j), t_{ij}'=t_{ji}$, we note that Property 1 represented by (\ref{eq:prop1dual}) and Property 2 represented by (\ref{eq:prop2dual}) in the dual problem are identical to Property 1 represented by (\ref{eq:prop1primal}) and Property 2 represented by (\ref{eq:prop2primal}) in the original problem. Therefore, feasibility of the original problem implies feasibility of the dual problem, and vice versa.\hfill\QED

\subsection{Proof of Theorem \ref{theorem:dualtree}}\label{proof:dualtree}
Consider a half-rate-infeasible multiple unicast topological interference management problem whose dual has either no cycles or no forks in each alignment set of its alignment graph. By Theorem \ref{theorem:tree}, the dual problem has symmetric capacity (DoF) of  $\frac{\Delta'}{2\Delta'+1}$ per message, where $\Delta'$ is the minimum conflict distance of the dual problem. Further, by the proof of achievability of Theorem \ref{theorem:tree}, the symmetric capacity is achievable by a linear scheme. Now, by Theorem \ref{theorem:duality}, any rate achievable by linear schemes in the dual problem is also achievable by linear schemes in the original problem. Therefore, the original problem can achieve at least a symmetric rate (DoF) of $\frac{\Delta'}{2\Delta'+1}$ per message. 

What remains to be shown is that $\frac{\Delta'}{2\Delta'+1}$ is also a symmetric capacity (DoF) outer bound for the original topological interference management problem. To show this, we now prove that the original problem cannot have a larger minimum conflict distance than the dual problem, i.e., $\Delta\leq \Delta'$.

To simplify the notation, let us make some observations. First, note that without loss of generality we can assume each destination desires only one message. If a destination desires multiple messages, one can create multiple copies of that destination (each connected to the same set of source nodes) such that each destination desires only one message. Next, since we are only interested in the alignment graph at this point, note that the alignment graph is not affected if we assume each message originates at a separate source. Again, if multiple messages originate at a source, we can create multiple copies of that source node (each connected to the same set of destinations), each of which is the source of only one message. Note that the latter assumption is not known to be without loss of generality for evaluating the capacity of a network. However, for the alignment and conflict graphs, which are our only concern right now, it involves no loss of generality, i.e., the alignment and conflict graphs are unaffected by this assumption. 

Based on these observations, we assume without loss of generality, that our $K$-unicast network has $K$ messages $W_1, W_2, \cdots, W_K$, originating at sources $S_1, S_2, \cdots, S_K$ and intended for destinations $D_1, D_2, \cdots, D_K$, respectively. In the dual network also there are $K$ messages,  $W_1, W_2, \cdots, W_K$, originating at sources $S_1', S_2', \cdots, S_K'$ and intended for destinations $D_1', D_2', \cdots, D_K'$, respectively. Recall that  source $S_i'$ in the dual network corresponds to destination $D_i$ in the original network and destination $D_j'$ in the dual network corresponds to source $S_j$ in the original network.

Without loss of generality, using the simplified notation, let us represent a minimum conflict distance path in the alignment graph of the dual problem  as follows.
\begin{eqnarray}
S_{j_1}'\stackrel{\small D_{j_2}'}{\longleftrightarrow} S_{j_3}'\stackrel{\small D_{j_4}'}{\longleftrightarrow} \cdots \stackrel{\small D_{j_{2\Delta'-2}}'}{\longleftrightarrow}S_{j_{2\Delta'-1}}'\stackrel{\small D_{j_{2\Delta'}}'}{\longleftrightarrow} S_{j_{2\Delta'+1}}'\label{eq:chaindual}
\end{eqnarray}
and
\begin{eqnarray}
t_{j_1j_{2\Delta'+1}}'&=&1
\end{eqnarray}
The chain in (\ref{eq:chaindual}) represents the edges in the alignment graph as follows:  There is an edge between $W_{j_1}$ and $W_{j_3}$ because they both cause interference at destination $D_{j_2}'$. There is an edge between $W_{j_3}$ and $W_{j_5}$ because both cause interference at destination $D_{j_4}'$, and so on. The conflict arises at the ends of the chain because destination $D_{j_1}'$ hears the source $S_{j_{2\Delta'+1}}'$ of the message $W_{j_{2\Delta'+1}}$. 

Translating back into the original network, this means that $W_{j_2}$ causes interference at destinations $D_{j_1}$ and $D_{j_3}$. $W_{j_4}$ causes interference at destinations $D_{j_3}$ and $D_{j_5}$, and so on. So, in the alignment graph of the original network there must be an edge between $W_{j_2}$ and $W_{j_4}$ because they are both heard by $D_{j_3}$, there must be an edge between $W_{j_4}$ and $W_{j_6}$ because they are both heard by $D_{j_5}$, $\cdots$, and finally (and most importantly) there must be an edge between $W_{j_{2\Delta'}}$ and $W_{j_1}$ because they are both heard by $D_{j_{2\Delta'+1}}$. There is also a conflict because $W_{j_2}$ is heard by $D_{j_1}$. This is represented as follows:

\begin{eqnarray}
S_{j_2}\stackrel{\small D_{j_3}}{\longleftrightarrow} S_{j_4}\stackrel{\small D_{j_5}}{\longleftrightarrow} \cdots \stackrel{\small D_{j_{2\Delta'-1}}}{\longleftrightarrow}S_{j_{2\Delta'}}\stackrel{\small D_{j_{2\Delta'+1}}}{\longleftrightarrow} S_{j_{1}}\label{eq:chain}
\end{eqnarray}
and
\begin{eqnarray}
t_{j_1j_{2}}&=&1
\end{eqnarray}
Thus we have an internal conflict in the original network of conflict distance $\Delta'$. The minimum internal conflict distance of the original network  is therefore bounded as $\Delta\leq \Delta'$. By Corollary \ref{theorem:slide}, this implies that the original network cannot have a symmetric capacity higher than $\frac{\Delta'}{2\Delta'+1}$.\hfill\QED

\subsection{Proof of Theorem \ref{theorem:onemimo}}\label{proof:onemimo}
The proof is almost identical to the proof of Theorem \ref{theorem:one}, so we only briefly summarize it here. We set the values of all channel matrices to be
\begin{eqnarray}
H_{ij}&=&\left(\sqrt{\mbox{SNR}\times\frac{\Gamma N_o}{P_j}}\right){\bf I}_{\Gamma}, ~~~~~\forall i\in\{1,2,\cdots, D\}, j\in\{1,2,\cdots, S\}
\end{eqnarray}
where ${\bf I}_\Gamma$ is the $\Gamma\times \Gamma$ identity matrix, and provide all messages $\mathcal{W}(S_j)$ such that $t_{ij}=0$ to destination $D_i$ so that it can remove the interference from $S_j$ introduced by the non-zero channel values. We then allow full cooperation between all sources and since the resulting channel is rank $\Gamma$, we replace it with a point to point MIMO Gaussian channel comprised of $\Gamma$ parallel channels of capacity $\log(1+S^2\mbox{SNR})$ each, whose transmitter has access to all messages and whose received signal is made available to all destinations. Using network equivalence theorem of \cite{Koetter_Effros_Medard} this MIMO link is replaced with a noise-free channel of capacity $\Gamma\log(1+S^2\mbox{SNR})$ which becomes the bottleneck link and the transformation to the index coding problem is complete. 

\subsection{Proof of Theorem \ref{theorem:linearmimo}}\label{proof:linearmimo}
To prove that the normalized DoF achievable in the SISO case are also achievable in the MIMO topological interference management problem, we assume a linear scheme exists for a SISO network and show how the same scheme can be applied over MIMO networks. Suppose precoding matrices ${\bf V}(W)$ and receiver combining matrices ${\bf U}_i(W)$ exist for the SISO setting that satisfy Property 1 (\ref{eq:timproperty1}) and Property 2 (\ref{eq:timproperty2}). We now describe the corresponding linear scheme for the MIMO setting.

Each message $W$ is split into $L(W)$ independent messages, $W(1), W(2), \cdots, W(L(W))$. For each $W(l)$, $X(W(l))$ is a $\Gamma \times 1$ vector containing $\Gamma$ symbols, that will be transmitted over $N$ channel uses along the precoding vector ${\bf V}(W(l))$ which represents the $l^{th}$ column of ${\bf V}(W)$.  Each $X(W(l))$ symbol is from an independent Gaussian codebook,  with power $\frac{P_j}{\Gamma|\mathcal{W}(S_j)|L(W)}$ where $W\in\mathcal{W}(S_j)$, and the columns of ${\bf V}(W)$ are scaled to have unit norm.

\noindent Over the $N$ channel uses, Source $S_j$ sends the $\Gamma N\times 1$ vector,
\begin{eqnarray}
{\bf X}_j&=&\sum_{W\in\mathcal{W}(S_j)}\sum_{l=1}^{L(W)}{\bf V}(W(l))\otimes{\bf X}(W(l)).
\end{eqnarray}
where $\otimes$ is the Kronecker product.
Destination $D_i$ receives the $\Gamma N\times 1$ vector,
\begin{eqnarray}
{\bf Y}_i&=&\sum_{j:t_{ij}=1}\sum_{W\in\mathcal{W}(S_j)}\sum_{l=1}^{L(W)}\left({\bf I}_N\otimes H_{ij}\right)({\bf V}(W(l))\otimes{\bf X}(W(l)))+{\bf Z}_i,\\
&=&\sum_{j:t_{ij}=1}\sum_{W\in\mathcal{W}(S_j)}\sum_{l=1}^{L(W)}{\bf V}(W(l))\otimes(H_{ij}{\bf X}(W(l)))+{\bf Z}_i,
\end{eqnarray}
and for each desired message $W'\in\mathcal{W}(D_i)\cap\mathcal{W}(S_j)$, $l'\in\{1,2,\cdots, L(W')\}$, projects the received signal vector ${\bf Y}_i$ into the ${\bf U}_i(W'(l'))\otimes {\bf I}_\Gamma$ space to obtain,
\begin{eqnarray}
\overline{{\bf Y}}_i(W'(l'))&=&({\bf U}_i(W'(l'))\otimes {\bf I}_\Gamma){\bf Y}_i\\
&=&\sum_{j:t_{ij}=1}\sum_{W\in\mathcal{W}(S_j)}\sum_{l=1}^{L(W)}({\bf U}_i(W'(l'))\otimes {\bf I}_\Gamma)({\bf V}(W(l))\otimes(H_{ij}{\bf X}(W(l))))\nonumber\\
&&+\underbrace{({\bf U}_i(W'(l'))\otimes {\bf I}_\Gamma){\bf Z}_i}_{\tilde{\bf Z}_i}\\
&=&\sum_{j:t_{ij}=1}\sum_{W\in\mathcal{W}(S_j)}\sum_{l=1}^{L(W)}({\bf U}_i(W(l')){\bf V}(W(l)))((H_{ij}{\bf X}(W(l))))+\tilde{\bf Z}_i\\
&=&H_{ij}{\bf X}(W'(l'))+\tilde{\bf Z}_i
\end{eqnarray}
where ${\bf U}_i(W'(l'))$ is the $l'^{th}$ row of the matrix
\begin{eqnarray}
[{\bf U}_i(W'){\bf V}(W')]^{-1}{\bf U}_i(W')
\end{eqnarray}
 the contributions from all other messages are eliminated  due to Property 1. Thus, the interference free $\Gamma\times\Gamma$ MIMO channel is available for each message, over which (normalized) $1/N$ DoF are achieved for each $W(l), l\in\{1,2,\cdots, L(W)\}$, so that $L(W)/N$ DoF are achieved per message $W\in\mathcal{W}$. 
\hfill\QED

\newpage

\bibliographystyle{IEEEtran}
\bibliography{Thesis}

\end{document}